\documentclass[twocolumn]{aastex631}
\usepackage{booktabs}
\usepackage{longtable}
\usepackage{tikz}
\usepackage{amsmath}
\DeclareSymbolFont{matha}{OML}{txm i}{m}{it}
\DeclareMathSymbol{\varv}{\mathord}{matha}{118}
\usepackage{graphicx}
\usepackage{appendix}

\usepackage{xcolor}

\newcommand{\Harvard}{Center for Astrophysics \textbar{} Harvard \& Smithsonian, 60 Garden Street, Cambridge, MA 02138-1516, USA}
\newcommand{\IAIFI}{The NSF AI Institute for Artificial Intelligence and Fundamental Interactions}
\newcommand{\Purdue}{Department of Physics and Astronomy, Purdue University, 525 Northwestern Avenue, West Lafayette, IN 47907-2036, USA}

\newcommand{\NI}{$\rm ^{56}Ni$}

\newcommand{\he}{$\rm ^{4}He$}
\newcommand{\mni}{$m_{\rm Ni}$}
\newcommand{\mhe}{$m_{\rm He}$}
\newcommand{\mop}{$m_{\rm bulk}$}
\newcommand{\velte}{$vel_{10}$}
\newcommand{\velt}{$vel_{25}$}
\newcommand{\velf}{$vel_{50}$}
\newcommand{\vels}{$vel_{75}$}
\newcommand{\veln}{$vel_{90}$}
\newcommand{\msun}{$\rm M_{\odot}$}

\newcommand{\mej}{$\rm m_{ejecta}$}

\newcommand{\sed}{$\texttt{SEDONA}$}
\newcommand{\Dp}{$\rm \eta_{vel}$}

\newcommand{\Rhp}{$\rm \eta_{He}$}
\newcommand{\Rnp}{$\rm \eta_{Ni}$}
\newcommand{\Ropp}{$\rm \eta_{bulk}$}

\newcommand{\gamn}{$r_{\rm 95}$}
\newcommand{\gamnh}{$r_{\rm 95, He}$}
\newcommand{\gamnn}{$r_{\rm 95, Ni}$}
\newcommand{\gamnop}{$r_{\rm 95, bulk}$}

\newcommand{\lc}{light curve}
\newcommand{\LCs}{Light Curves}
\newcommand{\lcs}{light curves}
\newcommand{\opacity}{bulk}

\newcommand{\tr}{$t_{rise}$}
\newcommand{\td}{$t_{decline}$}

\begin{document}

\title{Radiative Transfer Modeling of Stripped-Envelope Supernovae. I: A Grid for Ejecta Parameter Inference}

\author[0000-0002-0840-6940]{S.~Karthik~Yadavalli}
\affiliation{\Harvard}

\author[0000-0002-5814-4061]{V.~Ashley~Villar}
\affiliation{\Harvard}
\affiliation{\IAIFI}

\author[0000-0002-1633-6495]{Abigail Polin}
\affiliation{\Purdue}

\author[0000-0002-3352-7437]{S. E. Woosley}\affiliation{Department of Astronomy and Astrophysics, University of California, Santa Cruz, CA 95064, USA}

\author[0000-0001-7081-0082]{Maria R. Drout}\affiliation{David A. Dunlap Department of Astronomy and Astrophysics, University of Toronto, 50 St. George Street, Toronto, Ontario, M5S 3H4, Canada}

\author[0000-0001-7676-948X]{Miranda Pikus}
\affiliation{\Purdue}

\begin{abstract}
We present $1$,$800$ multiwavelength Type Ib/c supernovae light curve models obtained by running the radiation transport code Sedona and varying the mass distribution, velocity profile, and abundance ejecta profiles of helium star progenitors.
To create a flexible but physically-informed grid, we use autoencoders to construct a representation of ejecta profiles derived from stellar evolution models.
We present simulated nearest-neighbor multiband light curves matches to SN~1994I, SN~2007gr, and iPTF13bvn to demonstrate that realistic light curves can be generated in our grid. 
We show that the ejecta velocity distribution, in particular, strongly influences the \lc, while variation in \NI~mixing alone has a limited impact on the bolometric light curve, even in extreme and unphysical mixing schemes; however, mixing can modestly impact color evolution.
Finally, we show that the \NI~mass, ejecta mass, and both the magnitude and structure of ejecta velocity distribution can be inferred from the multiband light curves, enabling improved inference over widely used semi-analytical models. 

\end{abstract}

\section{Introduction}
\label{sec:intro}

Stripped-envelope supernovae (SESNe) arise from massive stars whose outer envelope has been at least partially stripped prior to explosion (see \citealt{Smartt2009} for a review). Hydrogen-free SESNe can be classified observationally as SNe~Ib (with spectroscopic helium signatures) or as SNe~Ic (without helium signatures), and are primarily powered by the decay of \NI. Some single-star progenitor systems have been proposed for SESNe \citep{Wheeler1985, Gaskell1986, Begelman1986, Woosley1993}, but interacting binary progenitors are likely more prevalent and persuasive \citep{nomoto1995evolution, podsiadlowski1992presupernova, Woosley1995, kobulnicky2007new}

SESNe, observationally, span a relatively large range of explosion energy, peak brightness, and \lc~shape, including subclasses such as broad-lined SNe~Ic (SNe~Ic-BL; \citealt{iwamoto1998, maeda2002explosive, modjaz2016}), SNe associated with long gamma-ray bursts (GRB-SNe; \citealt{dessart2012, Cano2013, modjaz2016}), superluminous SNe (SLSNe; \citealt{Dessart2012-Magnetar, Wang2015, Nicholl2017, Nicholl2021}), narrow-lined SNe~Ibn and SNe~Icn \citep{Maeda2022, Pellegrino2022, Davis2023}, ultrastripped SNe \citep{tauris2015ultra, de2018hot, sawada2022energy}, and luminous SNe \citep{Gomez2022}. Crucially, how much of the SESN \lc~diversity can be explained by \NI~radioactive decay (as opposed to other engines) is not currently known.



 
A frequently used technique for modeling the \lc~of SNe~Ibc is the semianalytical ``Arnett Model'' \citep{Arnett1982} which approximates the full radiative diffusion process through a one-zone, gray opacity model. This model explicitly assumes that the SN ejecta is in thermal equilibrium with energy inputted by \NI~decay or, equivalently, that the spatial dependence of the energy profile is fixed, and only the normalization changes with time (i.e., ``self-similarity''; see assumption four and Equation 13 of \citealt{Arnett1982}). 

Because \NI~mixing is caused by explosion asymmetry and by Rayleigh-Taylor instabilities \citep{dessart2012, dessart2015, dessart2016, woosley2019, ertl2020, Woosley2021}, there is no reason to expect the \NI~to be distributed such that the ejecta is continuously in thermal equilibrium (see \citealt{Khatami2019} for the drawbacks of the self-similarity assumption). This simplifying assumption causes the Arnett model to generally overestimate \mni~for SESNe \citep{dessart2015, dessart2016, Khatami2019}.

One-dimensional Monte Carlo radiative transfer (MCRT) codes offer better accuracy than semianalytical models while remaining computationally feasible for simulating populations of SN \lcs. MCRT codes track photon packets through ejecta described with a layer-by-layer prescription of velocity, density, and chemical abundances using up to $10^6$ atomic line transitions. As such, these codes provide more flexibility in describing ejecta profiles and allow for studies of influences of \NI~mixing in the ejecta. 

How the distribution of \NI~in the ejecta influences the \lc~has been studied extensively in literature using MCRT methods \citep{dessart2012, dessart2015, dessart2016, Khatami2019, woosley2019, ertl2020, Moriya2020, Woosley2021}. In particular, models with \NI~mixed farther into shallower ejecta layers are bluer prior to peak, have a more gradual rise, have an earlier peak, have a higher photospheric velocity during peak, and more rapidly transition to the nebular phase \citep{dessart2012, Moriya2020, Woosley2021}. However, to date many MCRT studies have been limited to relatively small ($\lesssim100$) model grids for a set of derived parameters. In coming years, as surveys like the Vera C. Rubin Observatory's Legacy Survey of Space and Time reveal large samples of SESN \lcs, broader grids are essential to constrain two important uncertainties: 1) the progenitor system(s) of SESNe and 2) the power source(s) of SESNe. Furthermore, such grids need to be embedded within inference routines to statistically infer SESN properties.  

Here, we use \sed~to systematically explore the physical parameters of a larger set of SESNe. Our models build on the works of \cite{woosley2019, ertl2020, Woosley2021}, who calculated a grid of 81 SESN \lcs~resulting from helium stars that were evolved from helium ignition to core collapse in \texttt{kepler} using the progenitor-agnostic mass-loss prescription from \cite{Yoon2017}. \cite{Woosley2021} (hereafter, W21) evolved the stellar models that successfully exploded using the MCRT code \sed~and presented relationships between generated \lcs~and the physical parameters of SESN ejecta. 

In this work, we generate ejecta profiles derived from the W21 grid, and examine how each SN parameter independently influences the observed \lc. We generate models by sampling from the distributions of physically relevant dimensions of the W21 models (e.g. \NI~mass, ejecta mass, and velocity gradient). We also motivate the next paper in this series, which will present an inference framework for SESN ejecta physical parameters using the results from MCRT. Because this work is observationally motivated, we present parameters and measurements that are easily translatable to observed \lcs, and we align all \lcs~so time zero is set to the time of $r$-filter peak. 

In Section~\ref{sec:sedona_exp}, we motivate the problem and describe our radiative transfer implementation. In Section~\ref{sec:lc_influences}, we present the influence of each physical parameter on the \lc. In Section~\ref{sec:grid}, we generate a representative grid of $1$,$000$ \lcs~sampled from the space of ejecta profiles. In Section~\ref{sec:Inference Machine}, we use the grid of $1$,$000$ \lcs~to infer SESN physical parameters from observed \lc~properties. In Section~\ref{sec:expanded_grid}, we generate another grid of $800$ \lcs~to explore the impact of \NI~mixing on the bolometric \lc. In Section~\ref{sec:conclusions}, we list our conclusions.

\section{Physical Motivation and Parameterization}
\label{sec:sedona_exp}
\subsection{Radiative Transfer Set Up}

We first describe the grid of models presented in W21. \cite{woosley2019} evolve a grid of mass-losing helium stars with masses at helium ignition of $1.6$\msun~to $120$\msun. The \cite{Yoon2017}  mass-loss prescription was used to simulate mass-loss during this phase until core collapse. Of the stellar models evolved in \cite{woosley2019}, helium stars with mass greater than $2.5$\msun~successfully exploded, resulting in pre-SN masses of $2$ to $60$\msun. Explosions which did not result as pulsational-pair instability SNe or as black holes were further simulated in the one-dimensional neutrino transport code PHOTB by \cite{ertl2020}. W21 then present the subsequent evolution of the explosions simulated in \cite{ertl2020} in the MCRT code \sed. The \mni~and explosion energy of each model in W21 is therefore self-consistently calculated for a neutrino-driven explosion. In the end, this set of models had helium star masses of $2.5$\msun-$40$\msun and pre-SN masses of $2.1$\msun-$19.6$\msun. Mixing of \NI~is simulated by using the ``boxcar averaging technique'', where the abundance profiles of all elements are convolved with a boxcar function with a set width out to a maximum velocity. 

In this work, we explore the space of SESN ejecta profiles by parameterizing and interpolating between the W21 model set. Following W21, we simulate all SESNe in the MCRT code \sed~\citep{Kasen2006}. \sed~ calculates the time-varying spectral energy distribution (SED) of expanding SN ejecta. \sed~constructs an SN ejecta profile from a layer-by-layer prescription of velocity, density, temperature, and chemical abundance. \sed~then evolves this system assuming homologous expansion (i.e. ejecta velocity is proportional to layer radius). The temperature structure is iteratively solved assuming thermal and radiative equilibrium from the given initial temperature profile estimate.

All models presented here are simulated in one dimension with a standard, limited set of chemical elements: 
$\rm ^{1}H, ^4He, ^{12}C, ^{14}N, ^{16}O, ^{20}Ne, ^{24}Mg, ^{28}Si, ^{32}S, ^{36}Ar, ^{40}Ca$,
$\rm ^{44}Ti, ^{48}Cr, ^{52}Fe, ^{56}Fe, ^{56}Co,~and~^{56}Ni$. As in W21, the radioactive decay of \NI~is the only relevant power source considered here (although a trivial amount of $\rm ^{48}Cr$ and $\rm ^{52}Fe$ is also tracked). Trace amounts of $\rm^1H$ are included in the radiative transfer modeling because trace amounts of $\rm^1H$ (the influence of which is negligible) are present immediately following the phase of stellar evolution and core collapse, even though these are SESN models. Finally, every model is simulated from initial time $t_{\rm min}~=4.0$~d until $t_{\rm max}~=90$~d with maximum timestep $dt = 0.2$~d. An SED is calculated over the frequency range $\nu_{\rm min} = 10^{13}~\rm Hz$ to $\nu_{\rm max}=10^{20}~\rm Hz$ sampled evenly in log-space by a spacing of $\log_{10} \left(1.002\right)$. We calculate \lcs~in Sloan Digital Sky Survey (SDSS) $ugriz$ filters for every model presented here \citep{SVO2012, SVO2020, SVO2024}\footnote{This research has made use of the Spanish Virtual Observatory (https://svo.cab.inta-csic.es) project funded by MCIN/AEI/10.13039/501100011033/ through grant PID2020-112949GB-I00.}. 

All models presented in this paper have been simulated under the local thermal equilibrium (LTE) assumption. Since the lowest energy excitation state of \he~is far higher than the typical blackbody energies produced in SESNe, modeling the influence of \he~on a SESN \lc~requires a non-local thermal equilibrium (NLTE) consideration of the radiation field. We do not expect NLTE effects to significantly alter the findings of this study, which focus on the broadband SED evolution.

We present the input parameter files, all \lcs, and SEDs for out grid on Zenodo \footnote{The Zenodo link will be provided when this manuscript is published.}. We note that $\sim10\%$ of the models simulated by \sed~developed nonphysical simulation artifacts in the generated \lc~at early timesteps. These artifacts result from a poor choice of initial conditions (e.g. initial temperature profile guess) or a poor choice of simulation parameters (e.g. particles per timestep, mass per simulated cell, or similar). The artifacts usually appear only as a brief early transient excess in the early part of the \lc~that quickly fade in subsequent timesteps as the simulation approaches the correct \lc~solution. As such, most of these artifacts do not impact any analysis presented here. A very small fraction ($\sim 1\%$) of the models developed artifacts that impacted analysis (usually because the early excess happens close to \lc~peak). These models chosen by eye and are removed from analysis in this study. We mark which models are contaminated by these unrecoverable artifacts in the data product listed above.

\subsection{Parameterizing Ejecta Profiles}
\label{subsec:Autoencoding}
Here we describe our parameterization of the distributions of SN ejecta mass, velocity, and abundances for input into \sed. To meaningfully explore the range of these distributions, the dimensionality of each needs to be reduced to only a few physically relevant parameters. Importantly, the \lc~produced by a lower-dimensionality encoding of an ejecta profile should be equivalent to the \lc~resulting from the original ejecta profile parameterization. 

\subsubsection{Elements Tracked}

We first explore which of the $17$ chemical element distributions need to be independently parameterized, i.e., identify which elements have nearly identical impacts on the SN light curves. Elements can influence the \lc~in two ways: 1) providing input energy via radioactive decay and 2) increasing the opacity of the ejecta. The radioactive decay of \NI~is the only power source considered here. All chemical elements contribute to opacity through the three sources of opacity we consider in this paper: 1) free-free interactions, 2) Thomson scattering, and 3) line-expansion opacity. Free-free interactions and Thomson scattering opacities are particularly enhanced at early times, when the ionization fraction tends to be higher. We isolate the individual contribution of each element's distribution to the \lc~by running multiple simulations where the mass distribution of each element is exchanged with another element (without changing the total ejecta mass) for several models in the W21 grid. We find all elements other than \NI~have a largely identical influence on the \lc. In the end, we independently parameterize the mass distributions of \he~and \NI, and we combine the mass distributions of all other elements into one: the ``bulk'' ejecta. We generate the bulk mass distribution by adding the average mass of each element that is not \he~or\NI~in each layer throughout the W21 grid. Though \he~does not influence the resulting \lc, we independently parameterize it as a point of comparison for a future study in which \he~interactions are accurately modeled in an NLTE implementation of \sed.

We confirm that parameterizing only the \he~mass distribution, \NI~mass distribution, and bulk mass distribution reproduces the \lcs~of models in the W21 grid (where the mass distributions of \textit{all} elements are parameterized). We find an absolute fractional difference of $\sim1\%$ in the bolometric \lc~between original model and the bulk-averaged model. We also find that the mean fractional difference in spectra from the original model and bulk-averaged model is $\sim5\%$ over the relevant wavelength range. However, we emphasize that the goal is to reproduce \lcs, rather than the detailed features in the spectrum; therefore, we do not inspect individual spectroscopic features. Finally, each mass and velocity distribution in this study is re-interpolated to have 256 layers, so that every model is parameterized with 256 cells in \sed.

\subsubsection{Mass and Velocity Distributions}
We next find a method to parameterize the shape of each mass and velocity distribution to further reduce the dimensionality of the ejecta space.  We find simple parametric models (e.g. Chebyshev polynomial expansion or similar) cannot reproduce the W21 model distributions without requiring an unacceptably large number of parameters. 

Instead, we encode mass and velocity distributions with a small set of parameters using autoencoders (AEs), a class of neural networks. A neural network is a machine learning algorithm that approximates a arbitrary function by using layers of linear combinations followed by simple, nonlinear activation functions. An AE is a kind of neural network that learns a compressed representation of the input set \citep{kramer1991}, where the space of compressed representation is known as the ``latent space''. An AE has an ``encoder'' portion where the input set is mapped into the latent space and a ``decoder'' portion where the latent representation is mapped to the original data space. 

We use four separate AEs to learn latent-space representations of the \he, \NI, bulk, and velocity distributions. All four AEs are constructed to have a latent dimensionality of one (i.e. the latent representation of each distribution is a scalar). The latent variable for all four distributions is constrained to the range $[0,1]$. Each AE has fully connected layers, with sigmoid activation functions throughout, and are trained for $20$,$000$ epochs using the adaptive gradient (Adagrad) optimizer at a learning rate of 0.01 with the mean squared error loss function. The AE architectures are:
\begin{itemize}
    \item Velocity AE: 256 $\rightarrow$ 512 $\rightarrow$ 16 $\rightarrow$ 1 $\rightarrow$ 16 $\rightarrow$ 64 $\rightarrow$ 128 $\rightarrow$ 512 $\rightarrow$ 256
    \item \he~Mass AE: 256 $\rightarrow$ 512 $\rightarrow$ 128 $\rightarrow$ 128 $\rightarrow$ 1 $\rightarrow$ 128 $\rightarrow$ 128 $\rightarrow$ 512 $\rightarrow$ 256
    \item \NI~Mass AE: 256 $\rightarrow$ 512 $\rightarrow$ 128 $\rightarrow$ 128 $\rightarrow$ 128 $\rightarrow$ 1 $\rightarrow$ 128 $\rightarrow$ 128 $\rightarrow$ 128 $\rightarrow$ 512 $\rightarrow$ 256
    \item \opacity~Mass AE: 256 $\rightarrow$ 512 $\rightarrow$ 16 $\rightarrow$ 1 $\rightarrow$ 16 $\rightarrow$ 64 $\rightarrow$ 128 $\rightarrow$ 512 $\rightarrow$ 256
\end{itemize}
where each number represents the number of neurons (or input/output values) in each layer.
The choices of AE architecture (i.e. number of layers and number of neurons per layer), activation function, number of epochs, learning rate, and optimizer were all made after a search over a grid of possible combinations of these ``hyperparameters''. 

While the AE captures the \textit{geometry} of these profiles, we still need to provide an overall scaling. For mass, we do so by multiplying the outputted values from the mass AE by a constant such that the integrated mass is equal to the prescribed total mass. Output distributions from the velocity AE are scaled by two extra parameters: the inner velocity of the ejecta profile and $\Delta$~vel, the difference between the inner and outer velocity of the ejecta profile. 

\subsubsection{Summary of Nine Ejecta Parameters Used}
We use nine dimensions to parameterize each ejecta profile (see Table~\ref{table:novenheim}). We describe the learned latent spaces of the four AEs in the following paragraphs. We choose the simulation range of each of the parameters to loosely represent the W21 range for that parameter. We note here that our upper limit  for \mni, $0.25$~\msun, is larger than the \mni~upper limit in \citep{ertl2020} and W21. As a result, some of the simulated \lcs~presented here have peak bolometric luminosity $>10^{42.6}$~erg/s, putting them outside the physically validated range from \cite{ertl2020}. The interpretation of models here that use \mni~in excess of $0.14$\msun and luminosities greater than  $10^{42.6}$~erg/s must therefore be considered ambiguous,
pending further study. The high inferred \NI~mass may be an indicator of additional central engine
activity (e.g. a magnetar) or an incomplete basis set from W21. Our overly luminous models are still useful for making a more accurate estimate for radioactive material required to power high-luminosity events, absent other engines.

\begin{figure*}[t]
    \centering
    \includegraphics[width=0.35\linewidth]{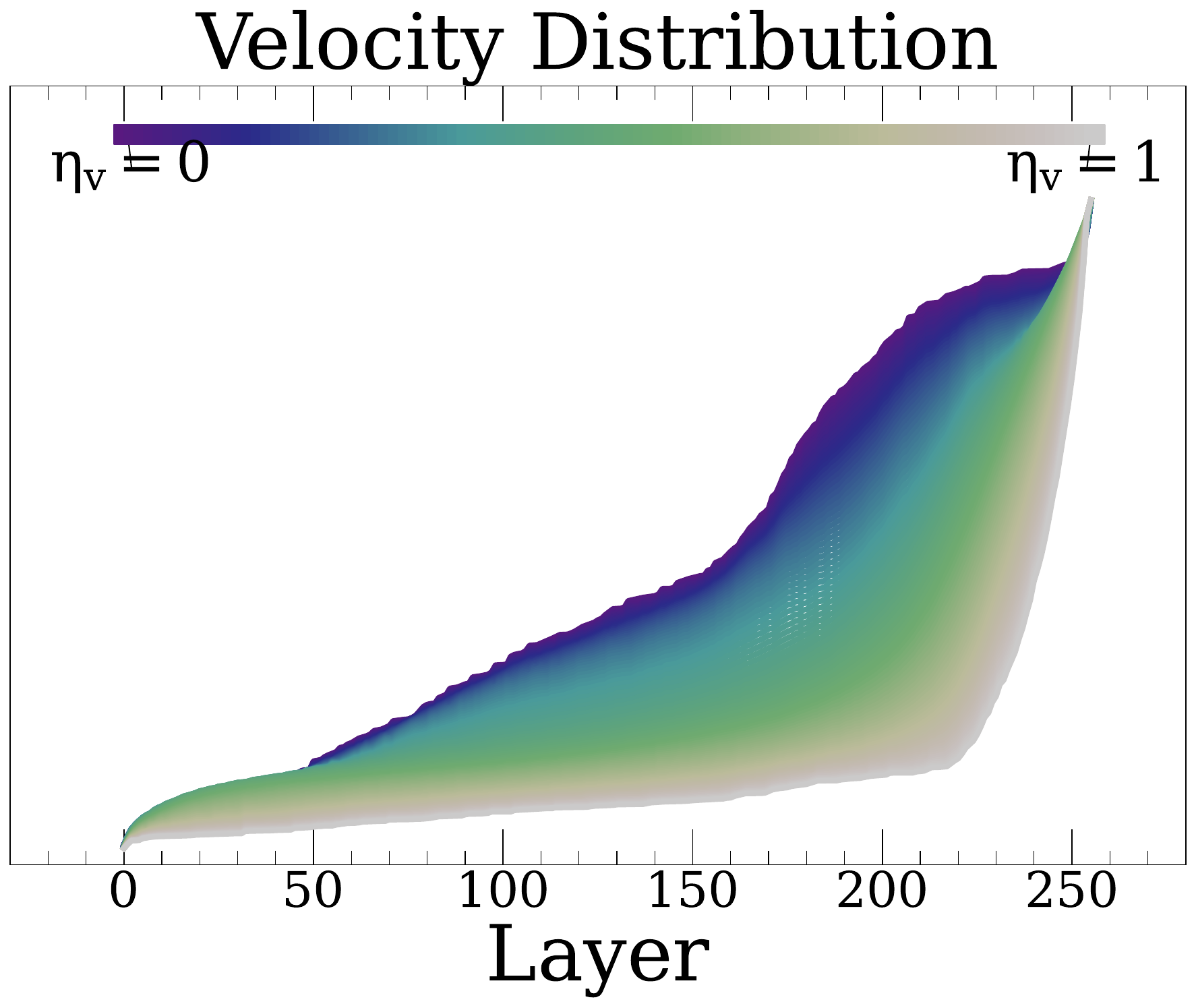}
    \hspace{0.5cm} 
    \includegraphics[width=0.4\linewidth]{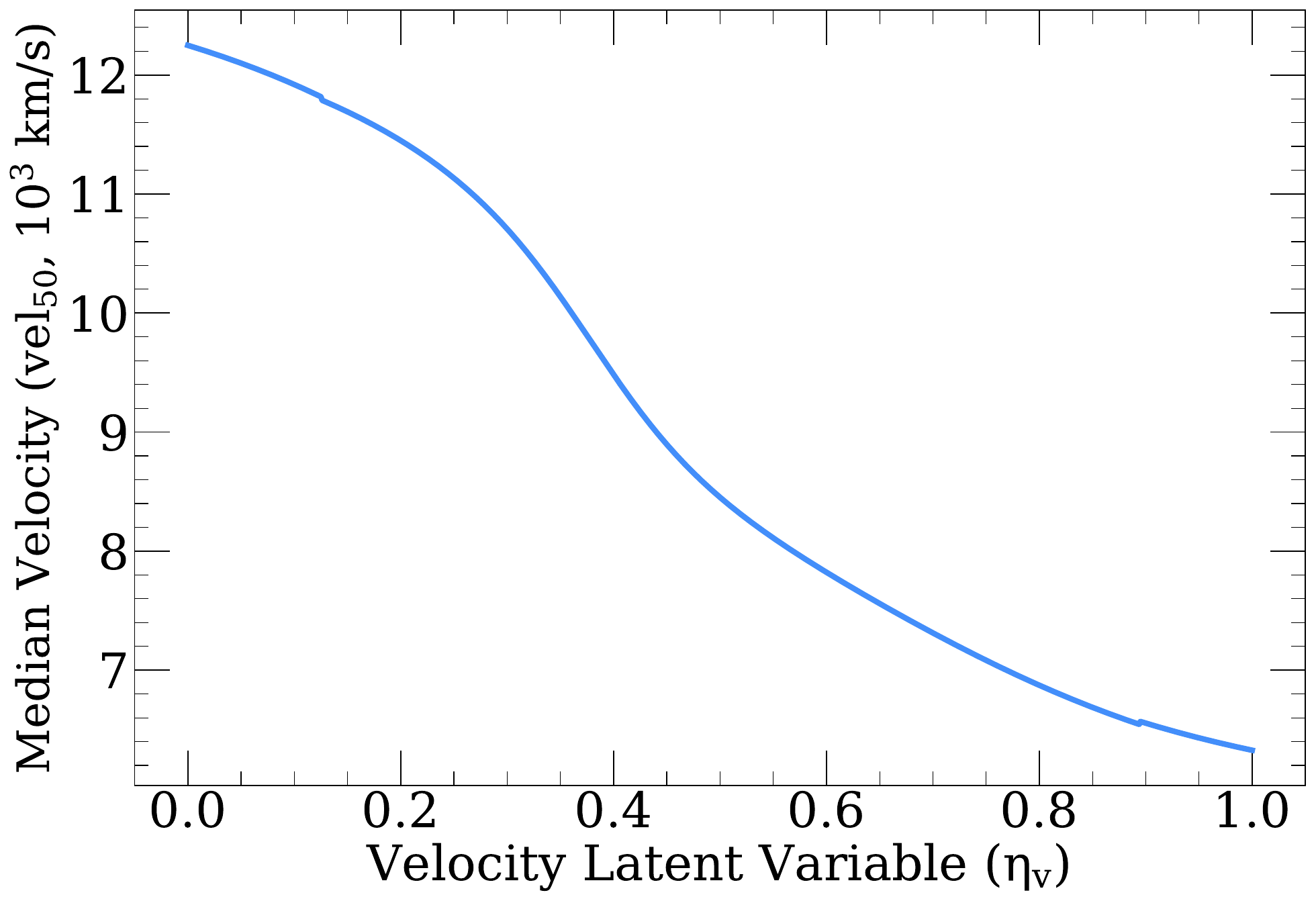}
    \includegraphics[width=0.35\linewidth]{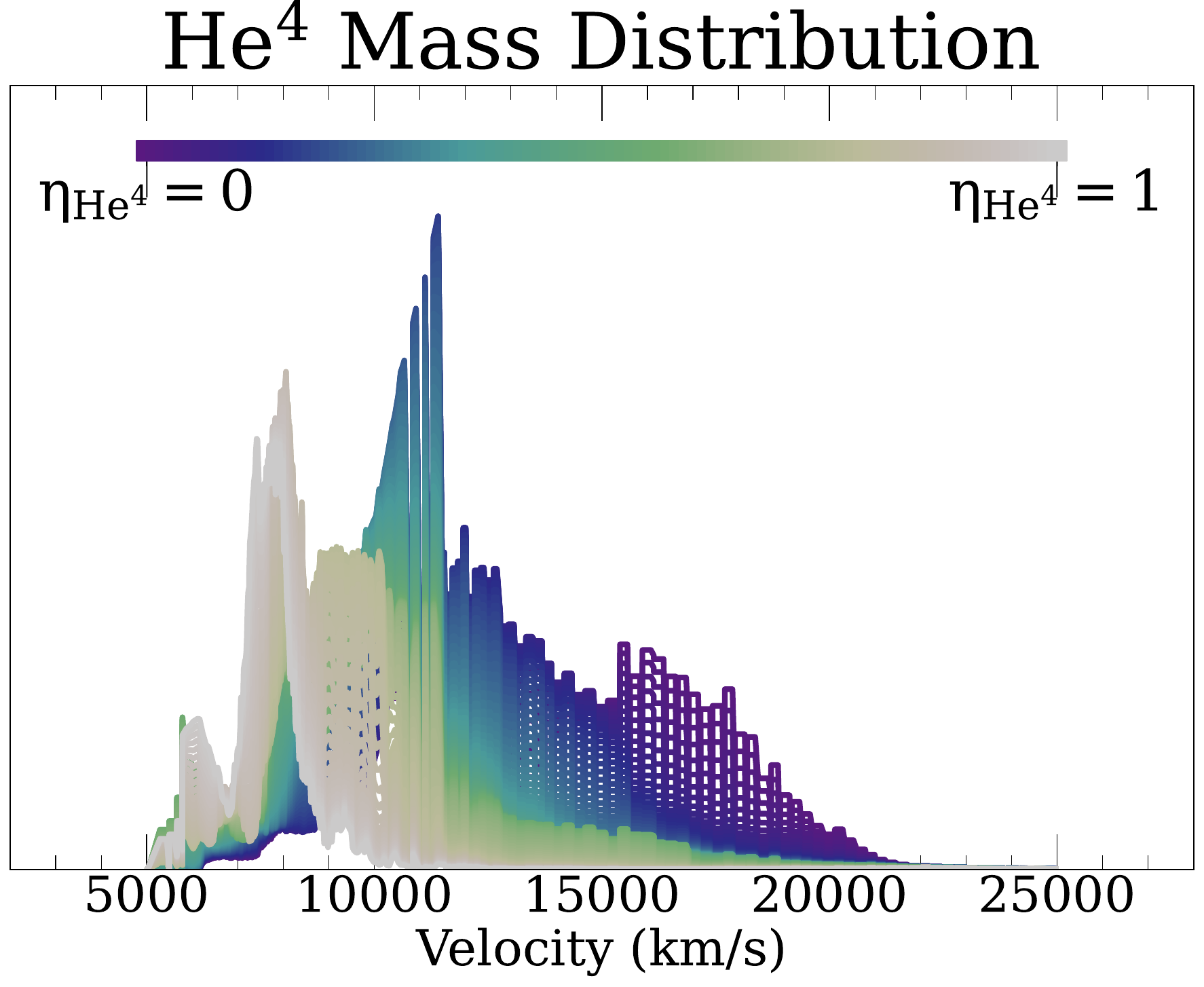}
    \hspace{0.5cm} 
    \includegraphics[width=0.4\linewidth]{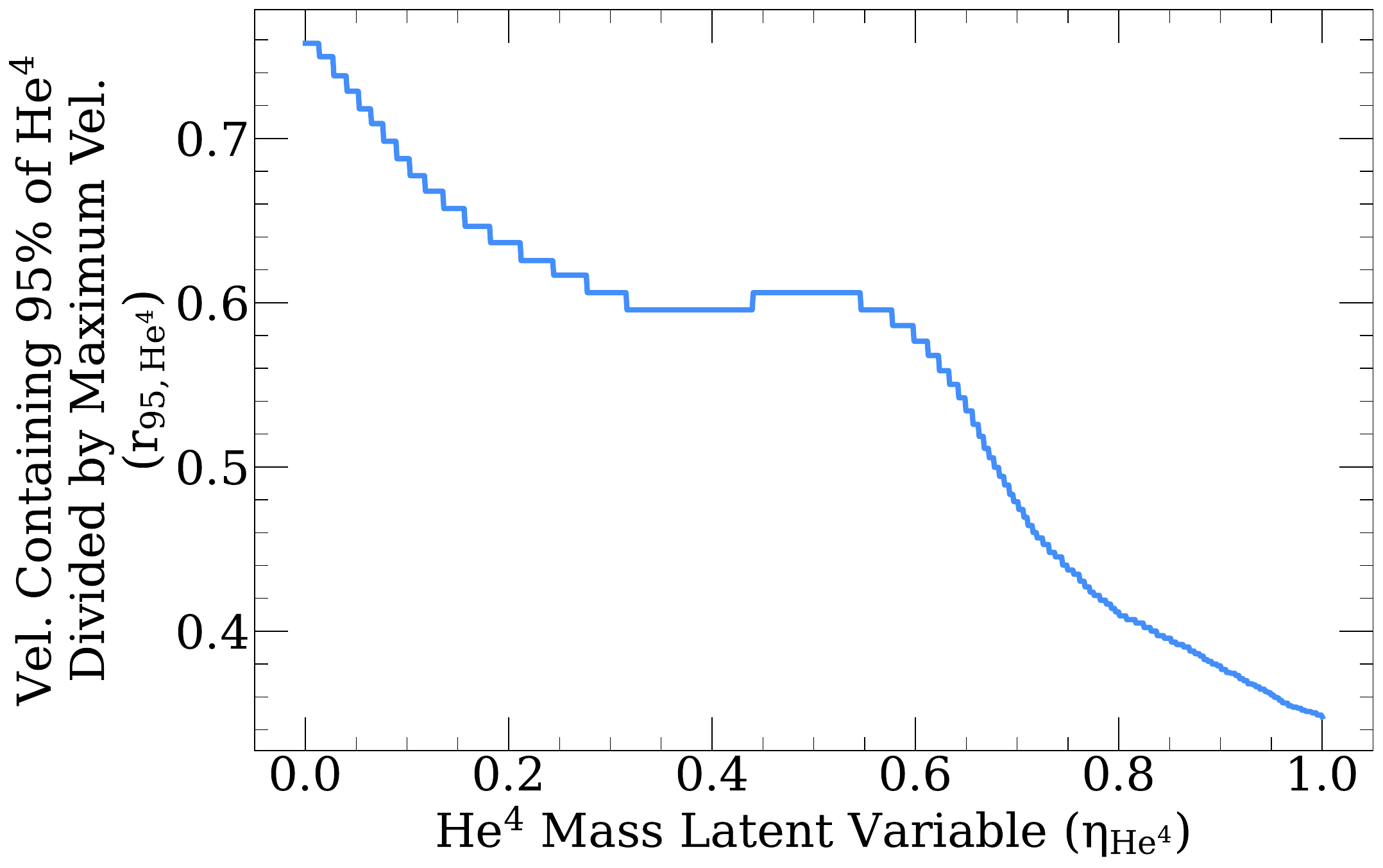}
    \includegraphics[width=0.35\linewidth]{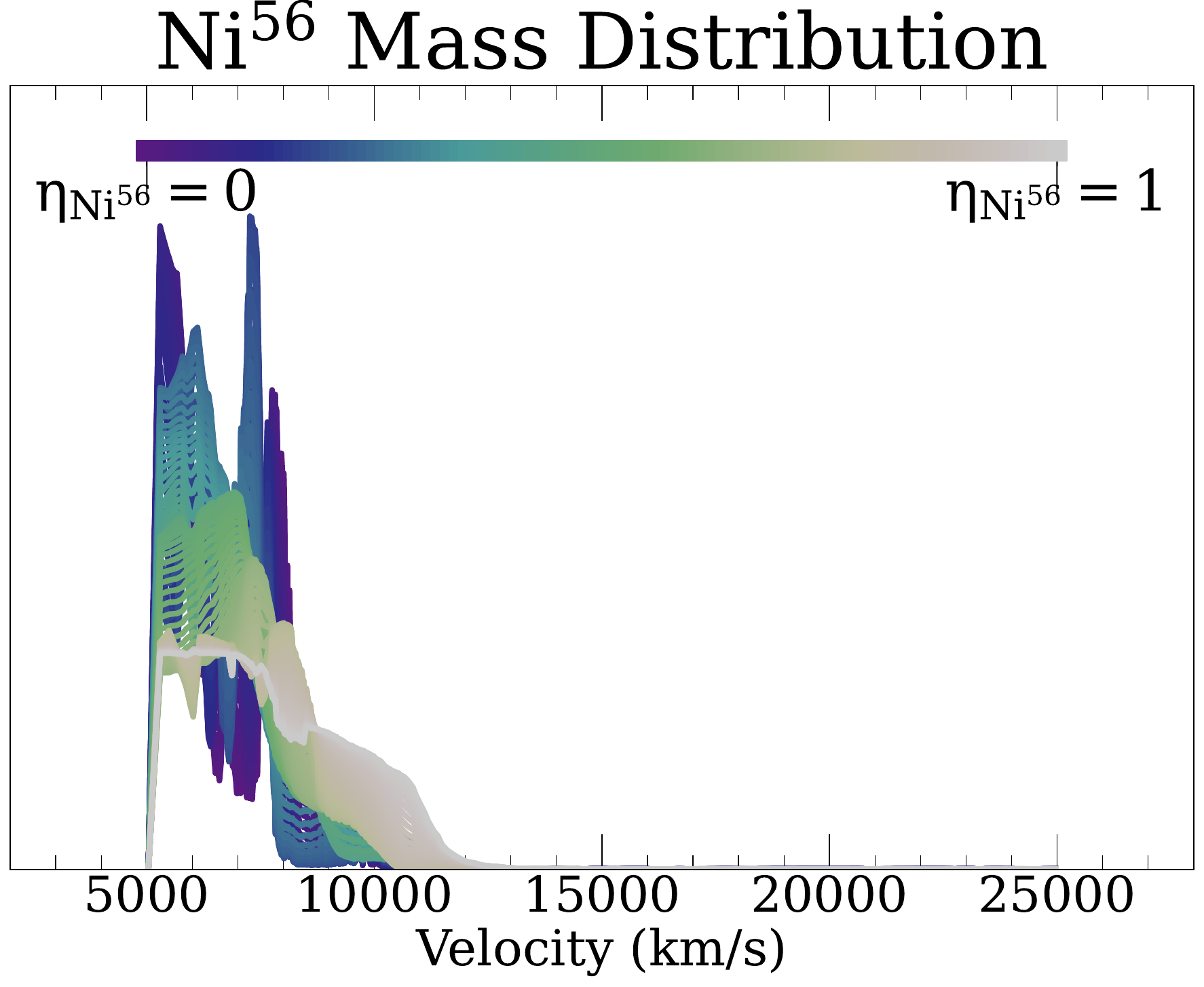}
    \hspace{0.5cm} 
    \includegraphics[width=0.4\linewidth]{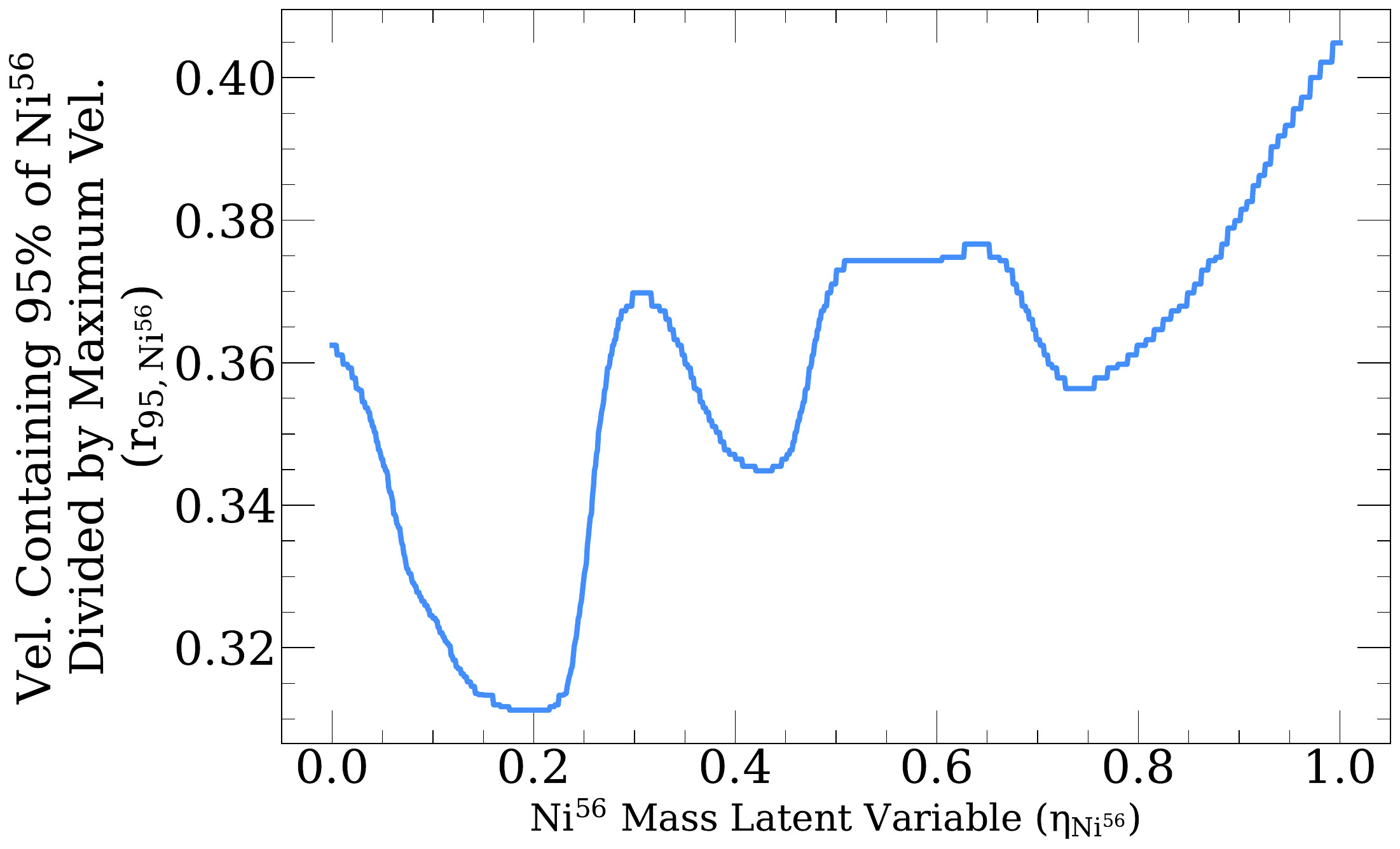}
    \includegraphics[width=0.35\linewidth]{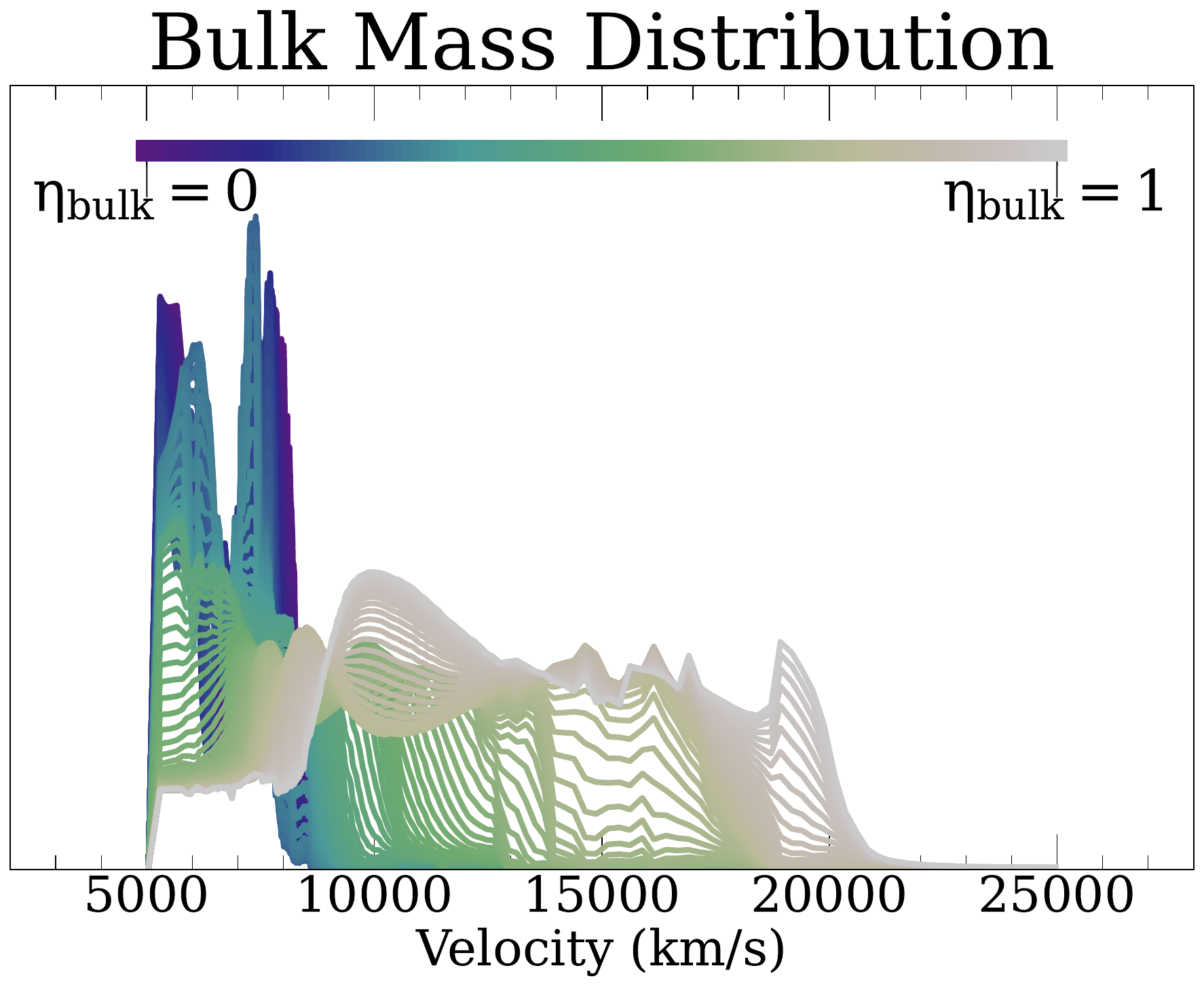}
    \hspace{0.5cm} 
    \includegraphics[width=0.4\linewidth]{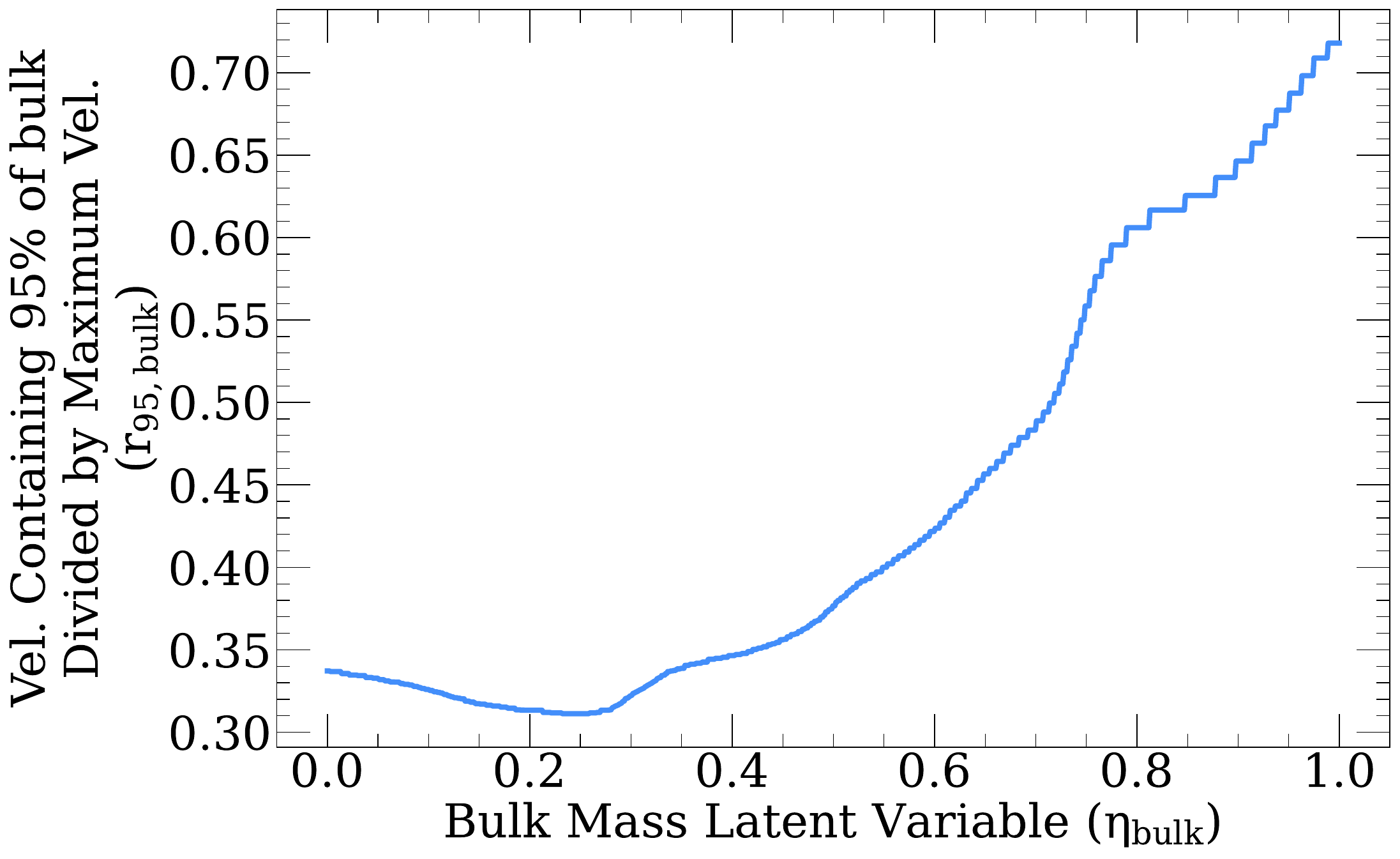}
    \caption{Range of velocity profiles (top left), \he~mass distributions (second from top, left), \NI~mass distributions (third from top, left), and \opacity~mass distributions (bottom left) from the W21 grid recreated by the corresponding autoencoder as the corresponding latent variable (\Dp~for velocity, \Rhp~for \he, \Rnp~for \NI, \Ropp~for \opacity) ranges from 0 to 1. The interpretable variable of each distribution (\velf, the median velocity, for the velocity distributions; and \gamn, the velocity that contains $95\%$ of the mass divided by the outer velocity for each mass distribution) is shown as a function of each latent variable in the right column. }
    \label{fig:Varying_Autoencoder}
\end{figure*}

In the top row of Figure~\ref{fig:Varying_Autoencoder}, we show the range of velocity distributions that we parameterize. We denote the latent variable of the velocity distribution as \Dp. At \Dp~$\simeq0$, the velocities increase smoother and gradually. As \Dp~increases to one, the profile becomes sharper in the outer layers and flatter in the inner layers.  We use the median ejecta velocity (\velf, top right panel of Figure~\ref{fig:Varying_Autoencoder}), as an interpretable analog for \Dp~to describe the velocity profile. When used in combination with the minimum and maximum velocities, \velf~characterizes how steeply velocity varies within the ejecta. As \Dp~increases, the inner velocity profile becomes flatter and the outer velocity profile becomes steeper, resulting in a monotonic decrease of \velf.

We denote the latent variables of the mass distributions of \he, \NI, and \opacity~as \Rhp, \Rnp, and \Ropp~respectively. Figure~\ref{fig:Varying_Autoencoder} shows a range of \he~(second row), \NI~(third row), and \opacity~(bottom row) mass distributions as \Rhp, \Rnp,and \Ropp~range from zero to one. As \Rhp~increases, the helium mass distribution moves monotonically inwards. In contrast with \he, the shapes of the \NI~and bulk distributions do not transform monotonically with their latent variables. As \Rnp~increases from zero to $\approx0.7$, the shape of the \NI~mass distribution moves inwards and outwards, oscillating three times. Beyond \Rnp~$\approx 0.7$, the \NI~mass distribution diffuses outwards. As \Rnp~increases, the \NI~mass distribution becomes smoother and flatter at the inner layers. Similarly, increasing \Ropp~from zero to $\sim0.3$ moves the opacity distribution inwards, after which increasing \Ropp~causes the opacity mass distribution to move outwards. As an interpretable parameter to describe these distributions, we introduce the value \gamn: the velocity that contains $95\%$ of the mass of a specific element divided by the maximum (outer) velocity of that model. The values \gamnh, \gamnn, and \gamnop~describe the extent to which \he, \NI, and \opacity~are mixed in the ejecta, respectively. We show how \gamnh~varies as a function of \Rhp, how \gamnn~varies as a function of \Rnp, and how \gamnop~varies as a function of \Ropp~in the second right, third right, and fourth right panels of Figure~\ref{fig:Varying_Autoencoder}, respectively. We present the range of values over our grid of \sed~models for each of our nine dimensions in Table~\ref{table:novenheim}.

We note here that other methods of parameterizing the \NI~mass distribution exist. For example, many studies evolve a range of stellar models and then generate novel extents of \NI~mixing by convolving a boxcar function (of some defined width in velocity or mass space) with the generated ejecta mass profile to imitate the effects of mixing due to Rayleigh-Taylor instabilities \citep{dessart2012, dessart2015, dessart2016, woosley2019, ertl2020, Woosley2021}. Other studies use analytical equations to prescribe more simplistic \NI~mass distributions \citep{Khatami2019, Moriya2020}. In this paper, we use \gamn~to parameterize the extent of mixing to maintain both 1) the relatively complicated \NI~mass distributions resulting from stellar evolution codes and 2) simple yet robust interpretability of the extent of \NI~mixing with just one parameter.

To characterize the precision of models generated by our AEs, we calculate the difference between \lcs~generated from encoded/decoded models and \lcs~from the W21 grid. We pass the velocity and mass profiles from models in the W21 grid through the corresponding AEs and then simulate \lcs~from the AE-reconstructed mass and velocity profiles. We then calculate the average of the fractional difference of the bolometric \lc~from the decoded ejecta profile and the \lc~from the W21 grid at all times. This fractional difference is less than $1\%$ for every model in the W21 grid, implying that the AE method of encoding ejecta profiles is sufficiently accurate for our purposes.

\begin{deluxetable*}{lll}
\tabletypesize{\footnotesize}
\tablecolumns{3}
\tablecaption{Description of the nine dimensions that parameterize the space of our simulated SESN ejecta profiles and the four interpretable parameters corresponding to the distributions. \label{table:novenheim}}
\tablehead{ \colhead{Parameter} & \colhead{Description} & \colhead{Simulation Range}}
\startdata
\Dp & Velocity Autoencoder Latent Variable & [0,1]\\
$\rm vel_{min}$ & Ejecta Profile Inner Velocity & [$150~\rm km/s$, $3000~\rm km/s$]\\
$\rm \Delta$~vel & Difference Between Outer Velocity and Inner Velocity & [$8000~\rm km/s$, $35~000~\rm km/s$]\\
\Rhp & \he~Distribution Autoencoder Latent Variable & [0,1]\\
\mhe & Total \he~Mass & [$2.5\times10^{-2}~\rm M_\odot$, $1~\rm M_\odot$] \\
\Rnp & \NI~Distribution Autoencoder Latent Variable & [0,1]\\
\mni & Total \NI~Mass & [$4\times10^{-3}~\rm M_\odot$, $0.25~\rm M_\odot$] \\
\Ropp & \rm ``Bulk'' Distribution Autoencoder Latent Variable & [0,1]\\
\mop & Total Bulk Mass & [$6\times10^{-3}~\rm M_\odot$, $8~\rm M_\odot$] \\
\hline
\velf & Velocity Distribution Interpretable Variable & [$690~\rm km/s$, $3050~\rm km/s$]\\
 & (= $50$th Percentile Velocity) & \\
 \gamnh & \he~Distribution Interpretable Variable & [0,1]\\
& (= Velocity Containing $95\%$ of \he~Mass Divided by Outer Velocity) & \\
\gamnn &  \NI~Distribution Interpretable Variable & [0,1]\\
& (= Velocity Containing $95\%$ of \NI~Mass Divided by Outer Velocity) & \\
\gamnop & ``Bulk'' Distribution Interpretable Variable & [0,1]\\
& (= Velocity Containing $95\%$ of Bulk Mass Divided by Outer Velocity) & \\
\enddata
\end{deluxetable*}

\section{Ejecta Profile Influence on Light Curves}
\label{sec:lc_influences}
\begin{figure*}[t]
    \centering
    \includegraphics[width=0.32\linewidth]{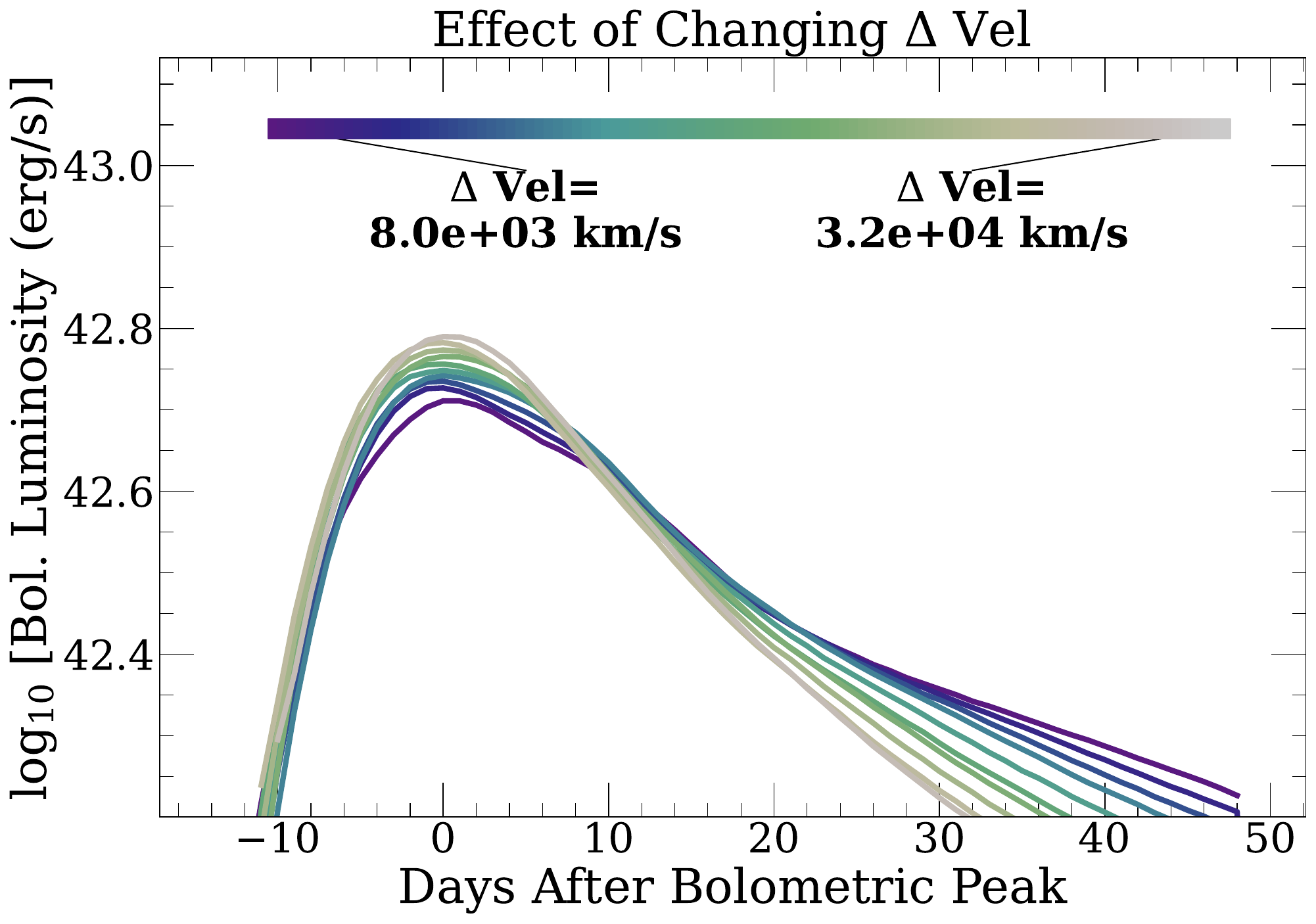}
    \includegraphics[width=0.32\linewidth]{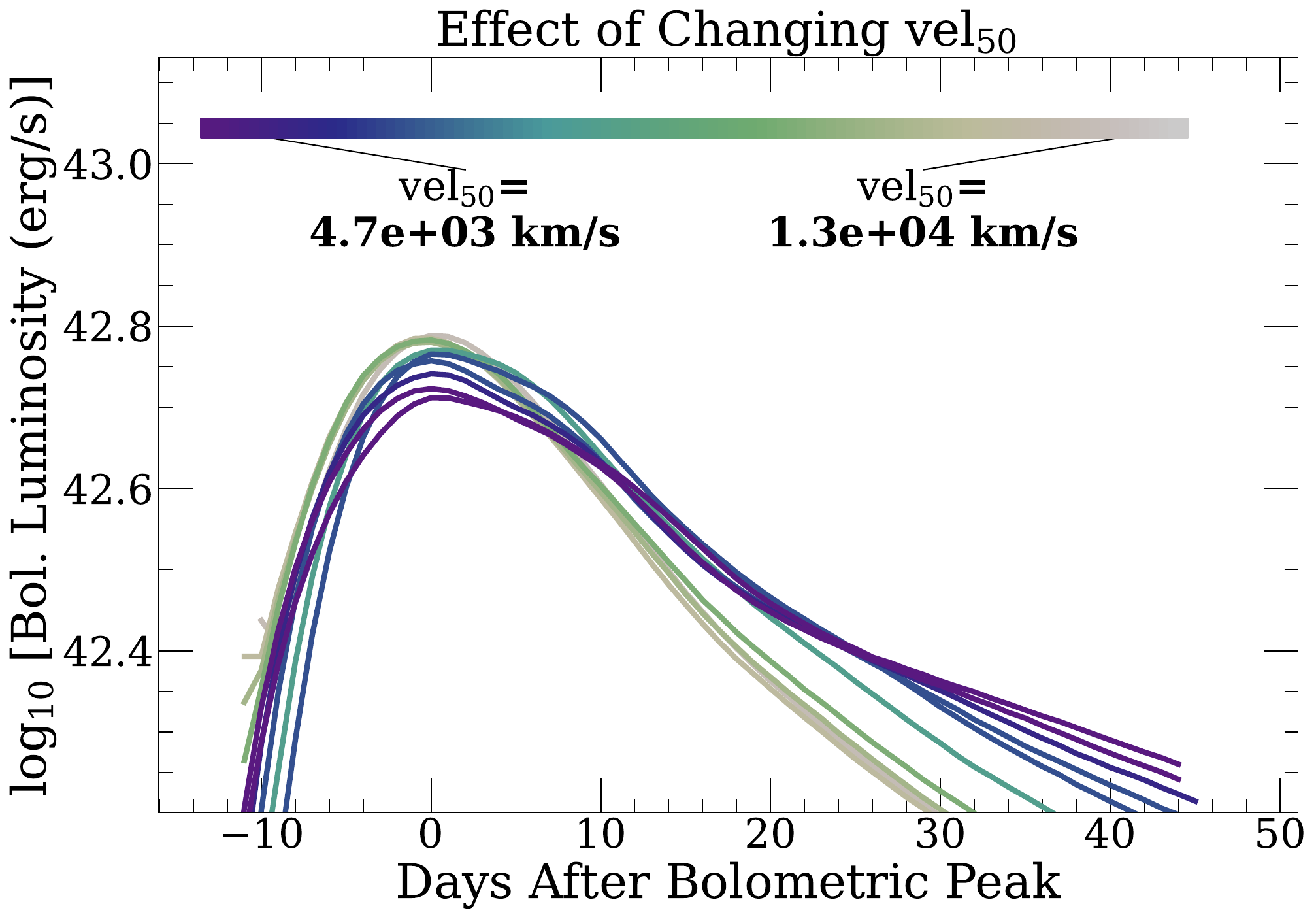}
    \includegraphics[width=0.32\linewidth]{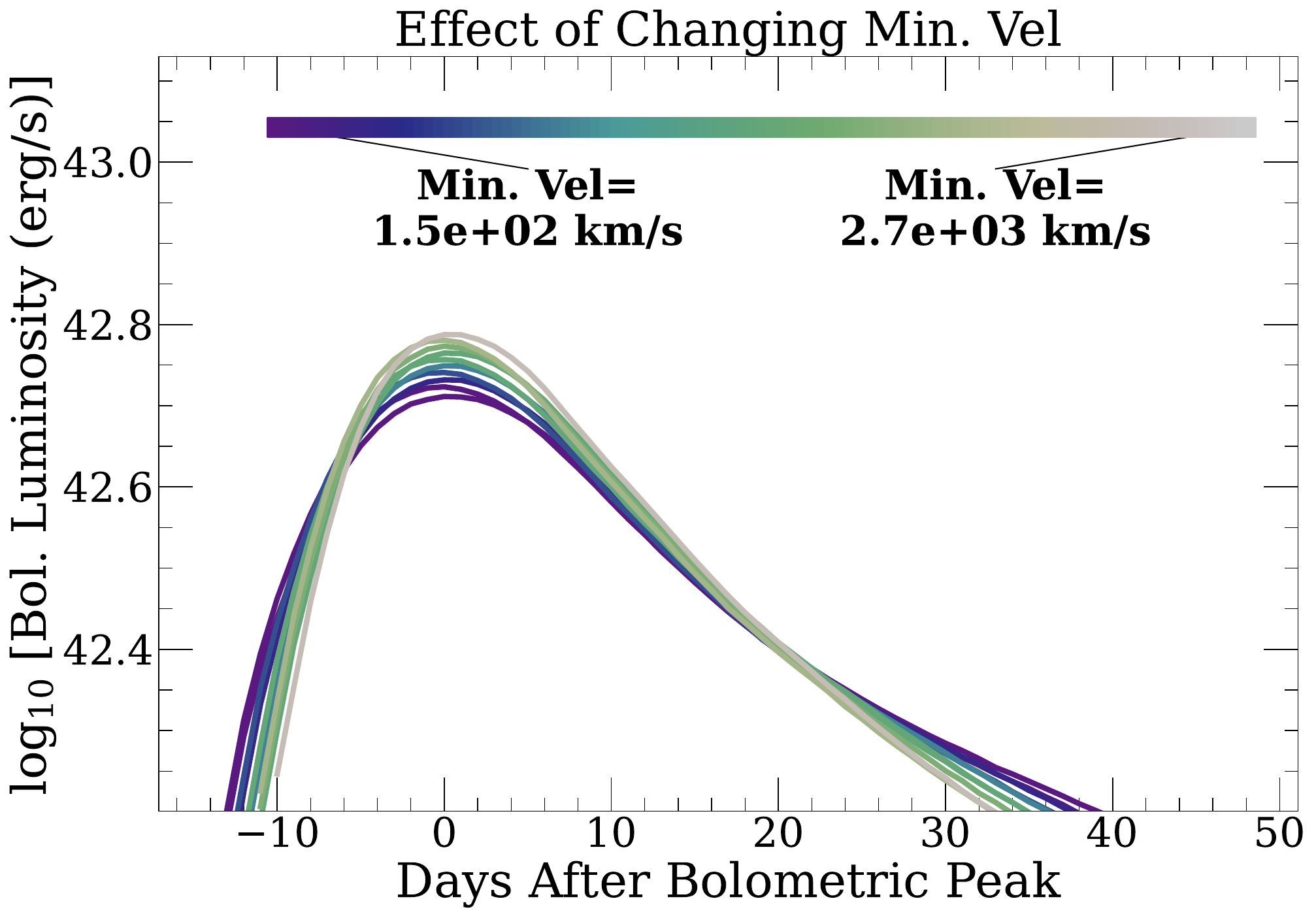}
    \includegraphics[width=0.32\linewidth]{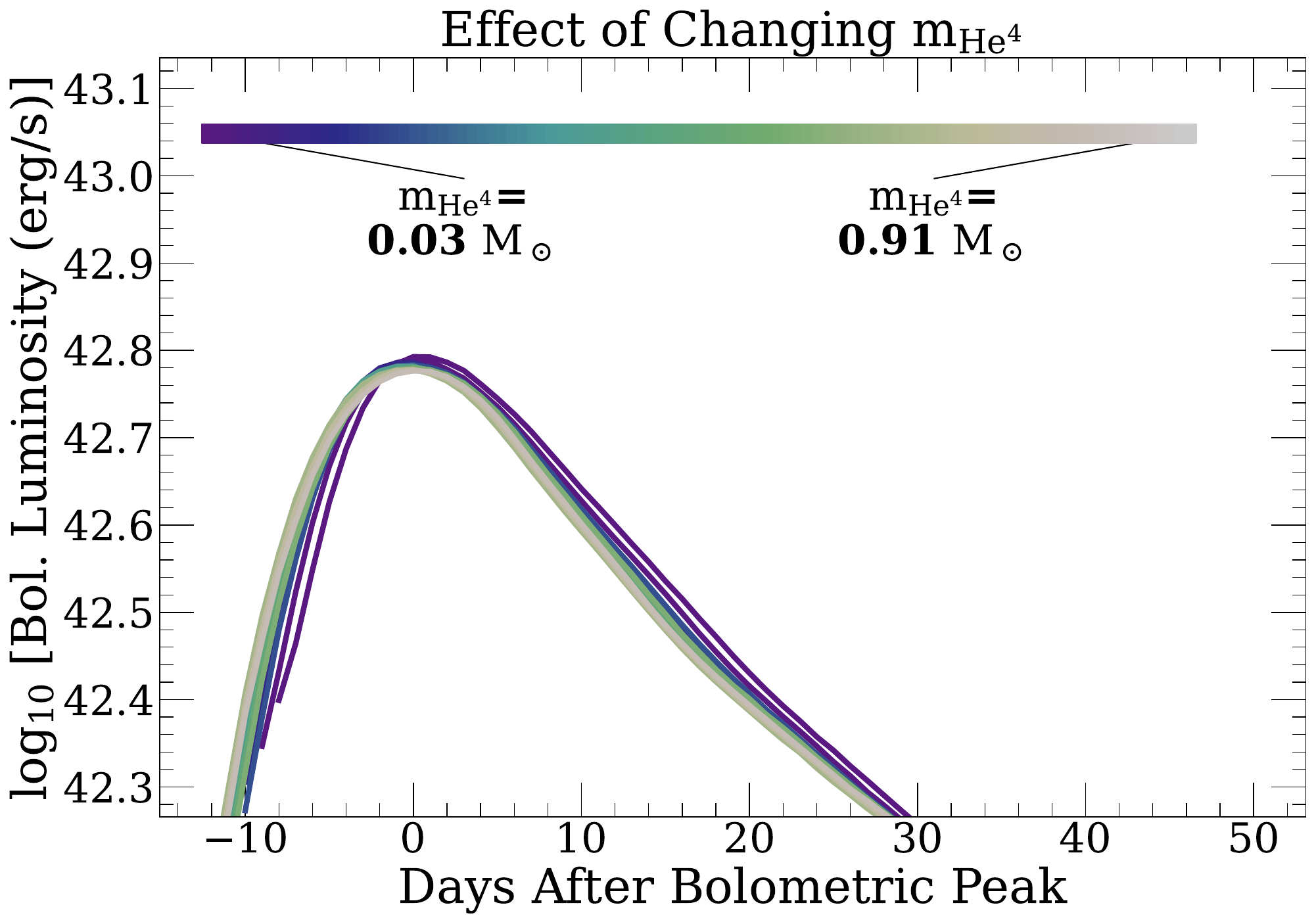}
    \includegraphics[width=0.32\linewidth]{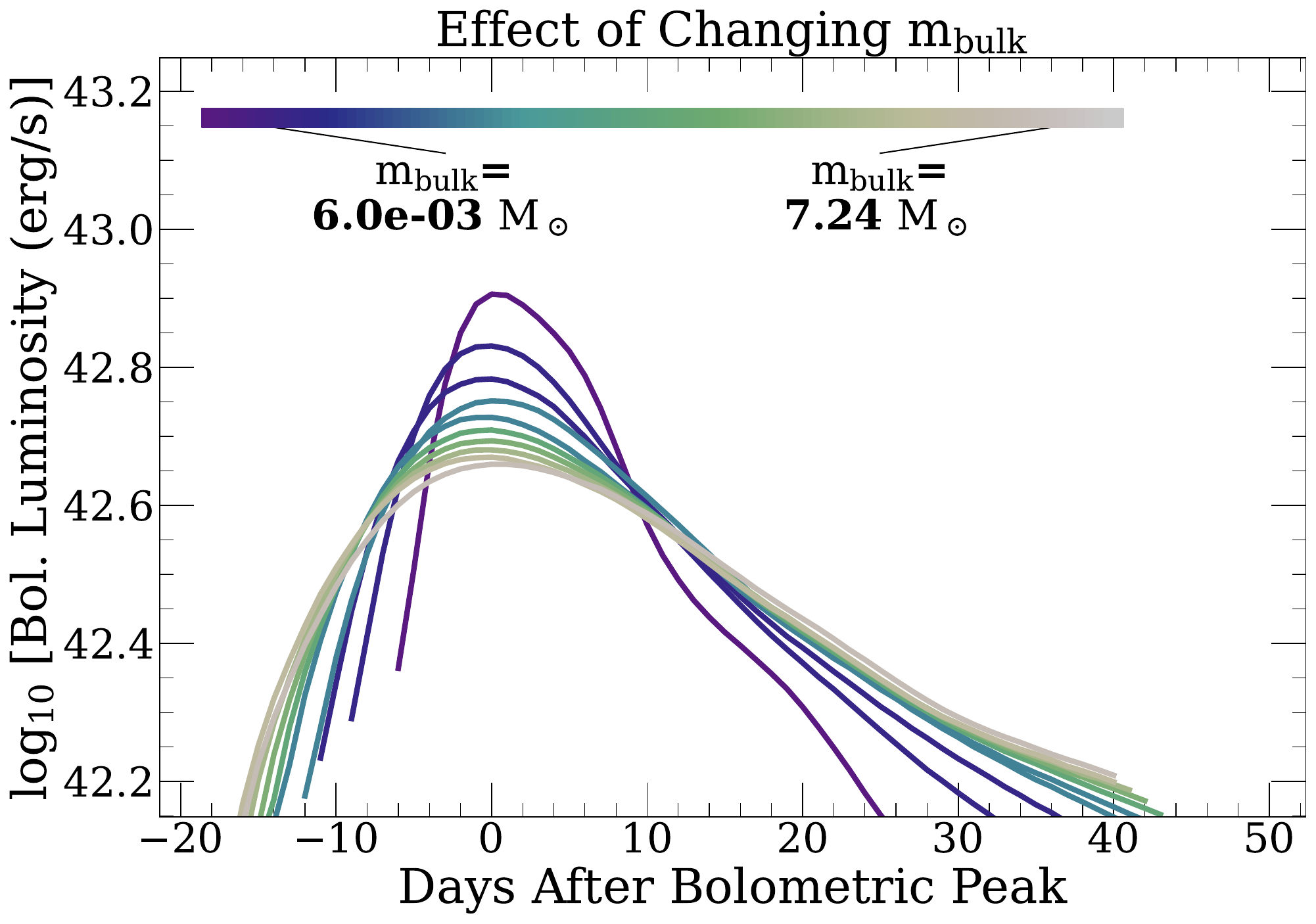}
    \includegraphics[width=0.32\linewidth]{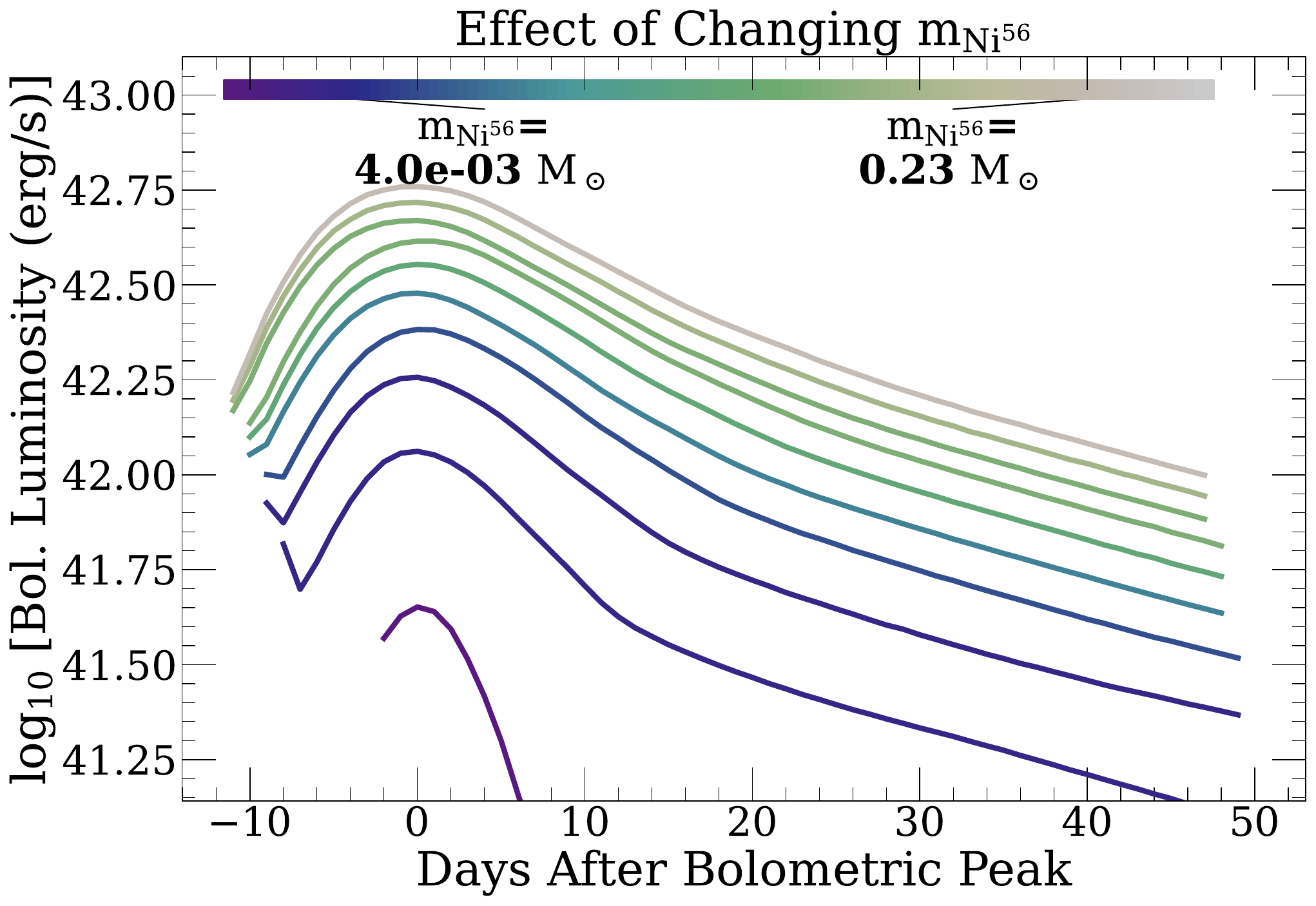}
    \includegraphics[width=0.32\linewidth]{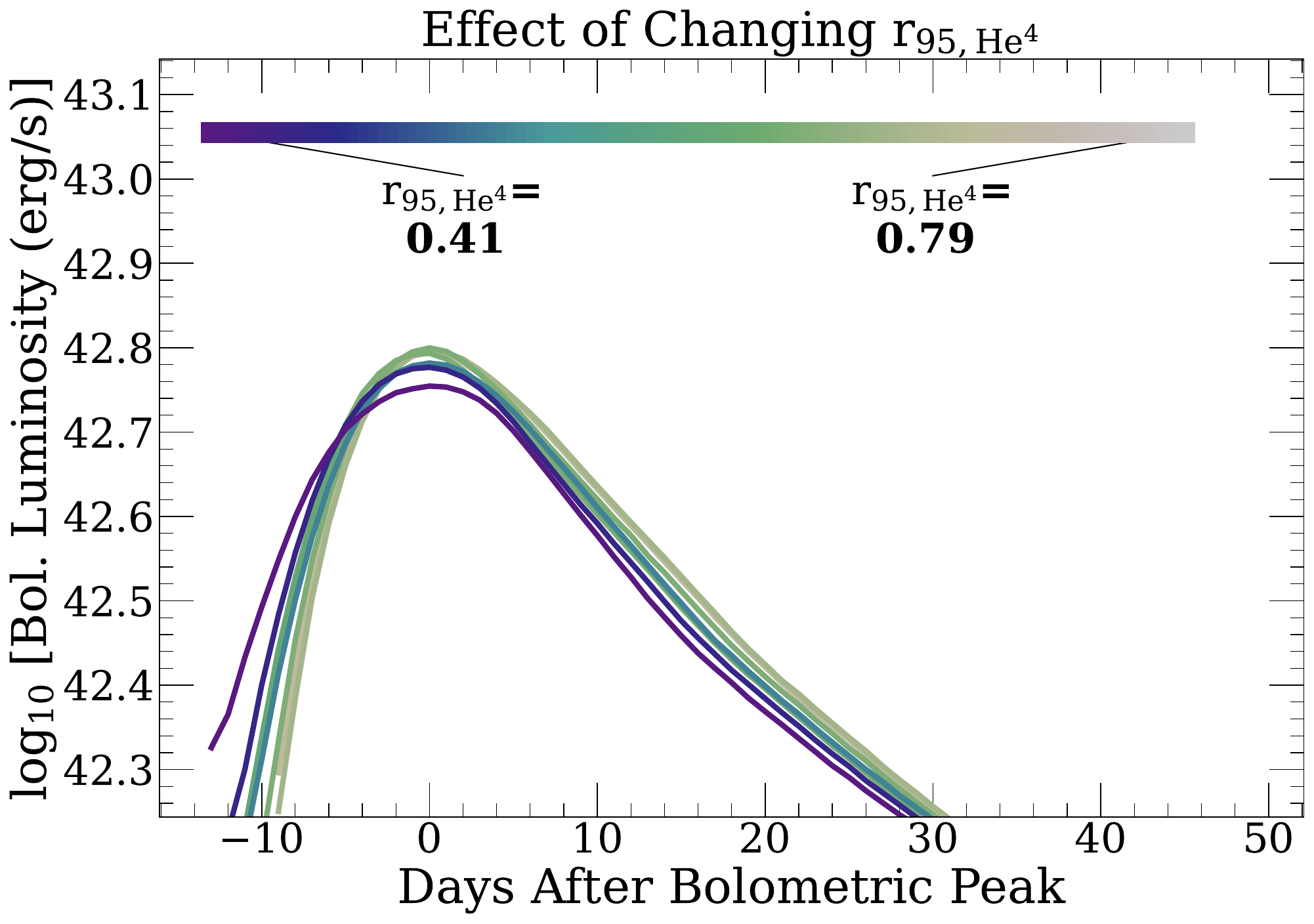}
    \includegraphics[width=0.32\linewidth]{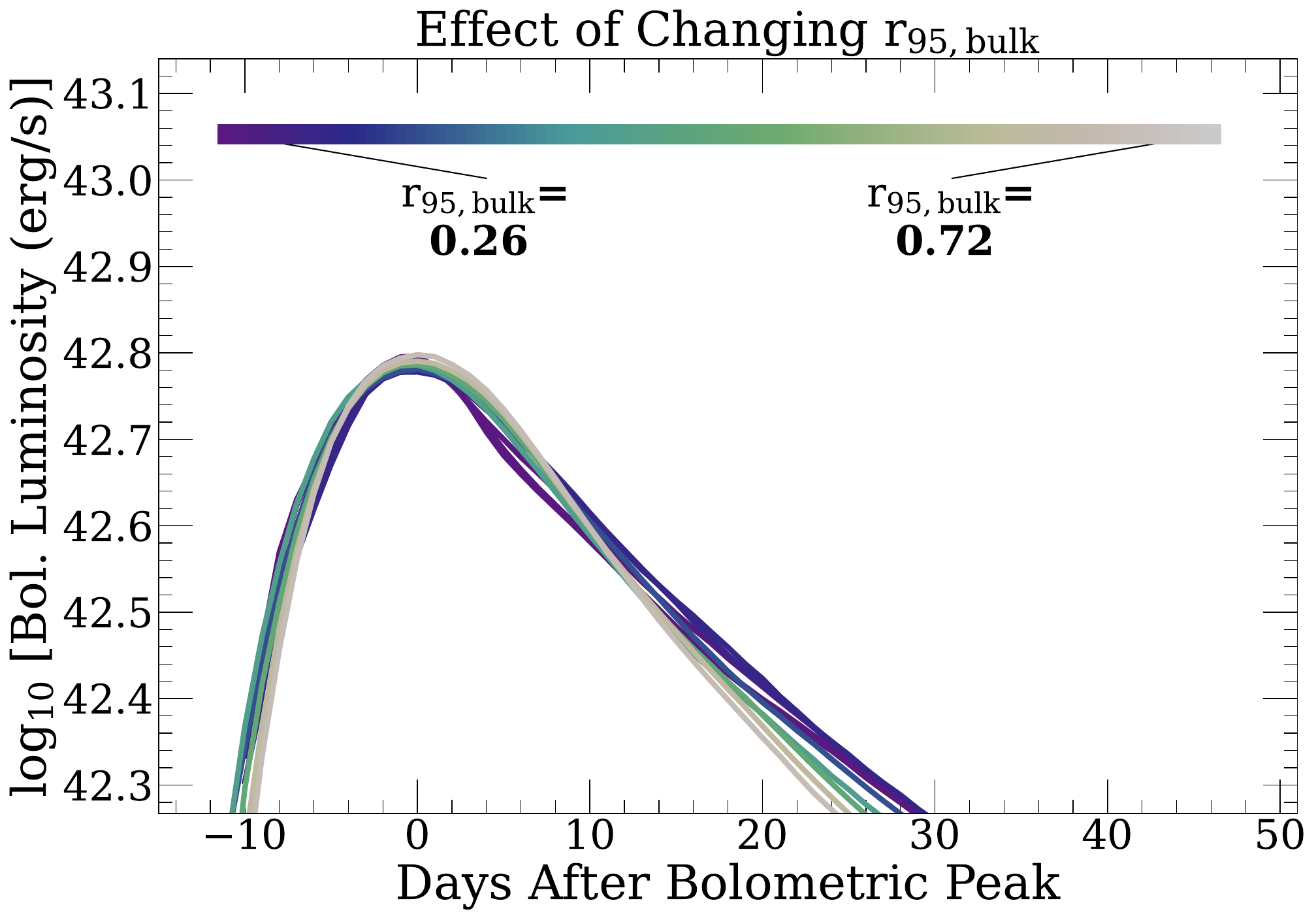}
    \includegraphics[width=0.32\linewidth]{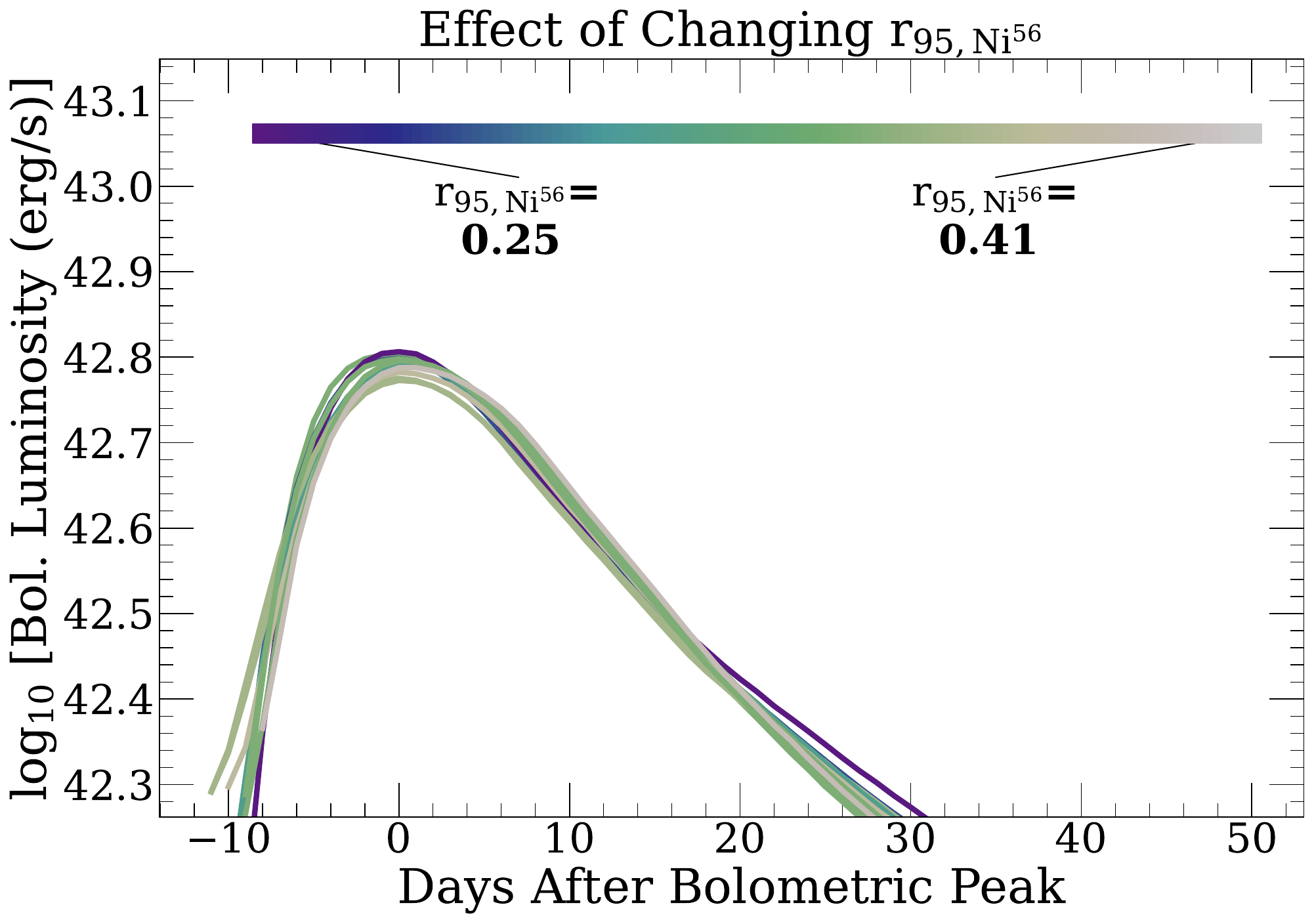}
    \caption{The influence on the bolometric \lc~ by each of the nine parameters, holding all other parameters constant. Increasing only the  $\Delta$~Vel (top left),  \velf~(top middle), or the minimum velocity (top right) while holding everything else constant causes a brighter bolometric peak. Changing only \mhe~(center left)  does not make a large impact on the bolometric \lc. Increasing $\rm m_{bulk}$ (center middle) decreases the peak luminosity and increases light curve duration. Increasing \mni~(center right) causes a brighter \lc~at all times. Changing how \he~is mixed (bottom left) does not influence the \lc~strongly. Mixing the bulk matter further out (bottom middle) causes a brighter \lc~with a shorter \tr. Mixing \NI~(bottom right) has a minimal effect on the \lc~(but see Section~\ref{sec:expanded_grid} for discussion about \NI~mixing).}
    \label{fig:perturbation_fig}
\end{figure*}

\begin{figure*}[t]
    \centering
    \includegraphics[width=0.32\linewidth]{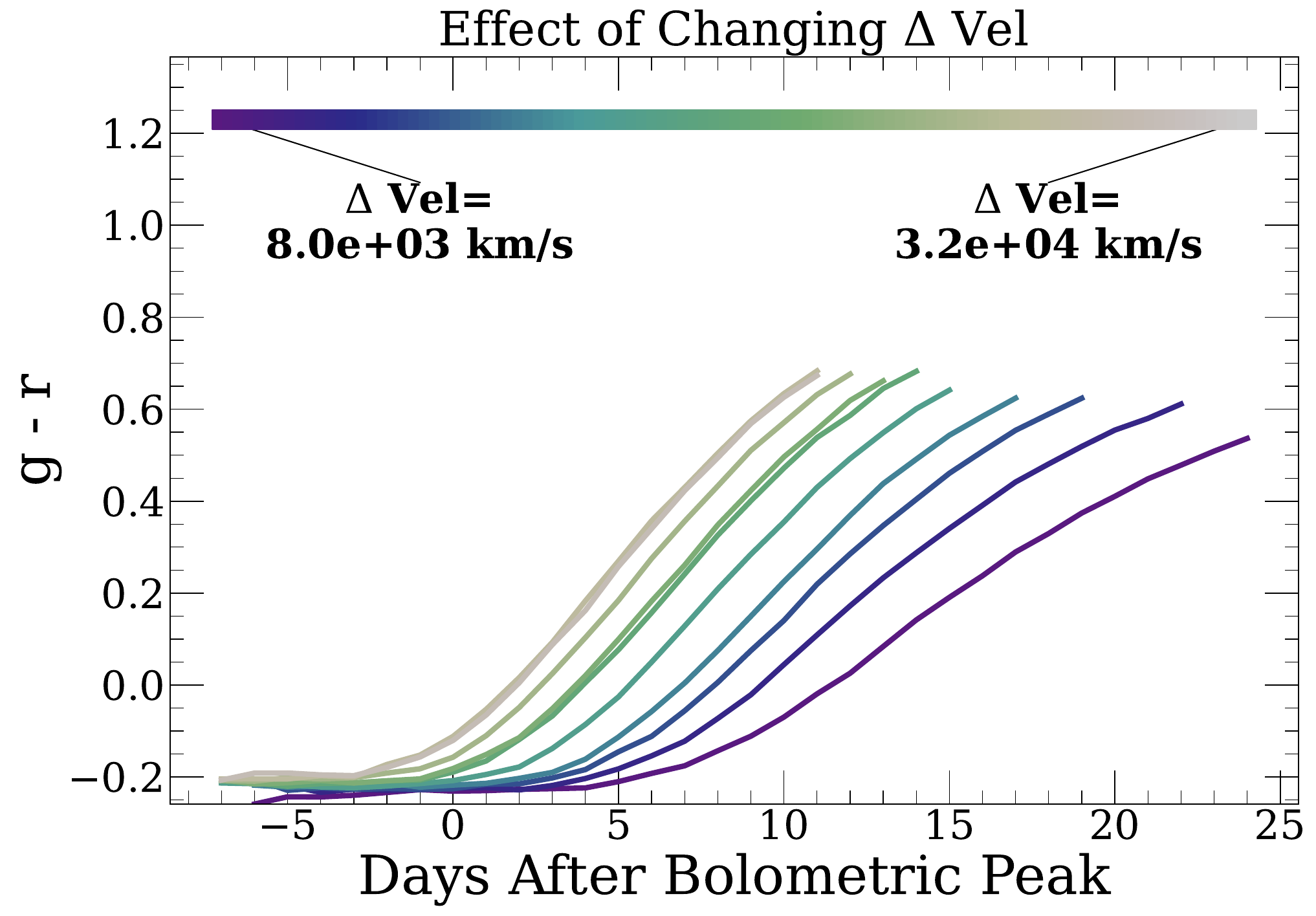}
    \includegraphics[width=0.32\linewidth]{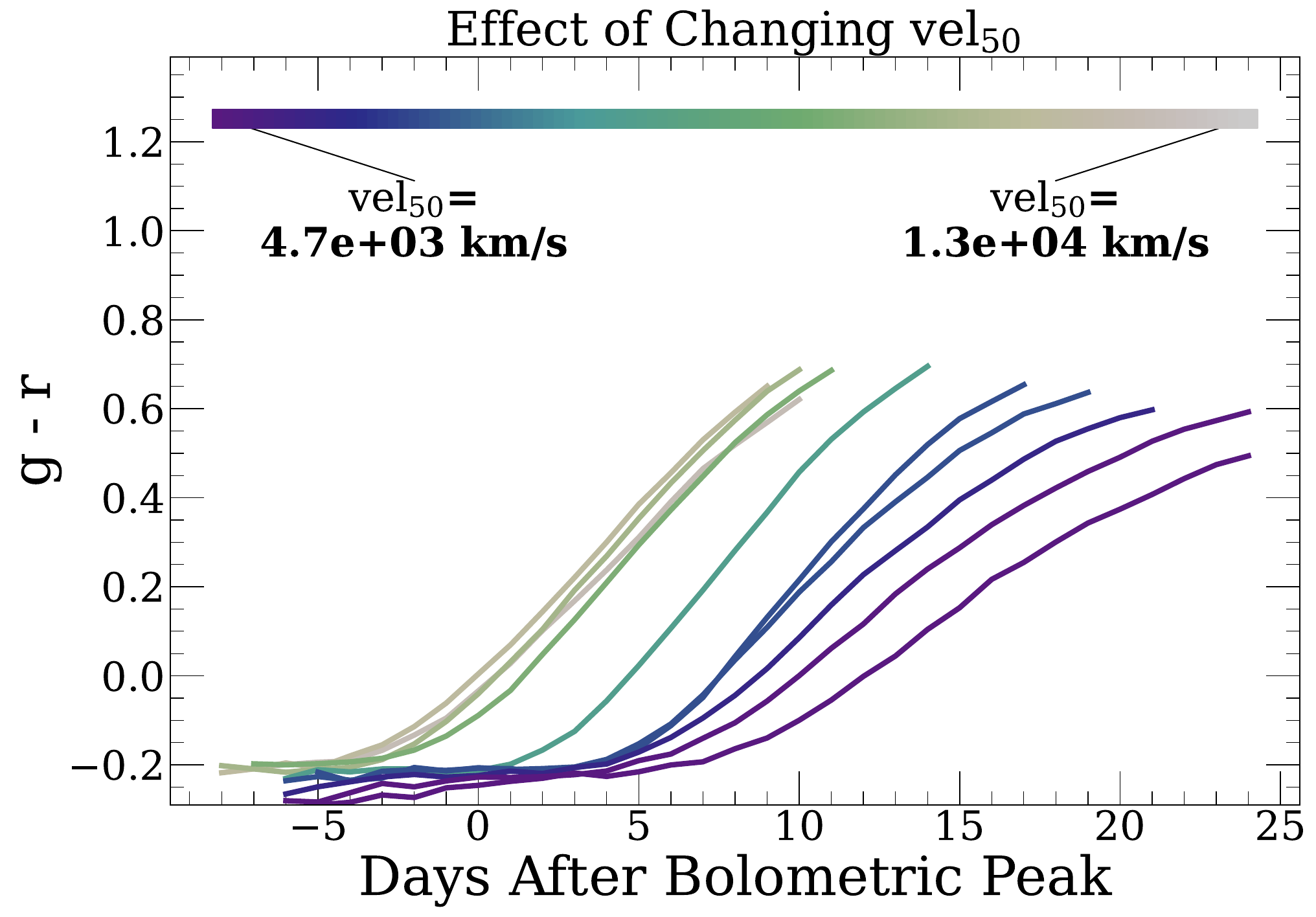}
    \includegraphics[width=0.32\linewidth]{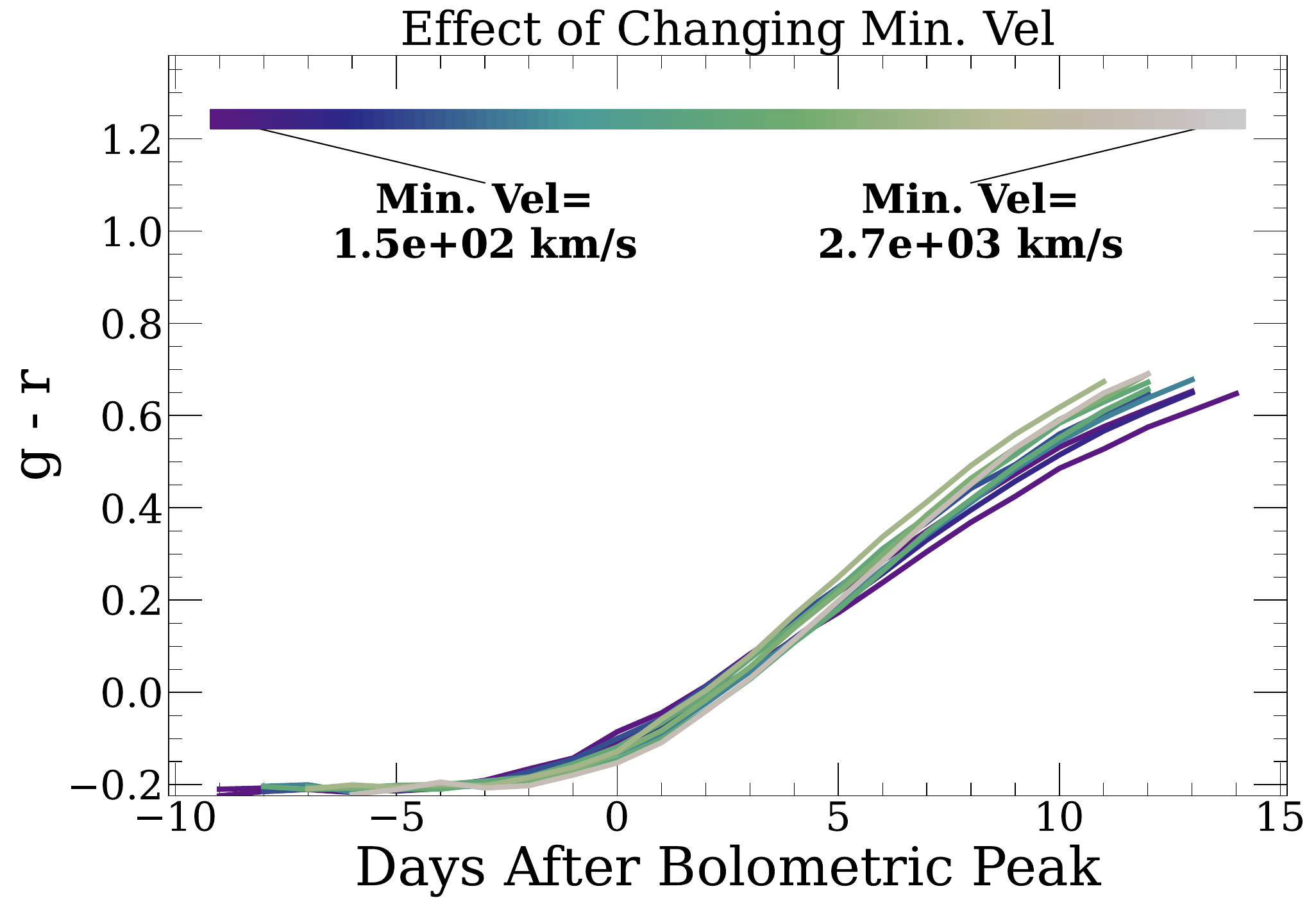}
    \includegraphics[width=0.32\linewidth]{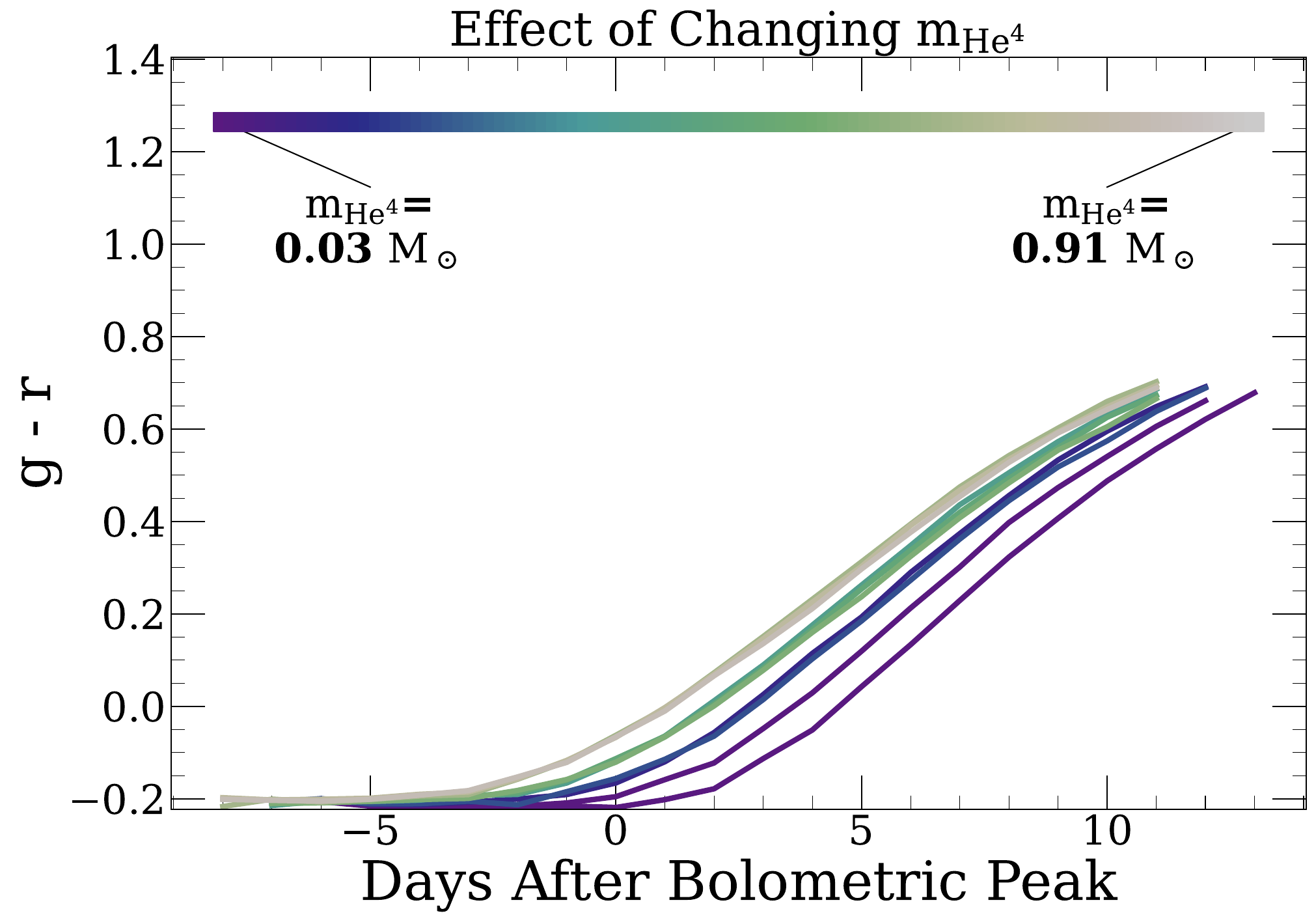}
    \includegraphics[width=0.32\linewidth]{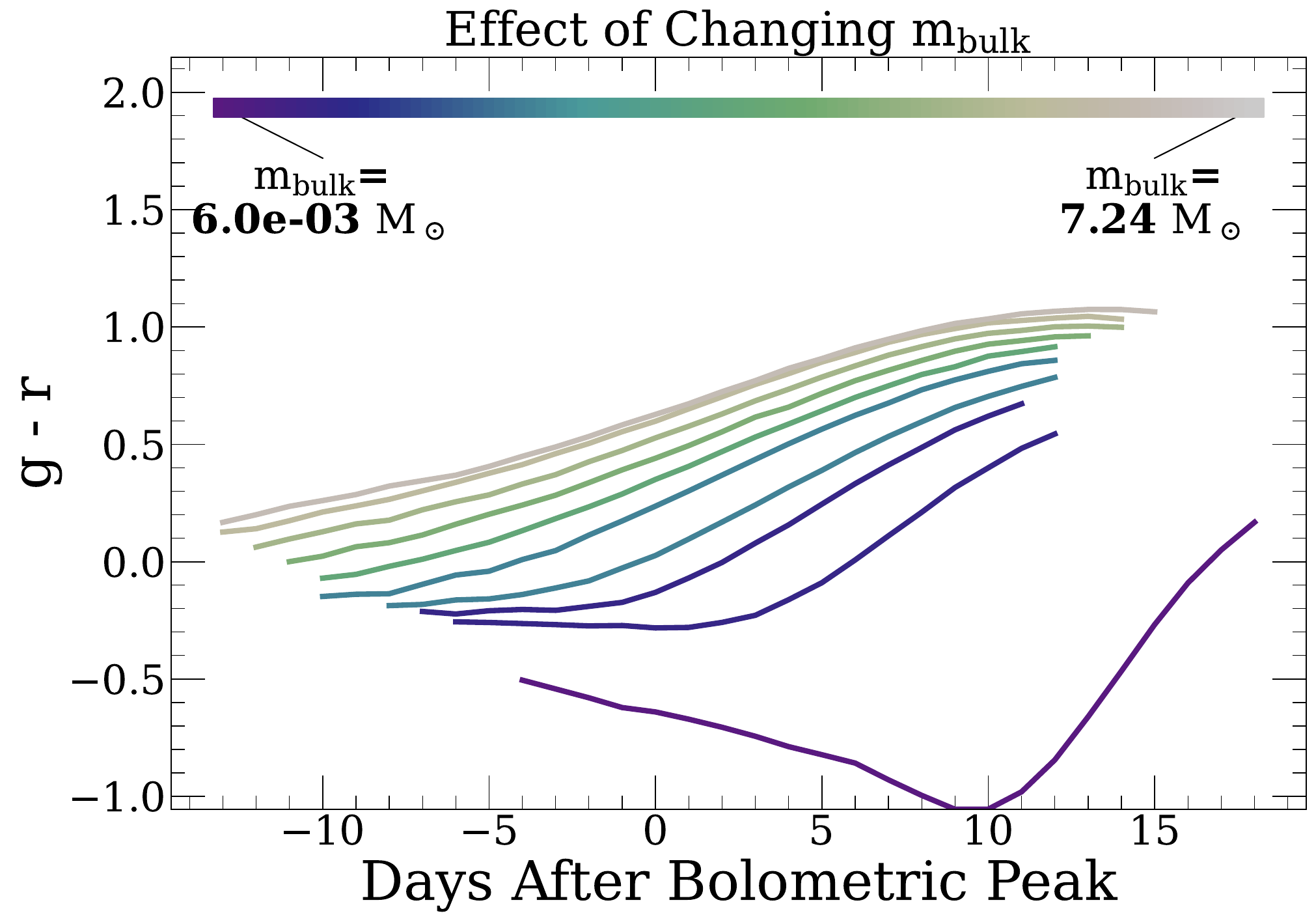}
    \includegraphics[width=0.32\linewidth]{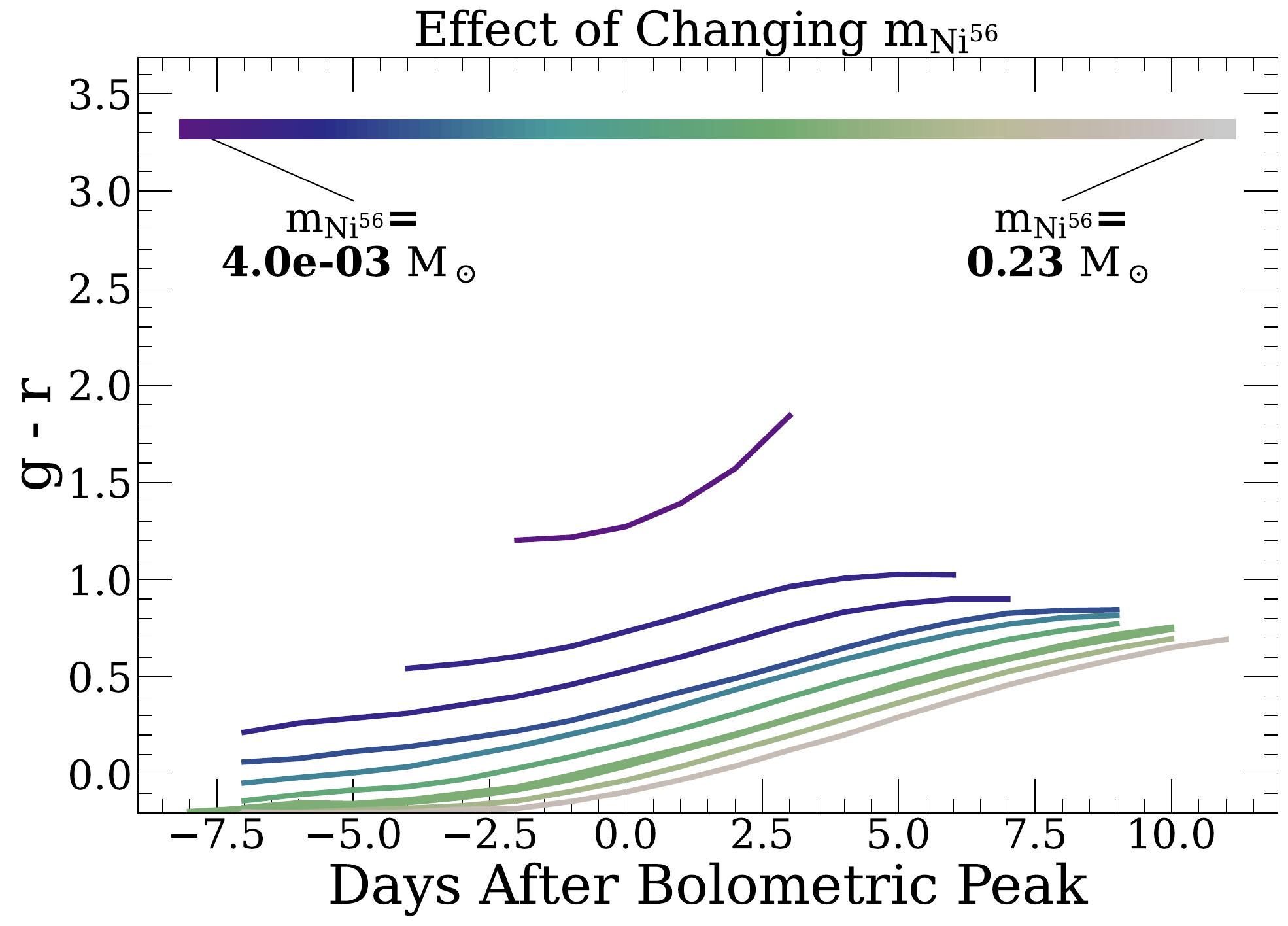}
    \includegraphics[width=0.32\linewidth]{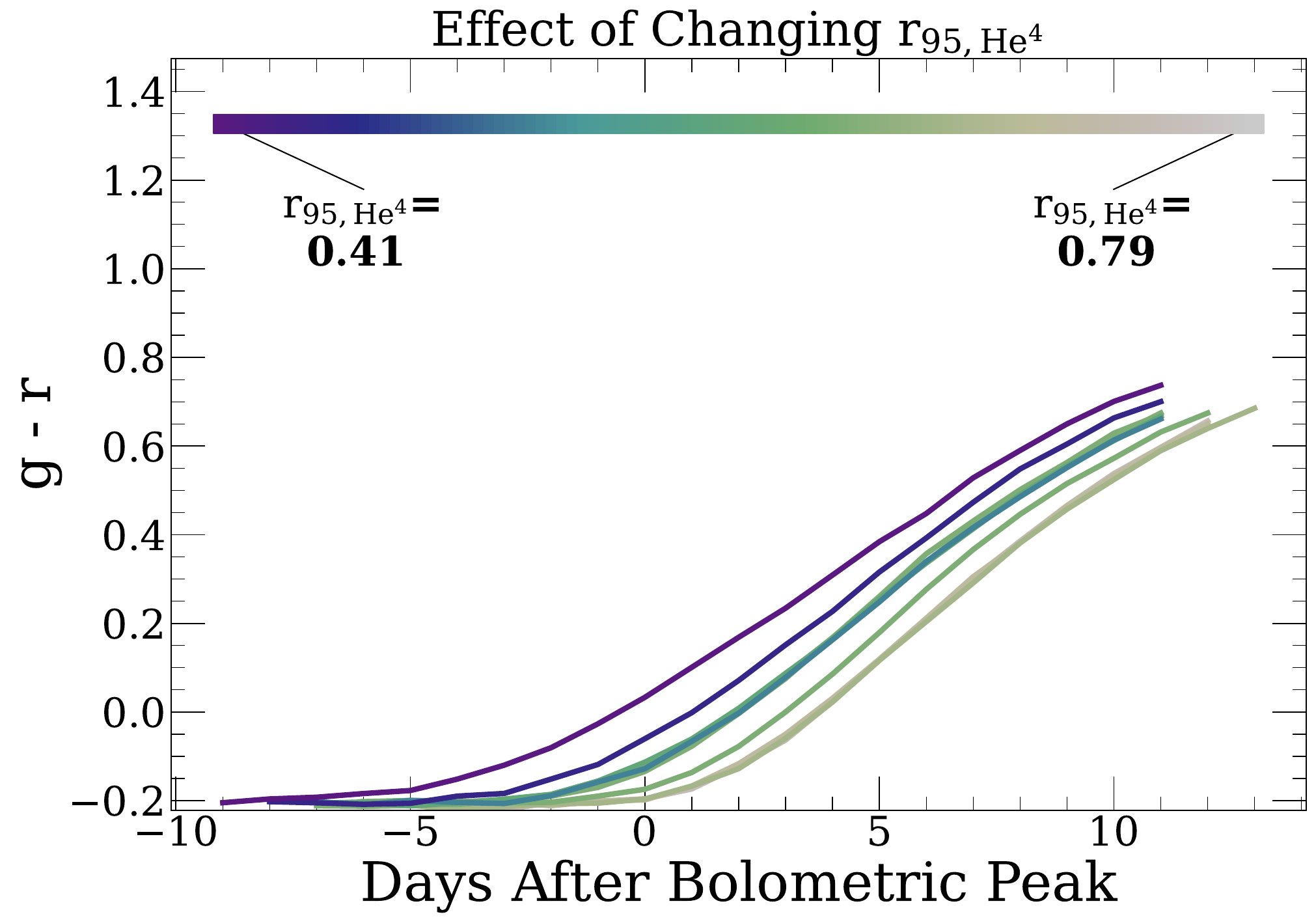}
    \includegraphics[width=0.32\linewidth]{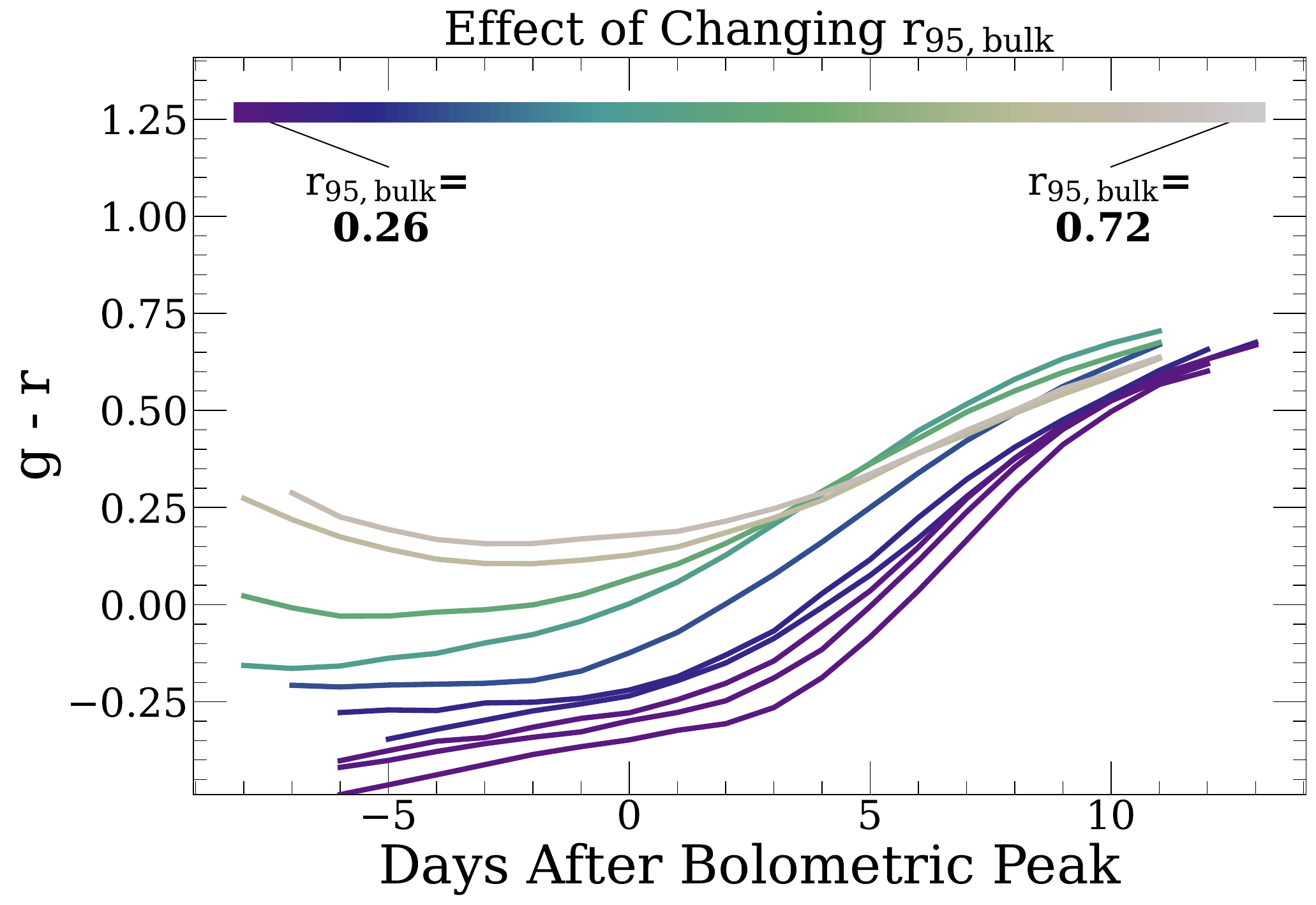}
    \includegraphics[width=0.32\linewidth]{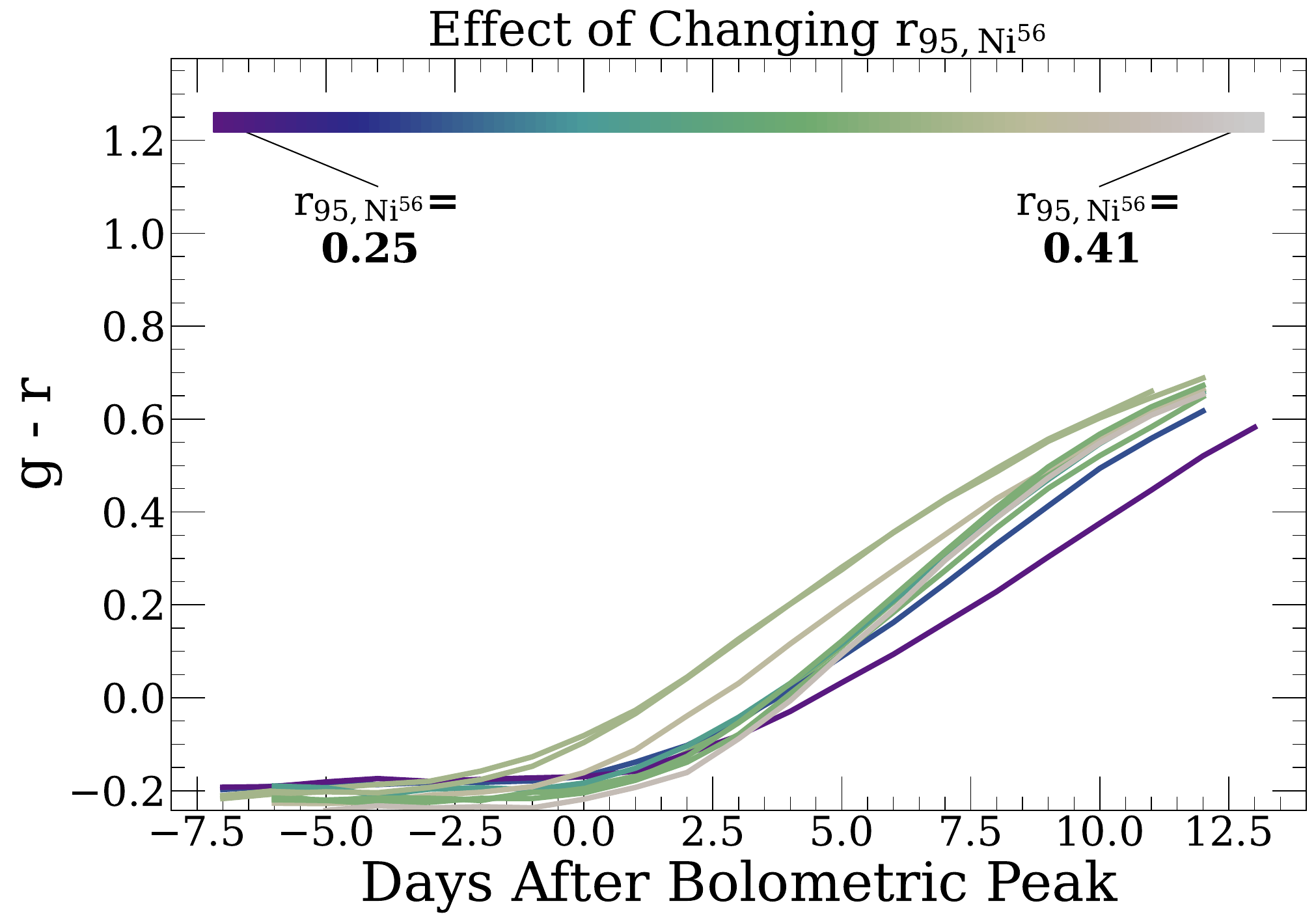}
    \caption{The influence on the $g$-$r$ color by each of the nine parameters, holding all other parameters constant. Increasing only $\Delta$~Vel (top left),  \velf~(top middle), or the minimum velocity (top right) causes the \lc~to evolve more quickly, reddening more quickly. Changing only \mhe~(center left) does not influence the color evolution. Increasing \mop~(center middle) tends to redden the \lc~because photons from \NI~decay must diffuse through more mass. Increasing \mni~tends to create a bluer \lc~because more photons from \NI~decay are present. Changing the mixing of \he~does not influence the color substantially. Mixing the bulk farther out creates a redder \lc~earlier in the evolution. Mixing just \NI~does not substantially change the color evolution.}
    \label{fig:perturbation_gr_color}
\end{figure*}

In this Section, we explore how each of the nine parameters detailed in Section~\ref{sec:sedona_exp} (see Table~\ref{table:novenheim}) influences the \lc~ through exploration of simple, measurable properties of the multiband light curves. In addition to the peak absolute magnitude, we parameterize the rise time (\tr; the amount of time for \lc~to brighten by one magnitude leading to the peak), and the decline time (\td; the amount of time for the \lc~to decline by one magnitude after peak). We additionally present the peak bolometric luminosity, bolometric \tr, bolometric \td, and the $g$-$r$ color as a function of the nine physical parameters. In Figures~\ref{fig:perturbation_fig} and \ref{fig:perturbation_gr_color}, we isolate the influence of each parameter on the \lc.

The top three panels of Figures~\ref{fig:perturbation_fig} and \ref{fig:perturbation_gr_color} show how variations in the ejecta velocity profile influences the \lc. Increasing the difference between outer velocity and inner velocity ($\Delta$~vel; top left panel), increasing the inner velocity (top center panel), and increasing \velf~(top right panel) while holding everything else constant all result in a brighter peak luminosity and a shorter \tr. Higher velocities cause the \lc~to evolve (and therefore redden/cool) more quickly with a shorter overall diffusion timescale, following intuition from the Arnett model. Of the three velocity parameters, the outer velocity and \velf~have the strongest influence on the \lc~ over our explored parameter ranges, with the inner velocity having a more modest effect.  

However, each velocity parameter influences the timing (and therefore color) of the \lc~differently. Changes to $\Delta$~vel do not change \tr~but decrease \td. Consequently, increasing the outer velocity results in redder colors at later times, while leaving early colors largely unaffected (top left panel of Figure~\ref{fig:perturbation_gr_color}). Increasing the inner velocity results in both a shorter \tr~and shorter \td, although the impact on both the bolometric light curve and colors is minimal. Increasing \velf~has an intermediary effect, where both \tr~and \td~are decreased, but \td~is more strongly influenced than \tr.

Because the physical size of the ejecta layers is set by the velocity profile, changing the velocity profile can be compared to mixing \textit{all} of the elements simultaneously, rather than just the \he, bulk, or \NI~distribution. Therefore, the influence of changing the velocity profile can be compared to the ``boxcar mixing method'' of \cite{dessart2012, dessart2015, Teffs2020, Woosley2021}, in which the ejecta elemental abundance profiles are convolved with a boxcar function to emulate mixing. There is a key difference between mixing by changing the velocity profile and boxcar mixing: our method of changing the velocity profile necessarily changes the density profile, while conserving the elemental relative abundance profiles; whereas boxcar mixing conserves both the density profile (see e.g. \citealt{dessart2012, Teffs2020}) and the velocity profile, while changing the relative abundance profiles. However, this effect is captured in the bottom three panels of Figures~\ref{fig:perturbation_fig} and \ref{fig:perturbation_gr_color}, and are explored later in this Section. 

Nevertheless, analogies between the two methods of mixing can be drawn. When $\Delta$~vel is increased, the inner velocity is unchanged and only the outer velocity is increased, causing the elements in the ejecta to be spread (i.e. \textit{mixed}) over a larger range of velocities. With little mixing (corresponding to low $\Delta$~vel), \td~is much larger than \tr. As the ejecta becomes more mixed (corresponding to increasing $\Delta$~vel), \td~decreases, creating a more symmetric \lc~about the peak. As such, the influence of increasing $\Delta$~vel on the \lc~can be understood as the result of two effects: 1) the overall velocity scaling of the ejecta is increased, causing the \lc~to be brighter at peak and to evolve more quickly, and 2) the ejecta is more mixed, creating a more symmetric \lc~about its peak. This trend of increasing \lc~symmetry via mixing is also found by \cite{dessart2012, Moriya2020, Teffs2020}. Changes to the inner velocity do not alter the extent of ejecta mixing, since the overall velocity range and distribution remain fixed. Therefore, the influence of the inner velocity on the \lc~reflects only the effect of rescaling the overall ejecta velocity. Although changing \velf~does not impact the inner or outer velocity, it does alter the  distribution of velocities in the ejecta. Increasing \velf~shifts more of the ejecta to higher velocities, analogous to enhanced ``mixing'' of the ejecta. In this sense, the influence of \velf~ on ejecta mixing is similar to that of the $\Delta$~vel.

The middle panels of Figures~\ref{fig:perturbation_fig} and \ref{fig:perturbation_gr_color} show how the total masses of \he, the bulk, and \NI~influence the \lc. As detailed in Section~\ref{sec:sedona_exp}, within the range of \he~masses in the W21 grid, variations in \he~mass (middle left panel) do not affect the \lc~appreciably. Increasing the \opacity~mass (middle center panel) results in a dimmer peak and a longer-lived \lc, consistent with expectations from the Arnett model. Both \tr~and \td~increase with larger \opacity~mass. In this case, photons from \NI~decay must diffuse through more mass, leading to delayed and cooler/redder emission at and beyond peak. Increasing the mass of \NI~(middle right panel) increases the luminosity of the \lc~at all phases, as expected. Similar to \opacity~mass, increasing the \NI~mass also results in a longer-lived transient due to an increased total ejecta mass.

W21 find comparable relationships to the ones we present in Figures~\ref{fig:perturbation_fig} and \ref{fig:perturbation_gr_color}. In contrast with our work, W21 map observed \lc~parameters to the physical parameters of the \textit{star} (e.g. pre-SN mass) prior to explosion, whereas we focus on the ejecta (e.g. \NI~mass and \opacity~mass). Nevertheless, comparisons can be made if the sum of \mhe, \mop, and \mni~is taken to be a proxy for pre-SN mass. W21 find that both \tr~and \td~both increase with larger pre-SN mass, consistent with our results. However, W21 find that peak luminosity increases with pre-SN mass at low pre-SN masses and decreases with pre-SN mass at larger pre-SN masses. This is because both \NI~mass and ejecta mass increase with increasing pre-SN mass, analogous to the trends shown in the middle row of Figure~\ref{fig:perturbation_fig}. W21 also find that models with higher pre-SN mass progenitors tend to be redder throughout their evolution,  in agreement with our findings.

The bottom panels of Figures~\ref{fig:perturbation_fig} and \ref{fig:perturbation_gr_color} illustrate how mixing of \he~(left), the bulk (center), and \NI~(right) impacts the light curves. As before, \he~has a minimal influence on the \lc. In contrast, a model in which the bulk is mixed farther out will exhibit brighter and later peaks. These models are significantly redder before \lc~peak, colors that gradually converge over time.  As \NI~is mixed farther out, \tr~is slightly decreased, while \td~and peak luminosities are largely unchanged. Compared to less-mixed models, strong \NI~mixing results in somewhat bluer colors before peak and redder colors after peak. Overall, mixing of \NI~has only modest effects on the light curve over the range of \NI~mass profiles explored in the W21 grid. 

That the bolometric \lc~is largely insensitive to \NI~mixing is well established and supported by several radiative transfer studies of SESNe \citep{dessart2012, dessart2015, Moriya2020, Woosley2021}. \cite{dessart2012} model mixing by a convolving the mass profile with a boxcar, coupling the mixing of both the bulk mass and \NI~mass. They find that more highly mixed models tend to have a longer \tr, but the peak luminosity and \td~are unchanged. Our analysis is consistent with this behavior (see bottom center and bottom right panels of Figure~\ref{fig:perturbation_fig}): when both the bulk and \NI~are mixed farther out, \tr~increases and the light curve is otherwise unchanged. \cite{Moriya2020}, in constrast, simulate models with varying \NI~mixing without mixing the bulk. They also find that the peak luminosity is largely unchanged (see e.g. their Figure 3). \cite{Moriya2020} also show that although the \lcs~of more mixed models peak earlier, differences in the bolometric light curve are minimal when aligned by peak time. Like us (see bottom right panel of Figure~\ref{fig:perturbation_gr_color}), they find that the most mixed models are typically bluer before peak and become redder than the the less mixed models after peak.

Taken altogether, we show that the observed light curve properties \tr, \td, peak luminosity, and color evolution are influenced by the nine physical parameters explored here in complex and often degenerate ways. Nevertheless, we show in Section~\ref{sec:Inference Machine} that it is possible to constrain many of these physical parameters using simple regression on the observed light curve features.

\section{Radiative Transfer Grid of 1,000 SESN Light Curves}
\label{sec:grid}

Using the \sed~setup described in Section~\ref{sec:sedona_exp}, we simulate $1$,$000$ SESN models by uniformly and independently sampling each of the nine physical parameters listed in Section~\ref{sec:sedona_exp}. In contrast to W21, we explicitly remove physical correlations between parameters by sampling each dimension independently. For example, \mop~and \mni~are positively correlated in the W21 grid as nickel production is tied to core mass in neutrino-driven explosions. Here, we intentionally break such correlations to isolate independent influence of each physical parameter on the \lc.  

\begin{figure*}
    \centering
    \includegraphics[width=0.49\linewidth]{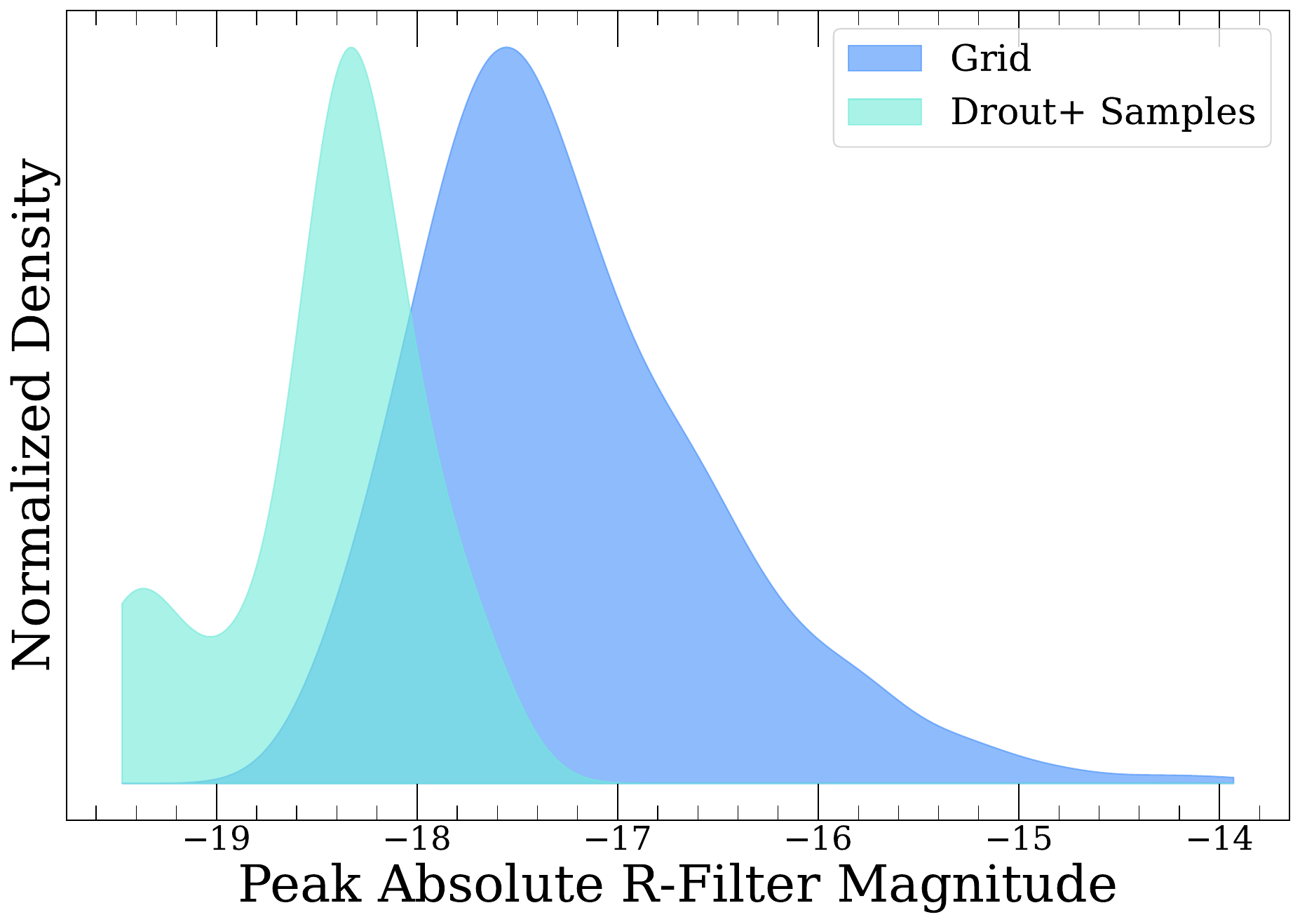}
    \includegraphics[width=0.49\linewidth]{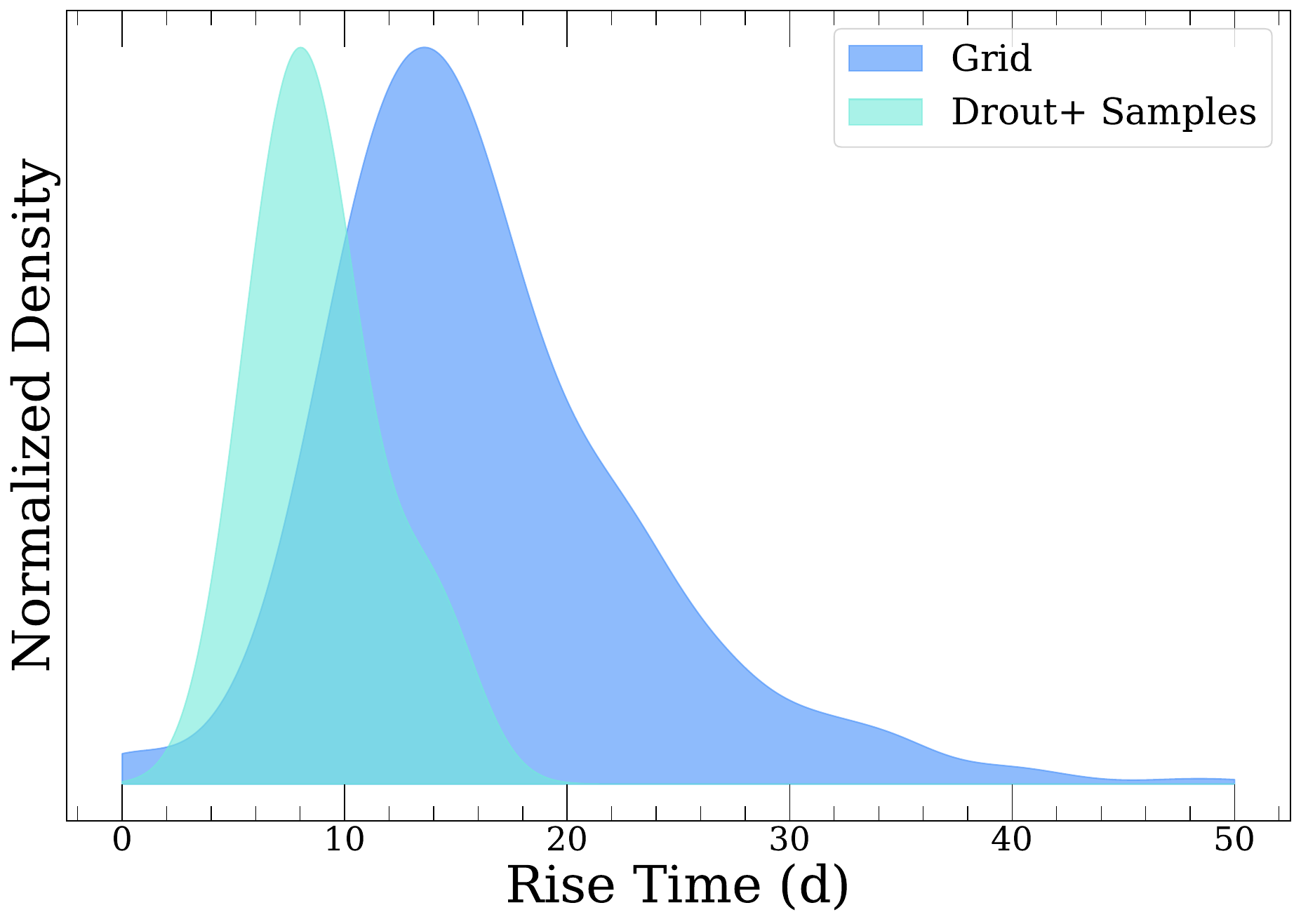}
    \includegraphics[width=0.49\linewidth]{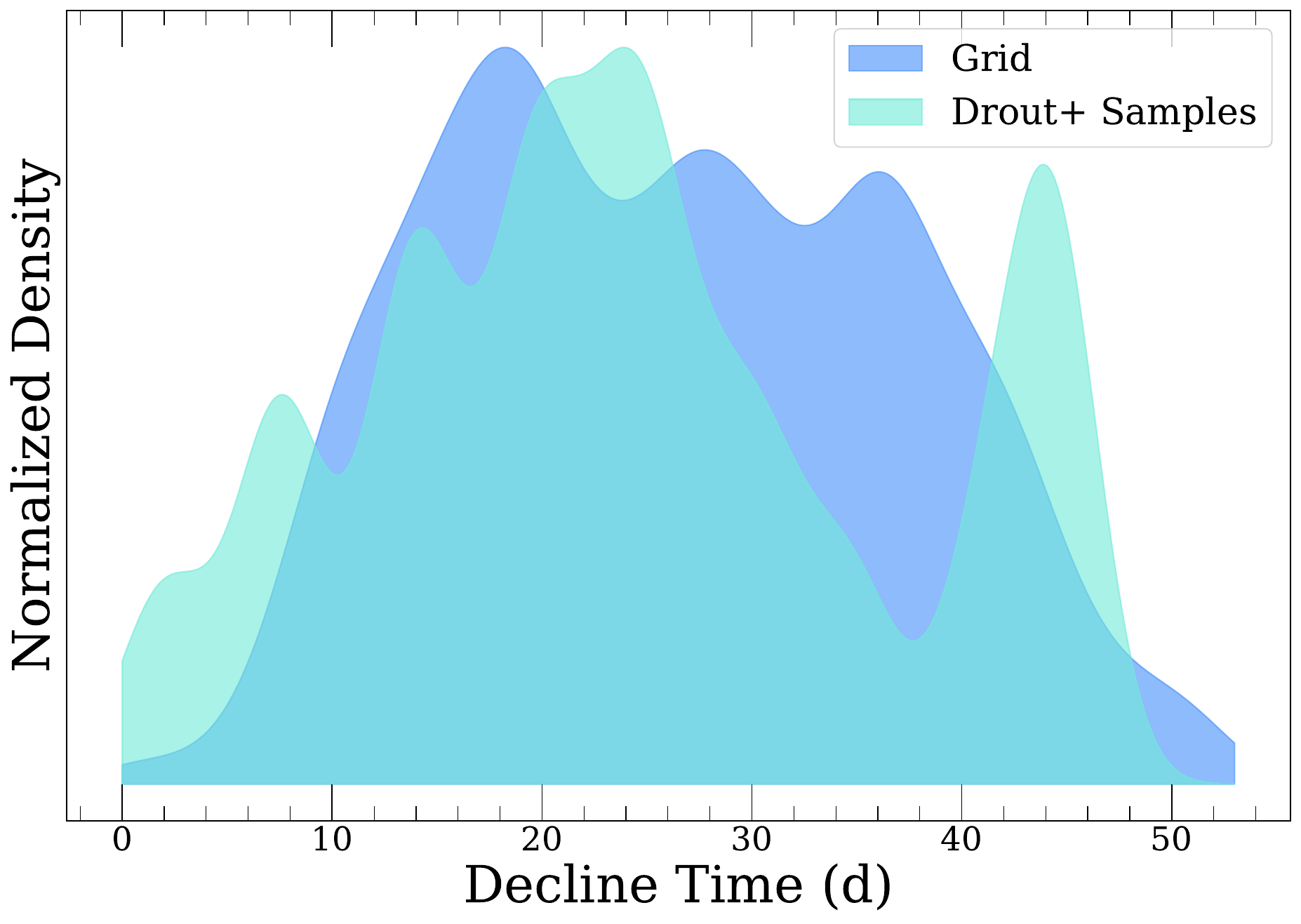}
    \caption{The range of peak $R$-filter magnitudes (top left), \tr~(top right), and \td~(bottom right) of the \lcs~simulated in our grid of $1$,$000$ models. The range of values from the \cite{Drout2011} sample of SESNe is also shown. }
    \label{fig:all_lcs}
\end{figure*}

We calculate full SEDs for all $1$,$000$ SESN models and present the SDSS $ugriz$ filter \lcs~for each model. Histograms of bolometric \lcs~properties are presented in Figure~\ref{fig:all_lcs}. We overlay distributions from the \cite{Drout2011} sample of SNe~Ib and SNe~Ic to show how our grid of \lcs~compares to an observed population of SESNe. The \cite{Drout2011} R-Filter peak magnitude distribution extends brighter than our distribution, partially because SNe~Ic-BL and GRB-SNe~Ic are included in the \cite{Drout2011} study, which likely have additional power sources. More recent samples of Type Ibc SNe (e.g., \citealt{Lyman2016, prentice2016, taddia2018carnegie}) show fainter typical peak magnitudes more inline with our grid. We emphasize, however, that our grid of models does not aim to reproduce the population of SESN (e.g., the Ibc luminosity function); rather, it is a representative grid for broadly exploring the physics of \NI~decay-powered SESNe.

\subsection{Reproducing Real Observed \LCs}
\label{subsec:overlay_real_ibc}

To access whether our model can produce realistic SESN multiband light curves, we compare the grid of $1$,$000$ generated \lcs~to \lcs~of three representative supernovae: SN1994I (a well studied SN~Ic with a relatively fast decline), SN2007gr (a widely-studied SN~Ic), and iPTF13bvn (the only SESN with a confirmed progenitor detection). Notably, \cite{afsariardchi2021} proposed that all three SNe may show evidence for excess peak bolometric luminosity relative to expectations from purely \NI-powered models, with the discrepancy being most pronounced for SN~1994I. For each event, we find the closest matching models from our grid using a chi-squared metric across the $BVRI$-band photometry. In each case, we are able to find a set of models that broadly reproduce the light curves, confirming that our grid is capable of generating realistic SESN \lcs. We emphasize that this is not a full fit to each \lc, which would require interpolation between models. Below, we discuss the objects and their matching models in more detail.

\begin{figure*}[ht!]
\centering
\begin{minipage}[t]{0.48\linewidth}
    \includegraphics[width=\linewidth]{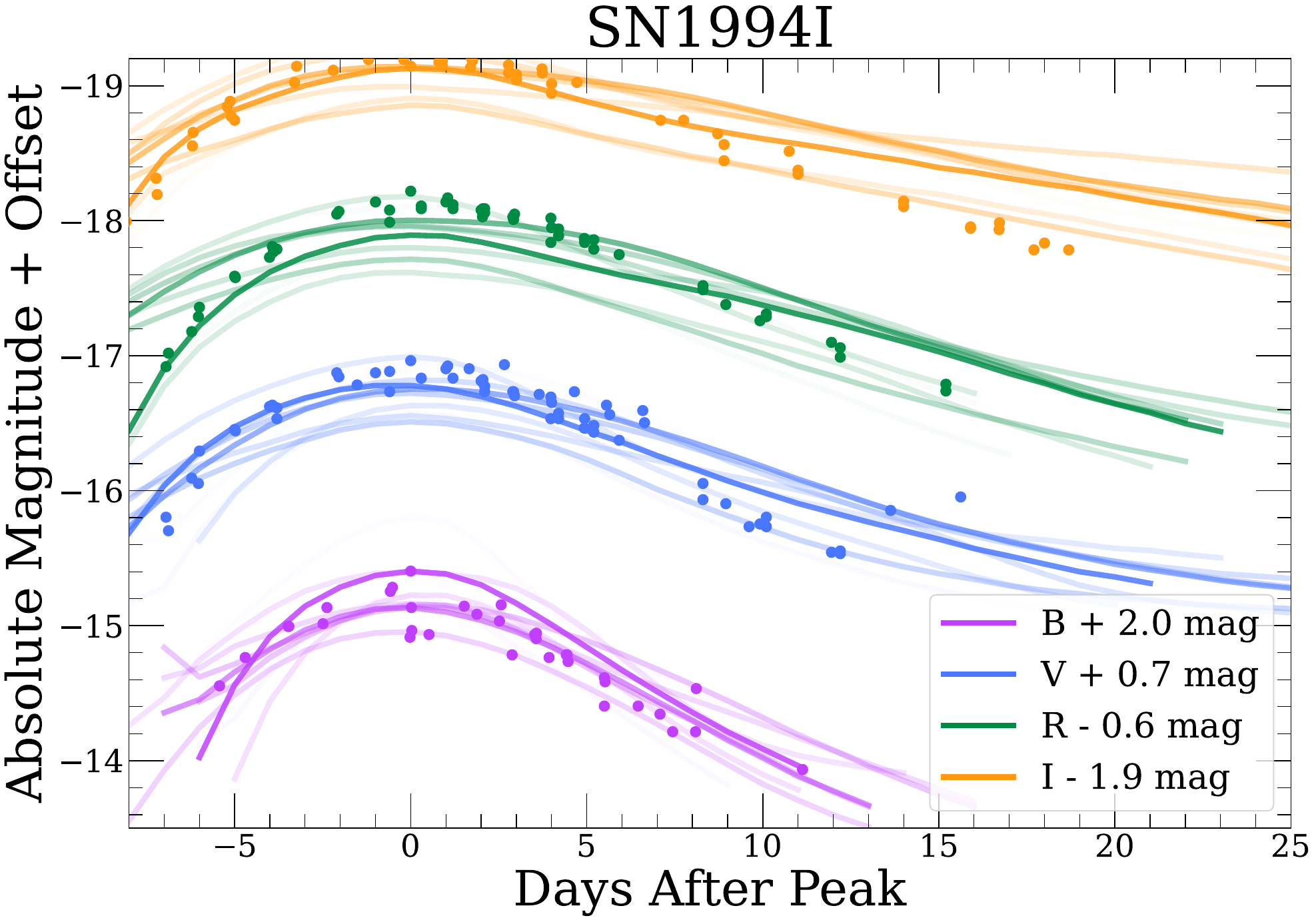}
\end{minipage}
\hfill
\begin{minipage}[t]{0.48\linewidth}
    \includegraphics[width=\linewidth]{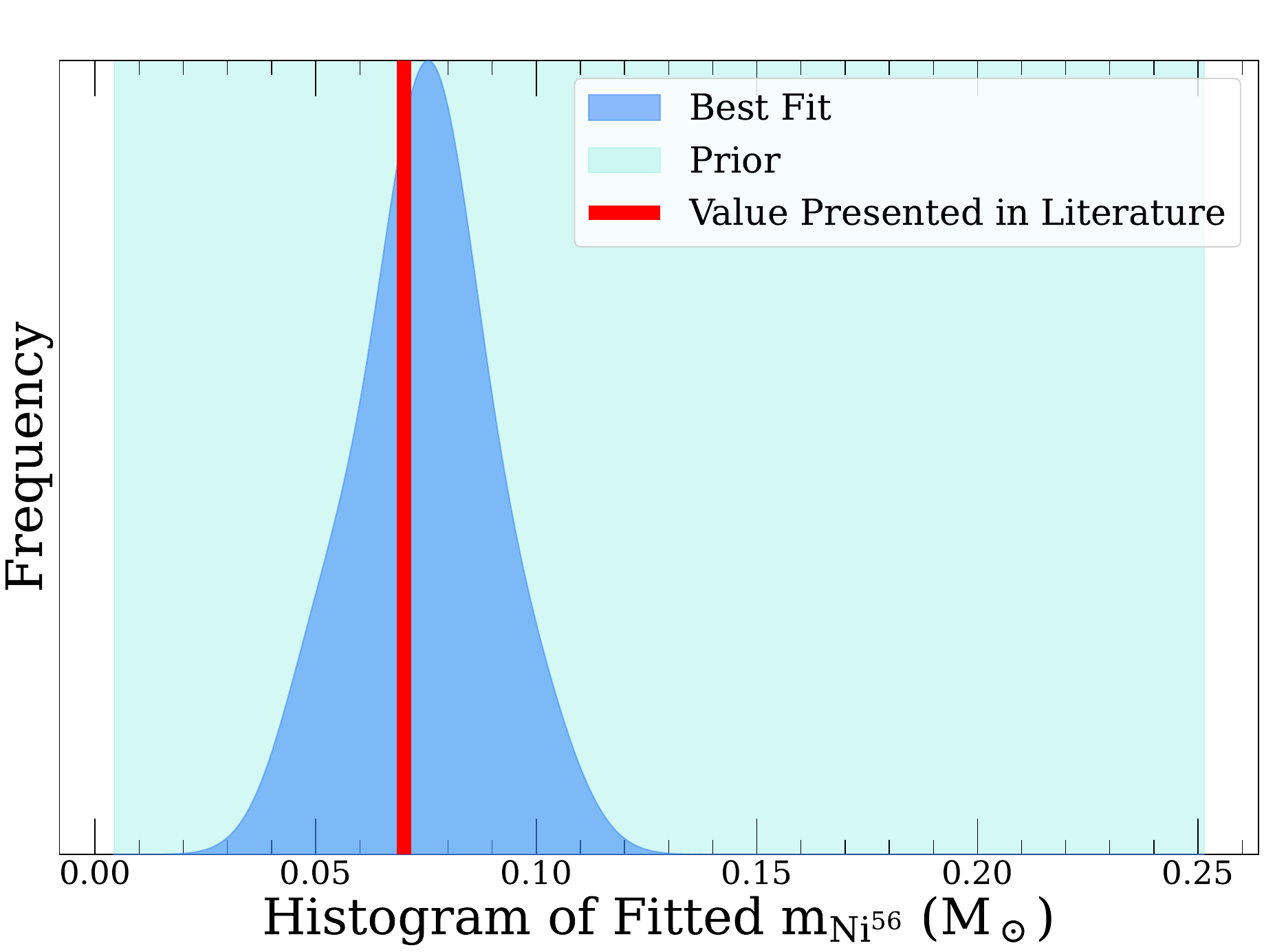}
\end{minipage}

\vspace{1em}  

\begin{minipage}[t]{0.48\linewidth}
    \includegraphics[width=\linewidth]{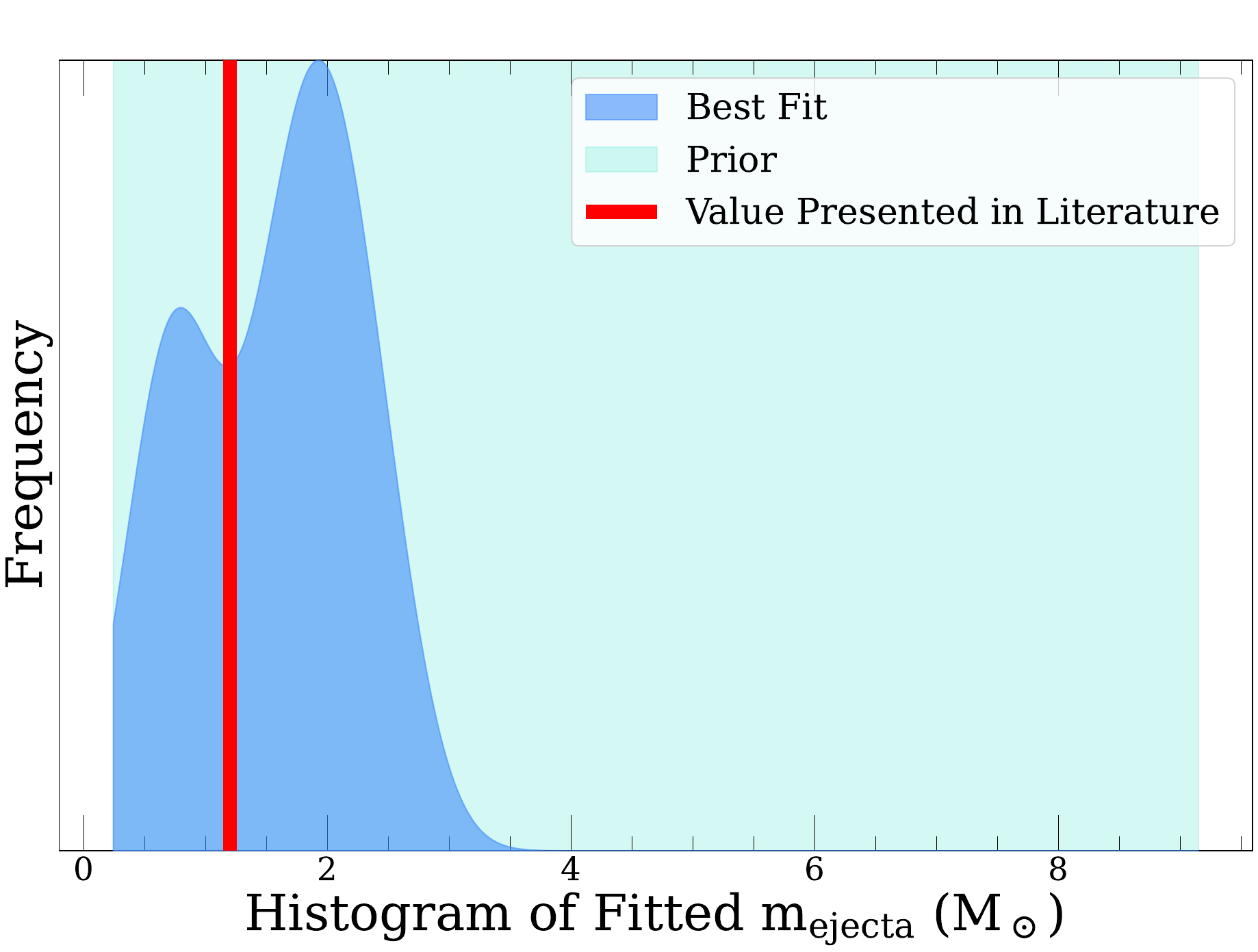}
\end{minipage}
\hfill
\begin{minipage}[t]{0.48\linewidth}
    \includegraphics[width=\linewidth]{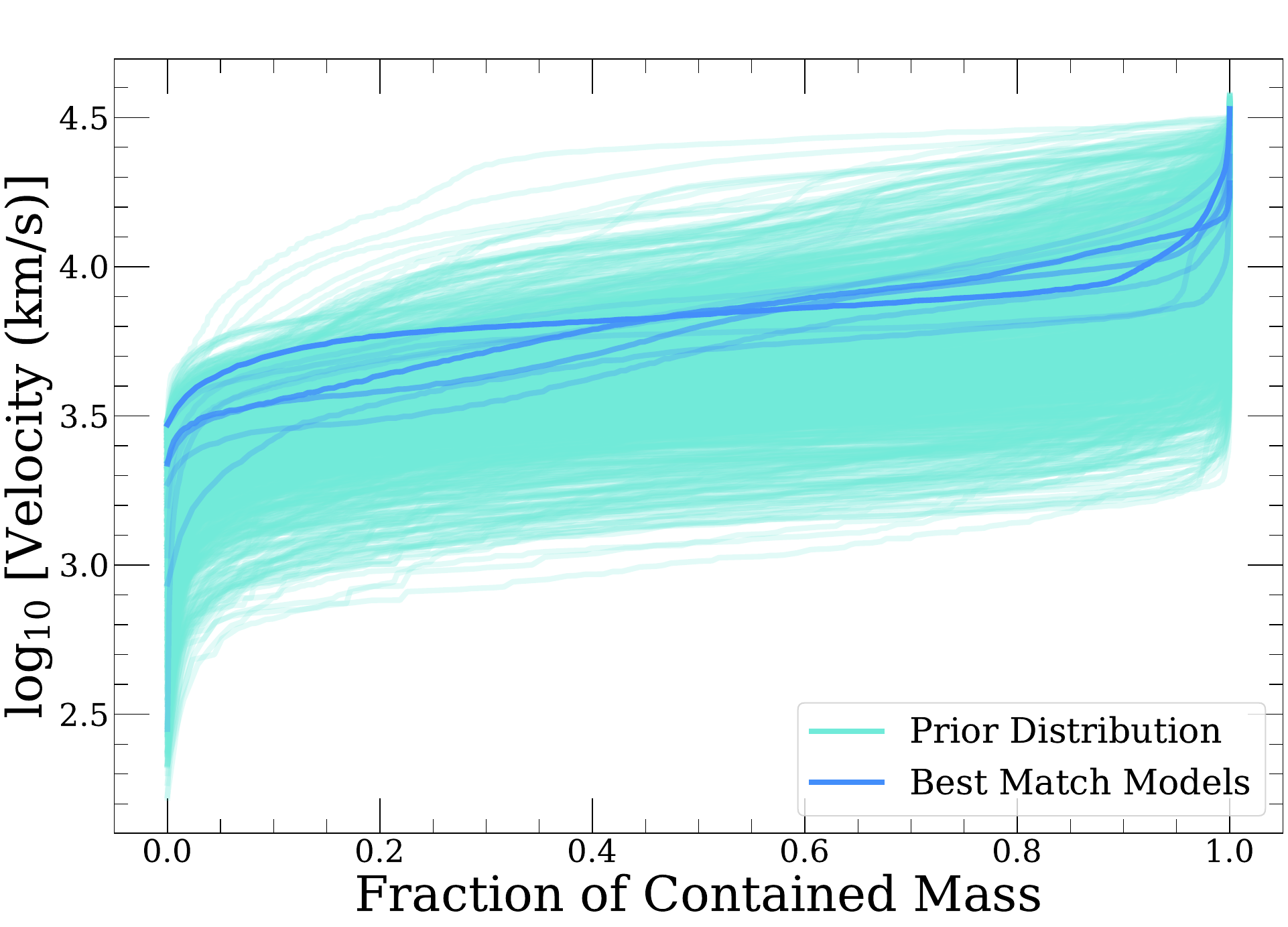}
\end{minipage}

\caption{A comparison between our generated grid of \lcs~to the SN~Ic SN~1994I \lc~in $BVRI$-bands. (top left) We show the closest several \lcs~to the \lc~of SN~1994I, and histograms of the \NI~mass values (top right), ejecta masses (bottom left), and velocity profile (bottom right) of the closest \lc~matches. Our range of \mni~values are consistent with the range of simulation-inferred literature values of the \mni~\citep{Sauer2006}.}
\label{fig:94I_matches}
\end{figure*}

\begin{figure*}[ht!]
    \centering
    \begin{minipage}[t]{0.48\linewidth}
        \includegraphics[width=\linewidth]{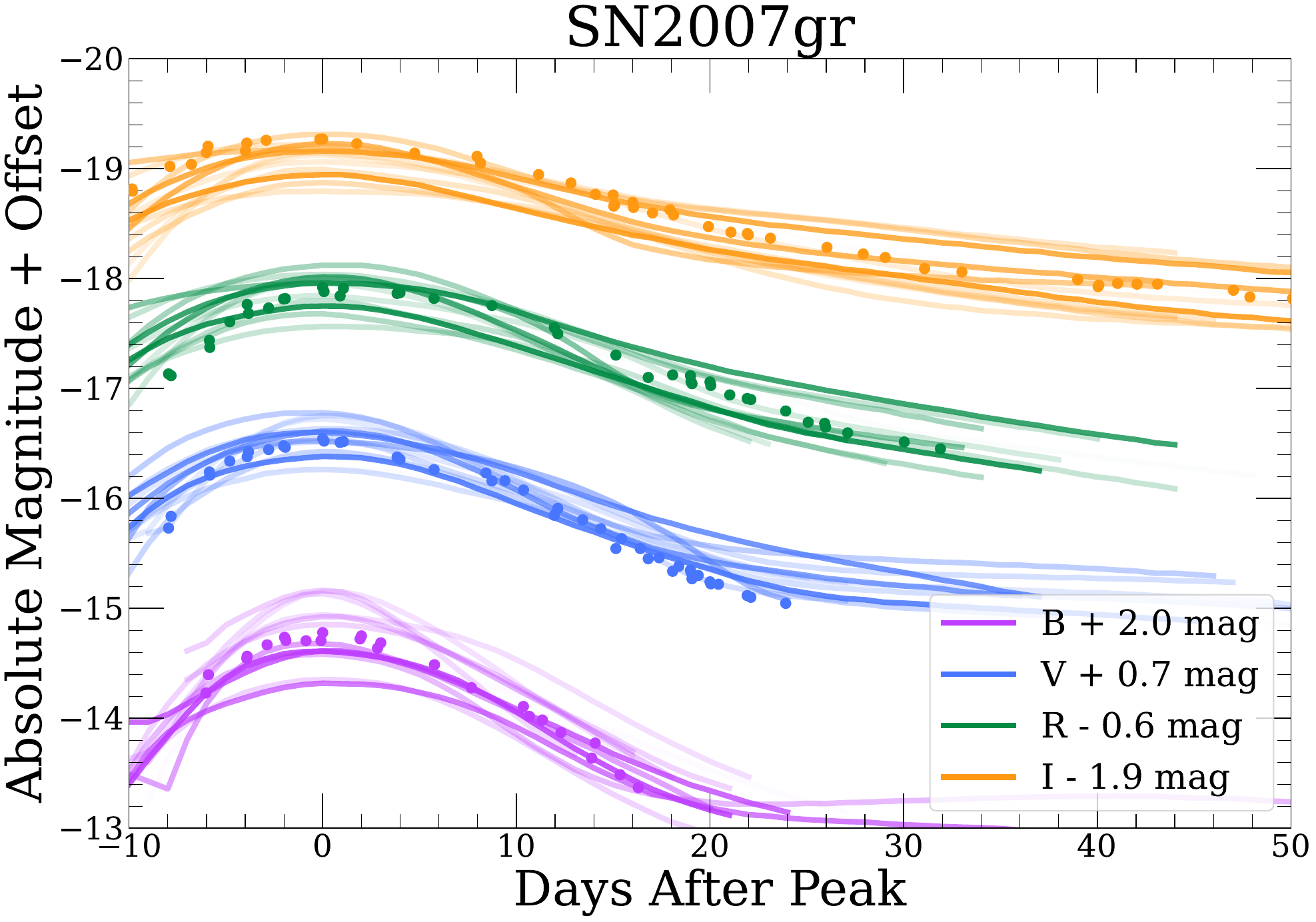}
    \end{minipage}
    \hfill
    \begin{minipage}[t]{0.48\linewidth}
        \includegraphics[width=\linewidth]{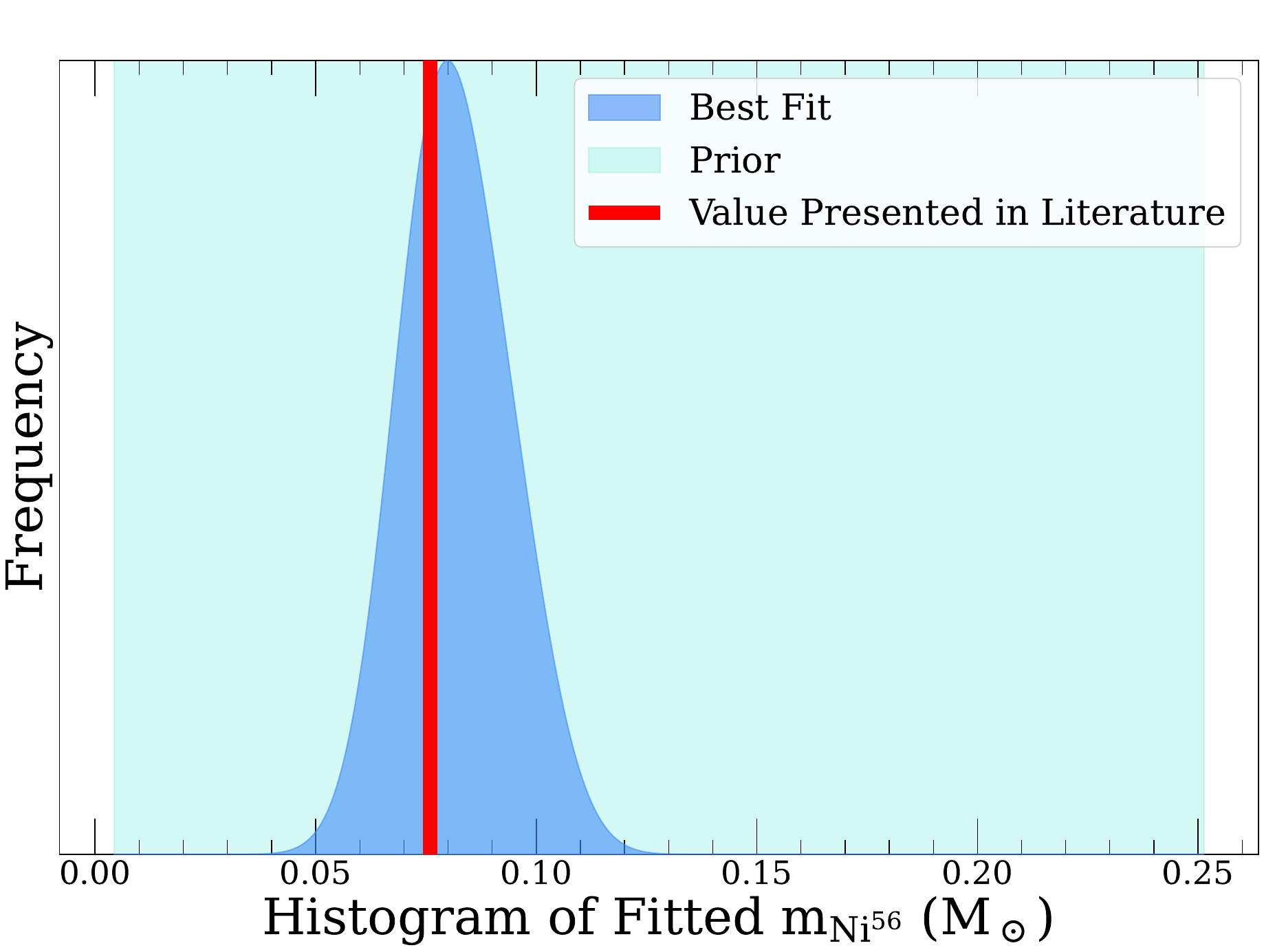}
    \end{minipage}

    \vspace{0.5em} 

    \begin{minipage}[t]{0.48\linewidth}
        \includegraphics[width=\linewidth]{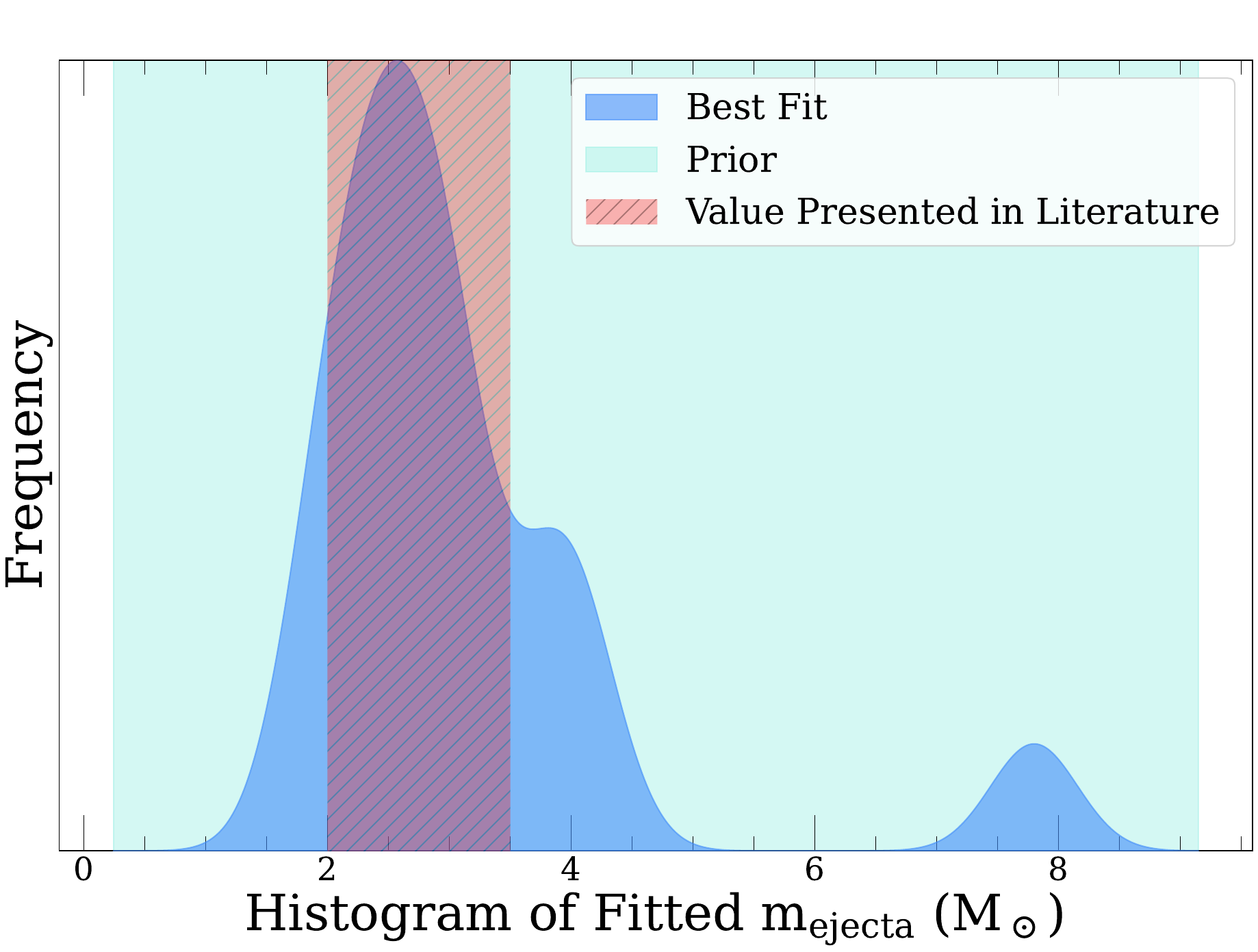}
    \end{minipage}
    \hfill
    \begin{minipage}[t]{0.48\linewidth}
        \includegraphics[width=\linewidth]{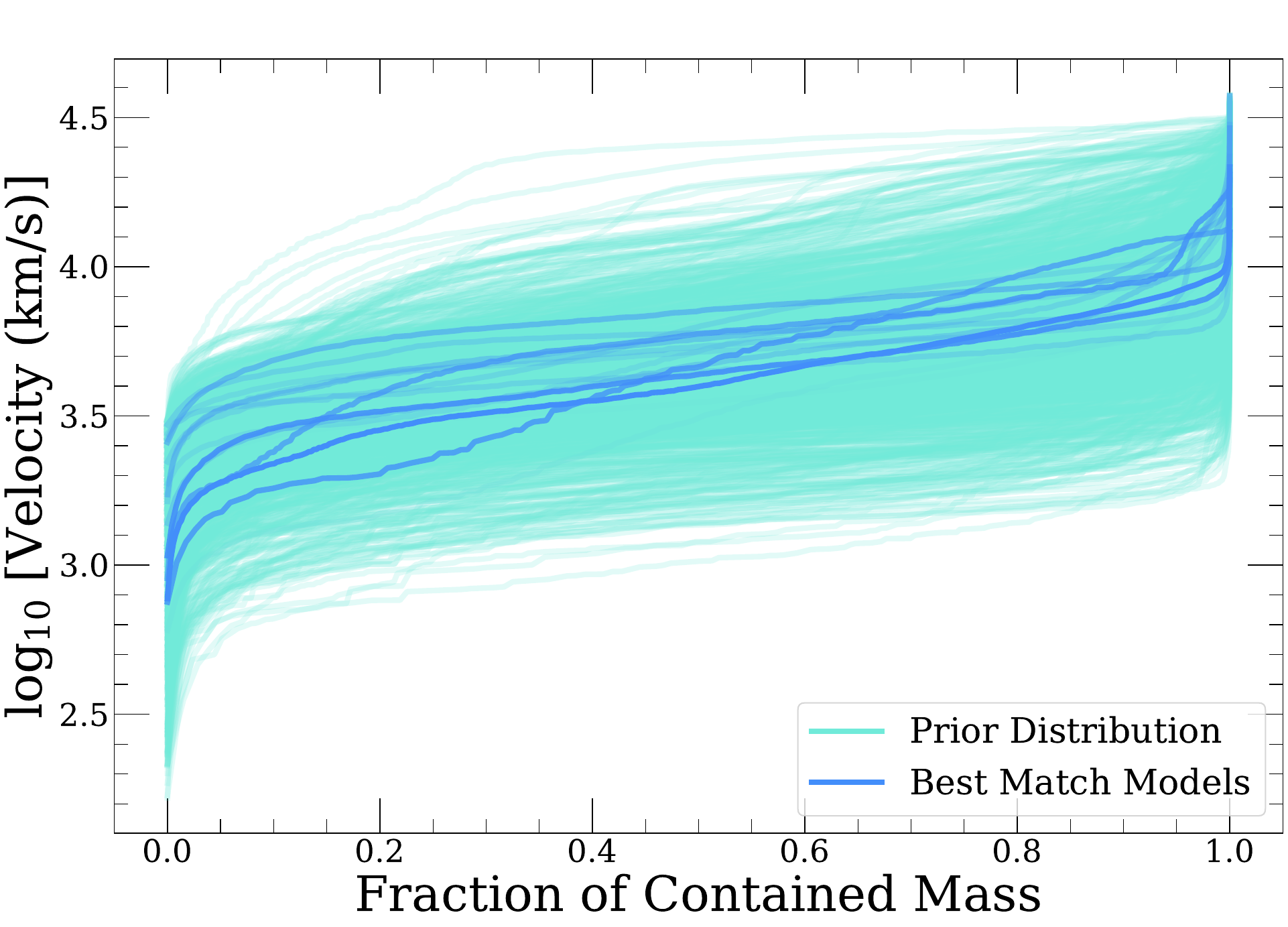}
    \end{minipage}

    \caption{A comparison between our generated grid of \lcs~to the SN~Ic SN2007gr \lc~in $BVRI$-bands. (top left) We show the closest several \lcs~to the \lc~of SN2007gr, and histograms of the \NI~mass values (top right), ejecta masses (bottom left), and velocity profile (bottom right) of the closest \lc~matches. Our range of nearest-fit \mni~values is consistent with the range of simulation-inferred literature values of the \mni~\citep{Hunter2009, Mazzali2010}.}
    \label{fig:07gr_matches}
\end{figure*}

\begin{figure*}[ht!]
    \centering

    \begin{minipage}[t]{0.48\linewidth}
        \includegraphics[width=\linewidth]{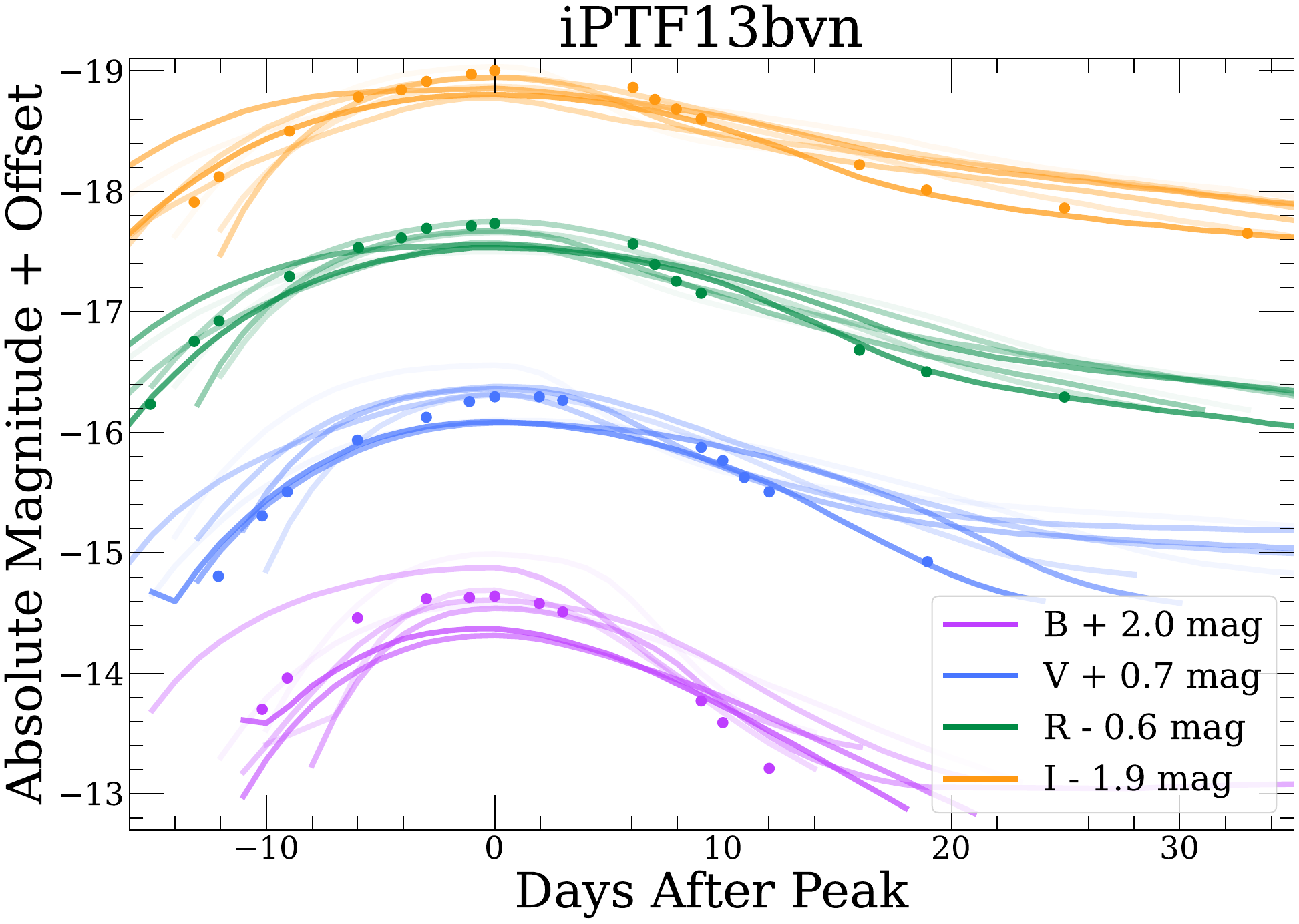}
    \end{minipage}
    \hfill
    \begin{minipage}[t]{0.48\linewidth}
        \includegraphics[width=\linewidth]{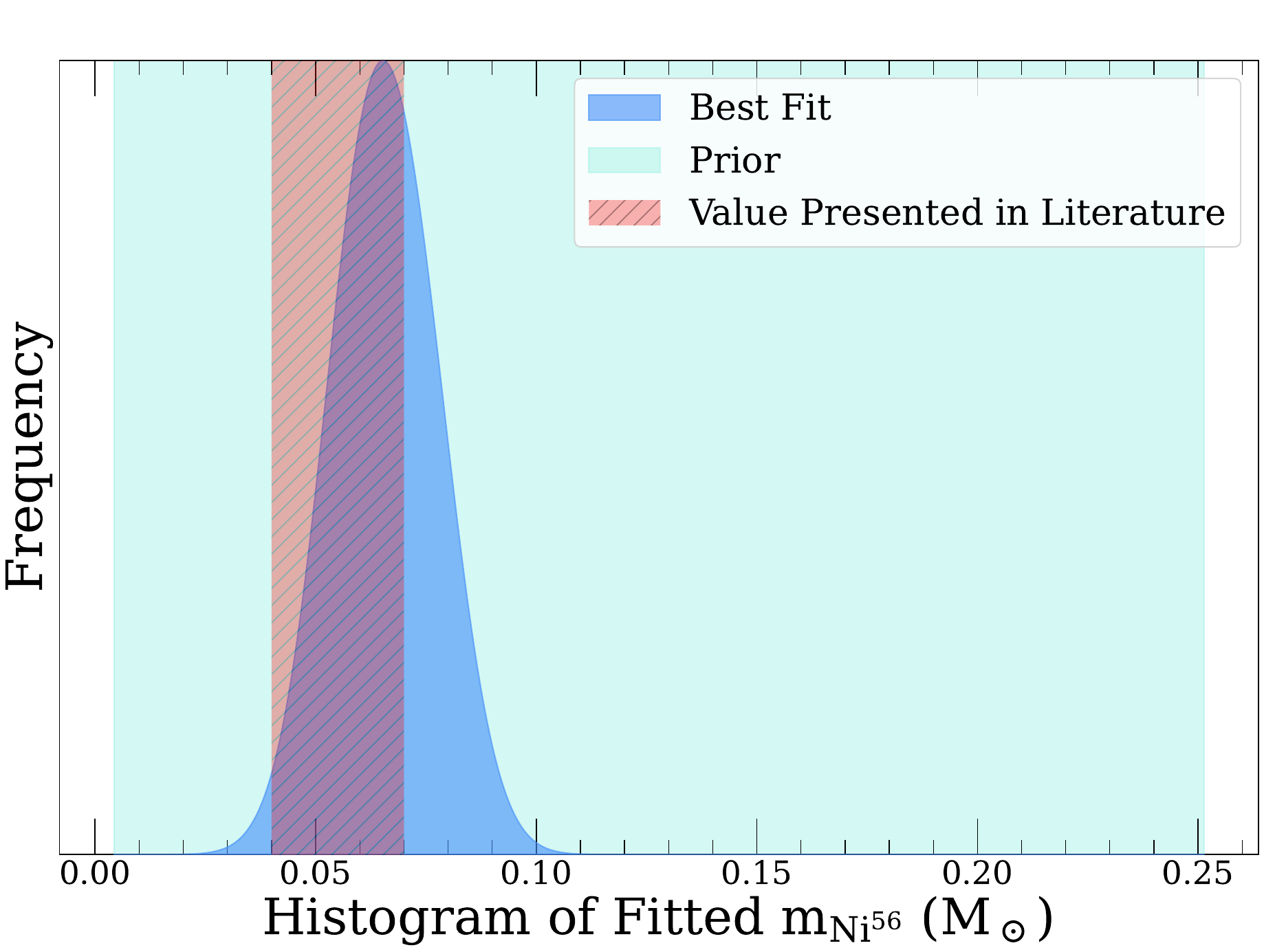}
    \end{minipage}

    \vspace{0.5em} 

    \begin{minipage}[t]{0.48\linewidth}
        \includegraphics[width=\linewidth]{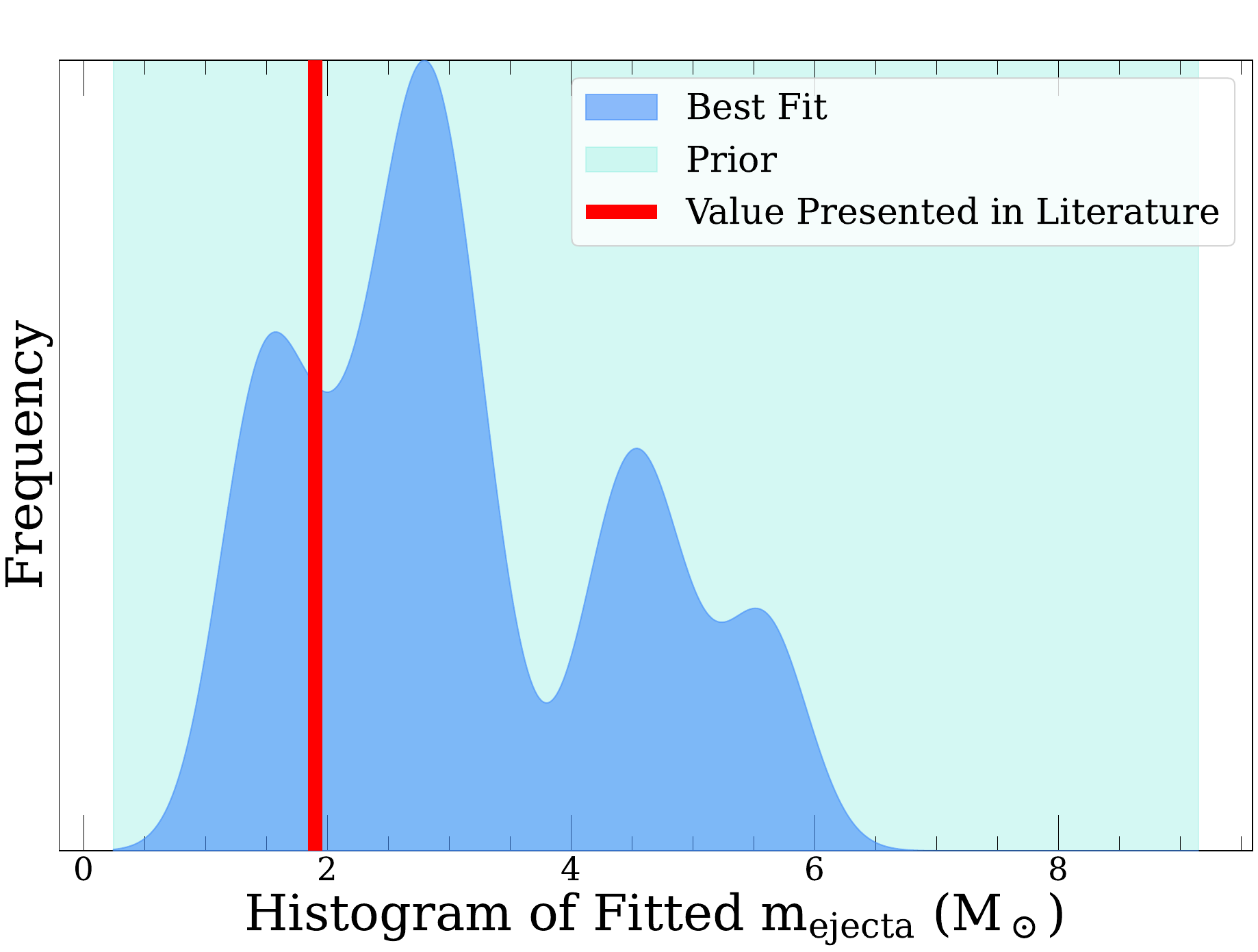}
    \end{minipage}
    \hfill
    \begin{minipage}[t]{0.48\linewidth}
        \includegraphics[width=\linewidth]{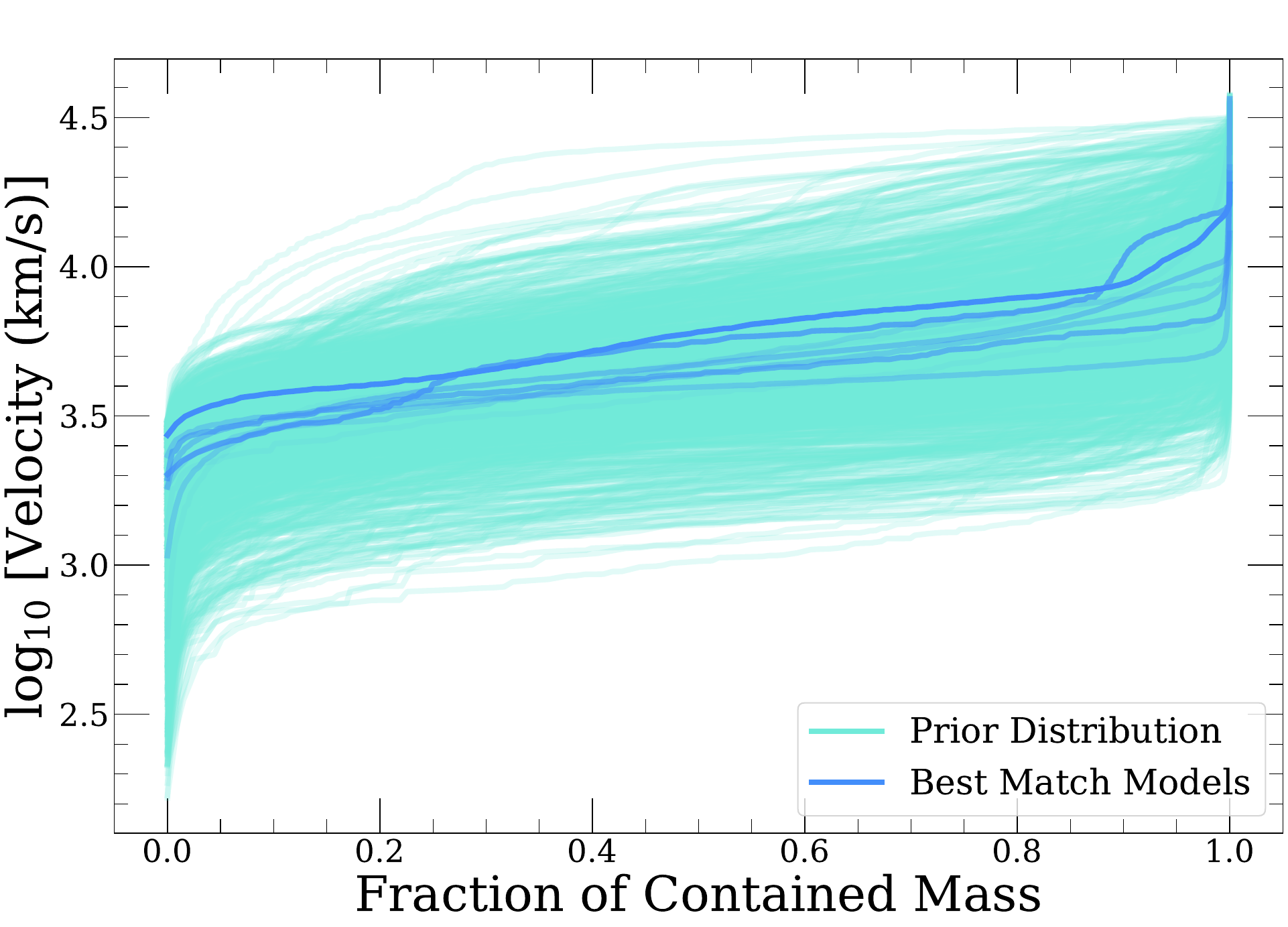}
    \end{minipage}

    \caption{A comparison between our generated grid of \lcs~to the SN~Ib iPTF13bvn \lc~in $BVRI$-bands. (top left) We show the closest several \lcs~to the \lc~of iPTF13bvn, and histograms of the \mni~values (top right), ejecta masses (bottom left), and velocity profile (bottom right) of the closest \lc~matches. Our range of \mni~values is consistent with the range of simulation-inferred literature values of \mni~\citep{Bersten2014, Fremling2014, Srivastav2014}.}
    \label{fig:13bvn_matches}
\end{figure*}

We begin by matching models to the \lc~of SN1994I. Among the best-matching models, the inferred \mni~values span $[0.05$,$0.1]$~\msun~(see Figure~\ref{fig:94I_matches}). This is consistent with \cite{afsariardchi2021}, who model the late-time \lc~tail of SN1994I to find an \mni$=0.08$. \cite{afsariardchi2021} also note that SN1994I's peak luminosity is nearly $50\%$ brighter than expectation for that \mni, based on MCRT models from\cite{dessart2016}. They propose that this may indicate a contribution from an additional power source (e.g. shock cooling, a central engine). However, we are able to find matches to the \lc~of this SESN within our grid of models powered only by \NI~decay. Our best matches also constrain \mej~for this SESN to $\leq3$~\msun. This is broadly consistent with spectral modeling by \cite{Sauer2006}, who infer $0.07$~\msun~of \NI~and $\approx1.2$~\msun of ejecta. Our best-fit models also have minimum velocities ($vel_0$) $\approx10^{3.5}$~km/s, with outermost velocities approaching $10^{4.5}$~km/s. It is not trivial to compare our inferred velocity distributions within the ejecta to e.g., single-line Doppler measurements, as those tend to trace specific line-forming regions. Connecting these profiles to light curve and spectroscopic velocity inference will be the focus of future work.

Next, we discuss our closest-match models to the \lc~of SN2007gr. The range of closest-match \mni~values and \mej~values we find for SN2007gr are $[0.06$,$0.12]$~\msun~and $[1,5]$~\msun, respectively (see Figure~\ref{fig:07gr_matches}). Both mass estimates agree with the values inferred by \cite{Hunter2009}, who find \mni~of $0.076$~\msun and an ejecta mass range of $2-3.5$~\msun via a two-component Arnett-like model originally presented in \cite{valenti2008broad}. Our best-match models imply that the velocity profile is steeper for SN2007gr than for SN1994I, with SN2007gr having lower inner velocities  of $\lesssim10^3.4$~km/s and outer velocities $\approx10^4.1$~km/s. Compared to SN1994I, the velocity is lower for SN2007gr throughout the ejecta. The similarity in ejecta profiles is consistent with line measurements from \cite{Hunter2009}. 

Finally, we present our closest-match models to the \lc~of iPTF13bvn. The best-matching model span \mni~values  $[0.04$,$0.075]$~\msun and ejecta masses $\lesssim6.2$~\msun (see Figure~\ref{fig:13bvn_matches}). Both  \cite{Bersten2014} and \cite{Fremling2014} modeled the bolometric \lc~via hydrodynamic simulations. \cite{Bersten2014} find a \mni~of $\simeq0.1$~\msun, while \cite{Fremling2014} find a somewhat lower \mni~of $0.04-0.07$~\msun. As discussed by \cite{Srivastav2014}, this difference is likely because \cite{Bersten2014} use adopted a higher E(B-V)$=0.17$ while \cite{Fremling2014} use an E(B-V)$=0.045$. Here, we use an E(B-V) value of 0.045, and find comparable results to those of \cite{Fremling2014}; however we note that the range of our best-match \mej~is quite broad. The velocity profiles of best-match models to iPTF13bvn are intermediate to those of SN1994I and SN2007gr. The inner velocities are $\approx10^{3.4}$~km/s and outer velocities are $10^{4.1}$~km/s. 

Modeling \lcs~of SESNe using MCRT is a promising approach to constraining  the physical properties of SESNe beyond the simplified inference from semianalytical models. Figures~\ref{fig:94I_matches}, \ref{fig:07gr_matches}, and \ref{fig:13bvn_matches} suggest that in addition to just \mni, both \mop~and the velocity profile may be constrained by light curves alone. In Section~\ref{sec:Inference Machine}, we explore how well each physical parameter can be constrained with simple, observable light curve features.

\subsection{Comparing Accuracy of Arnett to Grid-based Inference of \mni}
\begin{figure*}
    \centering
    \includegraphics[width=8cm, height = 6cm]{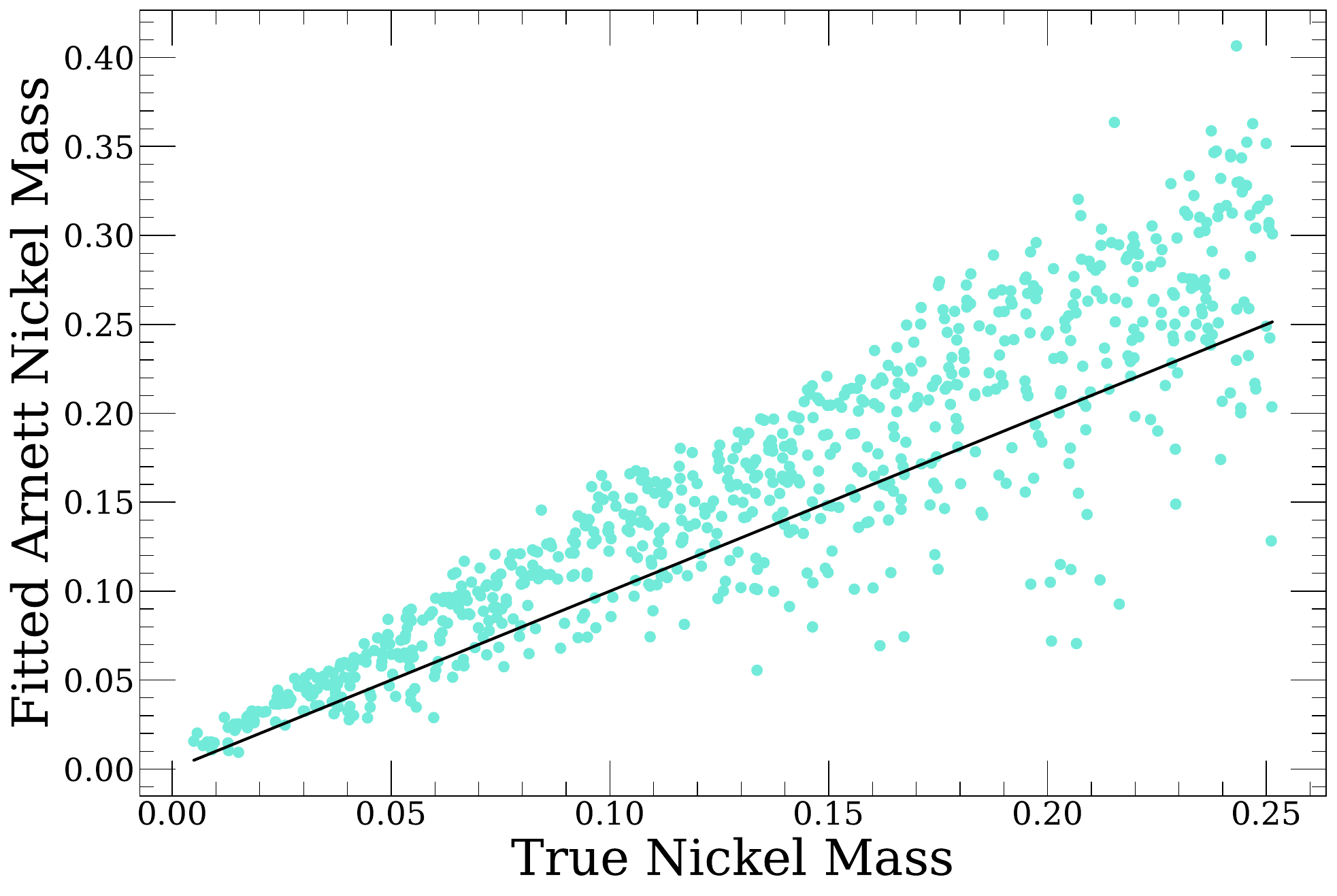}
    \includegraphics[width=8cm, height = 6cm]{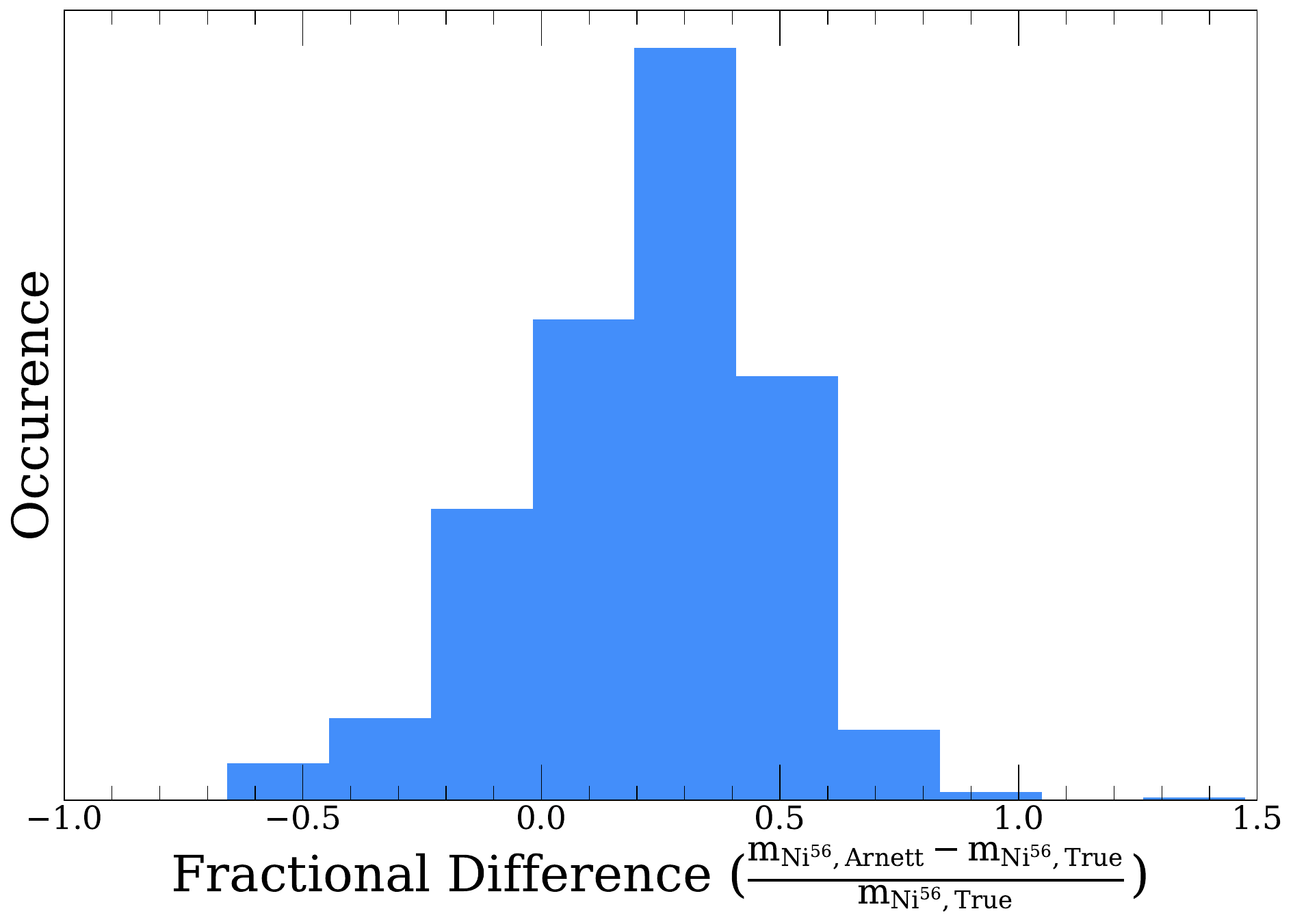}
    \caption{The values of \mni~derived from the Arnett model compared to the true \mni~(left panel) and the fractional difference between the two are shown (right panel). The Arnett model overestimates the \mni~of our models by, on average, $\simeq25\%$.}
    \label{fig:Arnett_Vs_real}
\end{figure*}

It is well-known that the Arnett model (and the associated ``Arnett rule") tends to overestimate the \mni~of \lcs~simulated by radiative transfer codes (e.g. \citealt{dessart2015, dessart2016, afsariardchi2021}). For example, in models presented by \cite{dessart2015} and \cite{dessart2016}, the Arnett rule overestimates the \mni~by $\approx50\%$. In Figure~\ref{fig:Arnett_Vs_real}, we compare the true \mni~and \mni~inferred from Arnett's rule using the peak luminosity. We find that Arnett's rule overestimates the \mni~by an average fractional difference 0.25 (right panel of Figure~\ref{fig:Arnett_Vs_real}), though this can be as high as $1$, or even as low as $-0.5$. The inaccuracy in the Arnett-derived \mni~is largest at largest \mni. Beyond this, we do not find obvious trends driving discrepency between our simulated models and those from Arnett's rule.

\section{Demonstrating Simple Physical Inference from the Grid}
\label{sec:Inference Machine}

In this Section, we explore which physical parameters can be constrained from the observed parameters of the \lc. This analysis motivates the second paper in this series, where we will present a heavily expanded inference methodology to be tested on a large sample of SESNe. 

\subsection{Regression Over Simple Observed Parameters}
\begin{figure*}
    \centering
    \includegraphics[width = 5.9cm, height = 4.15cm]{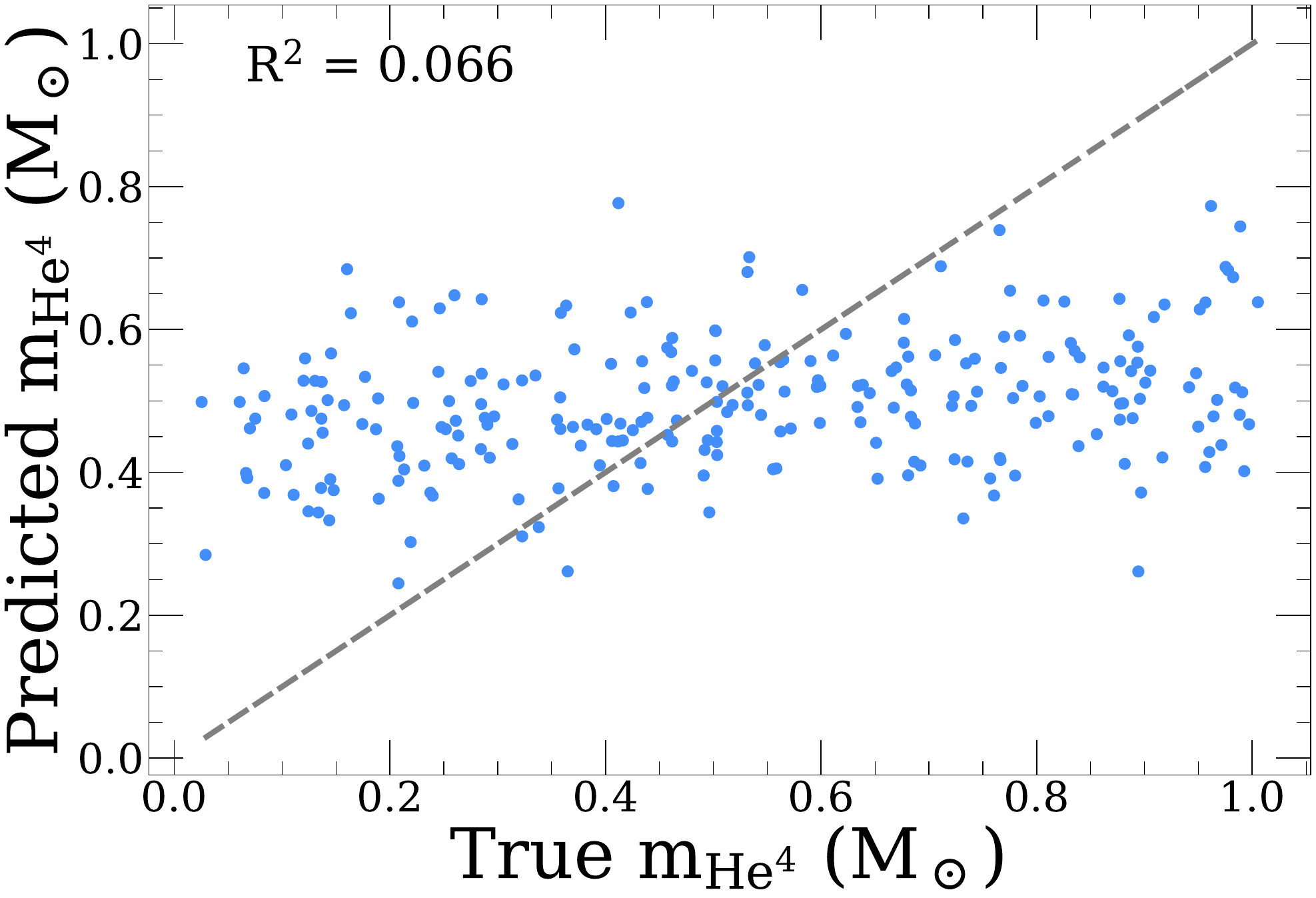}
    \includegraphics[width = 5.9cm, height = 4.15cm]{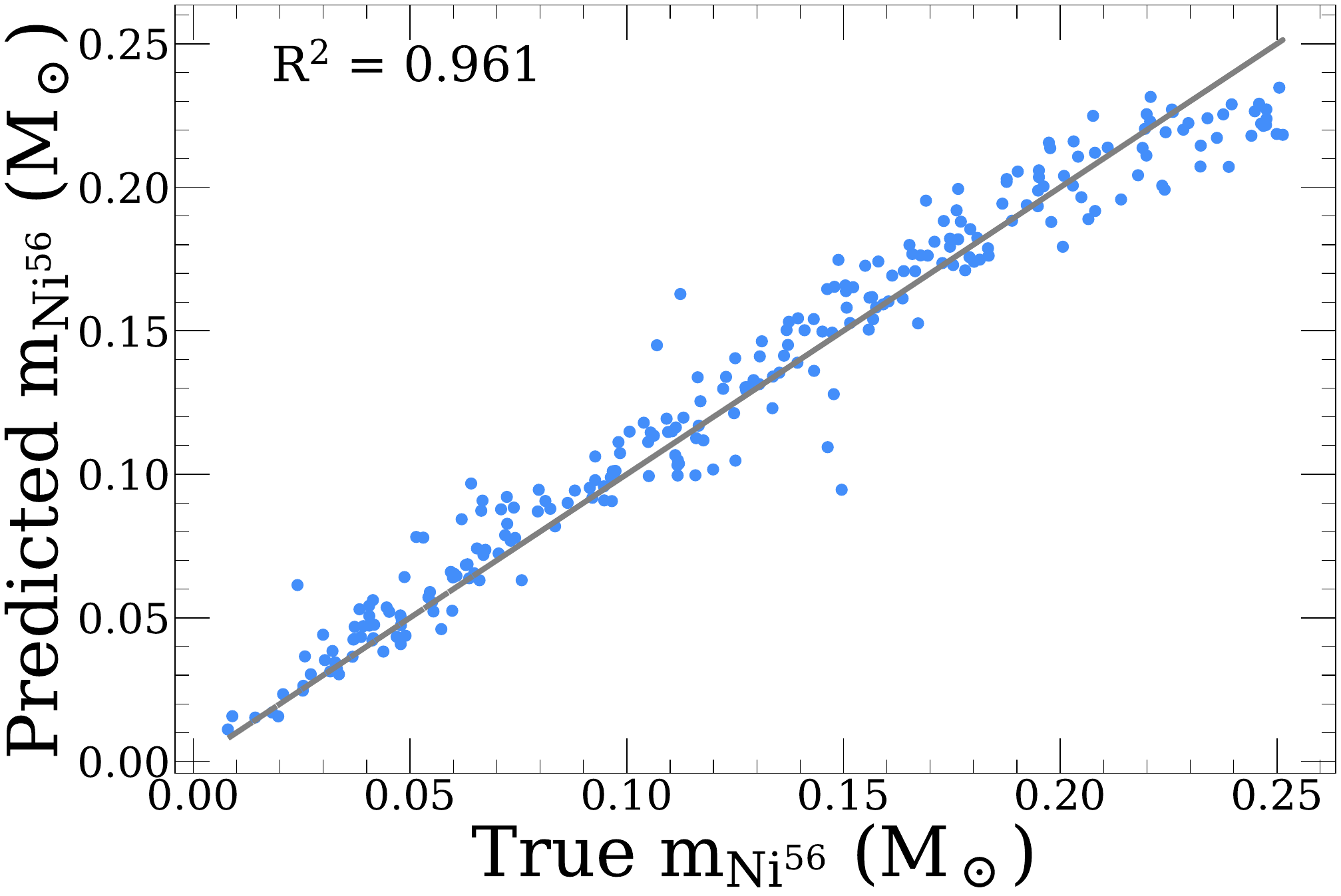}
    \includegraphics[width = 5.9cm, height = 4.15cm]{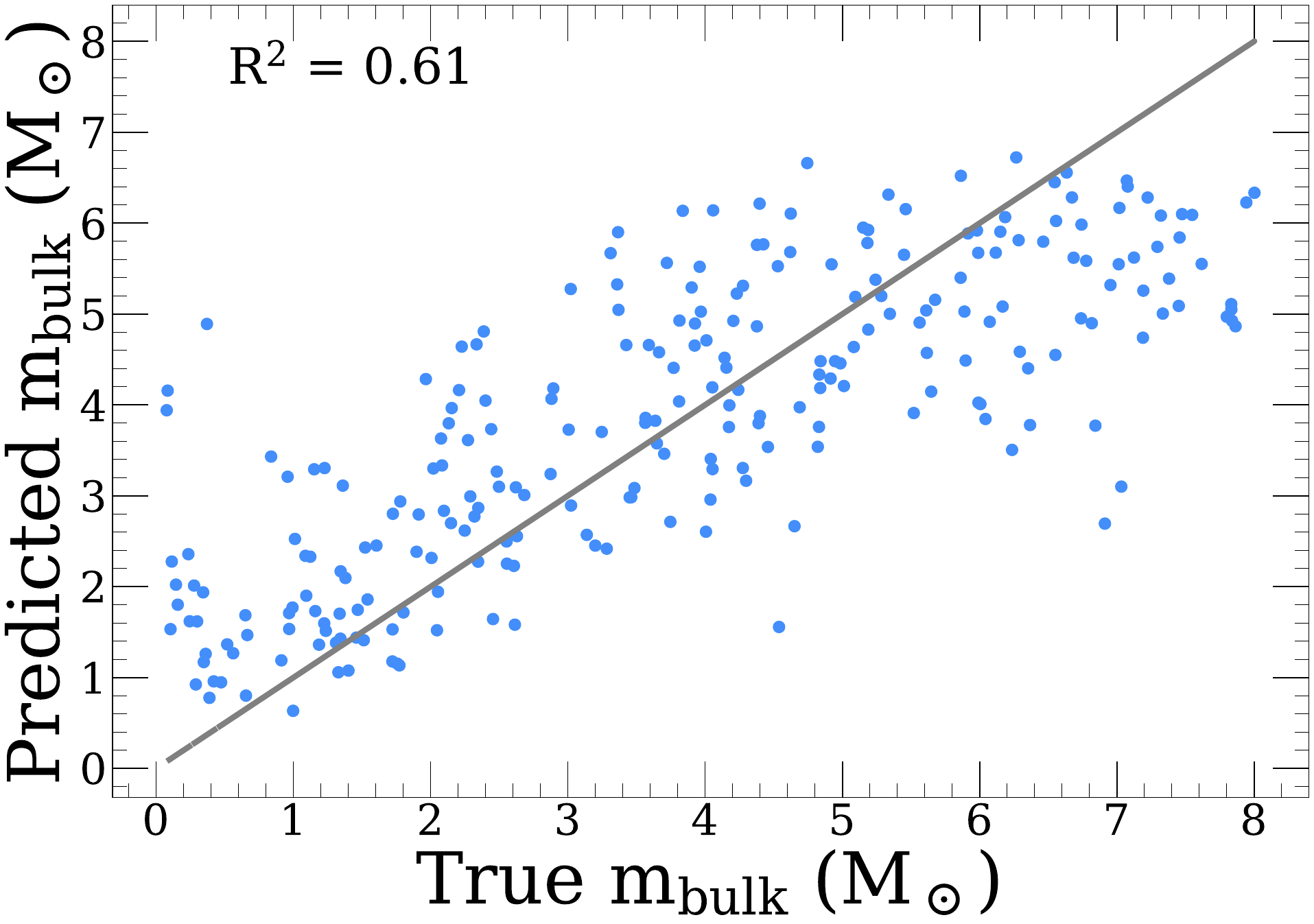}
    \includegraphics[width = 5.9cm, height = 4.15cm]{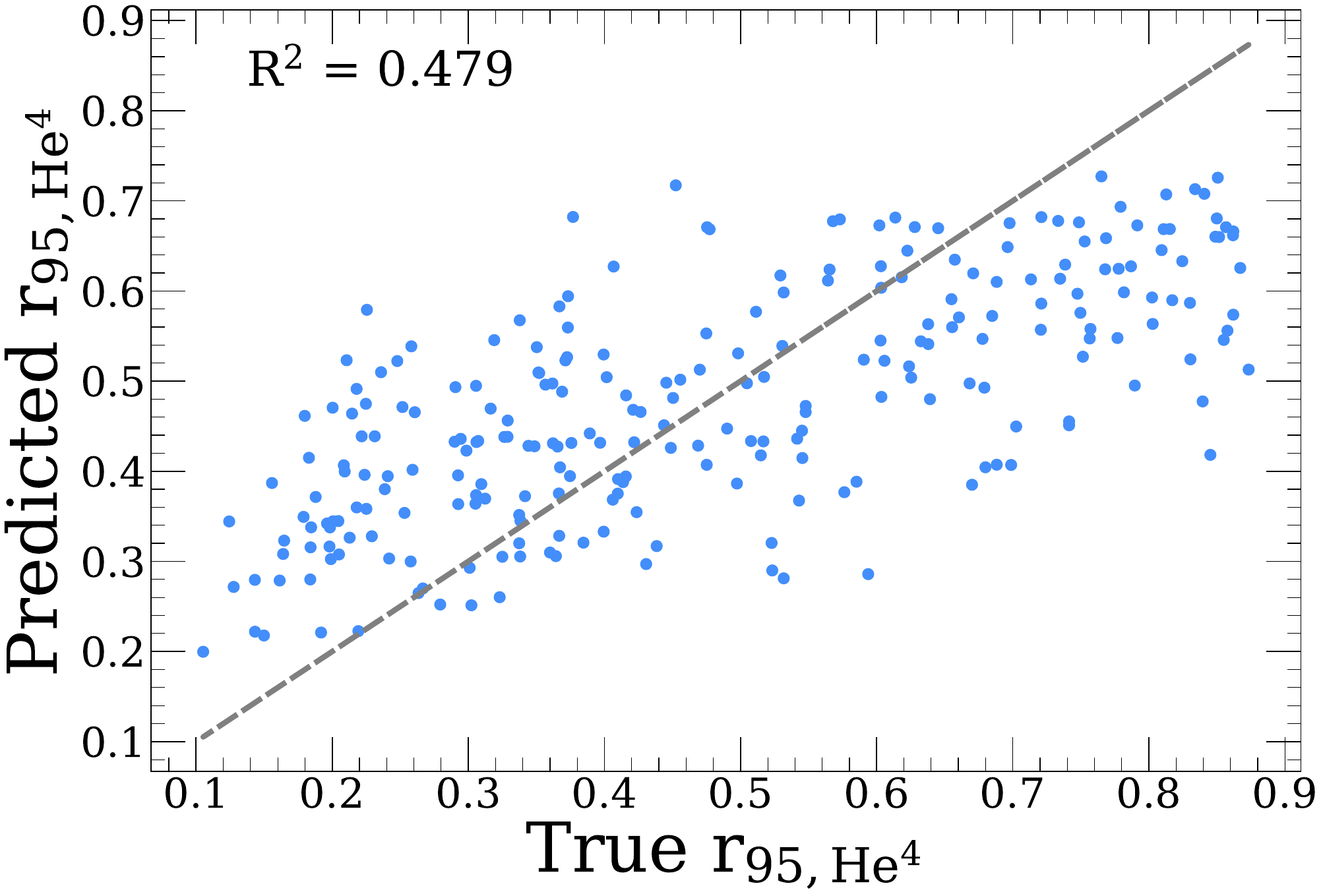}
    \includegraphics[width = 5.9cm, height = 4.15cm]{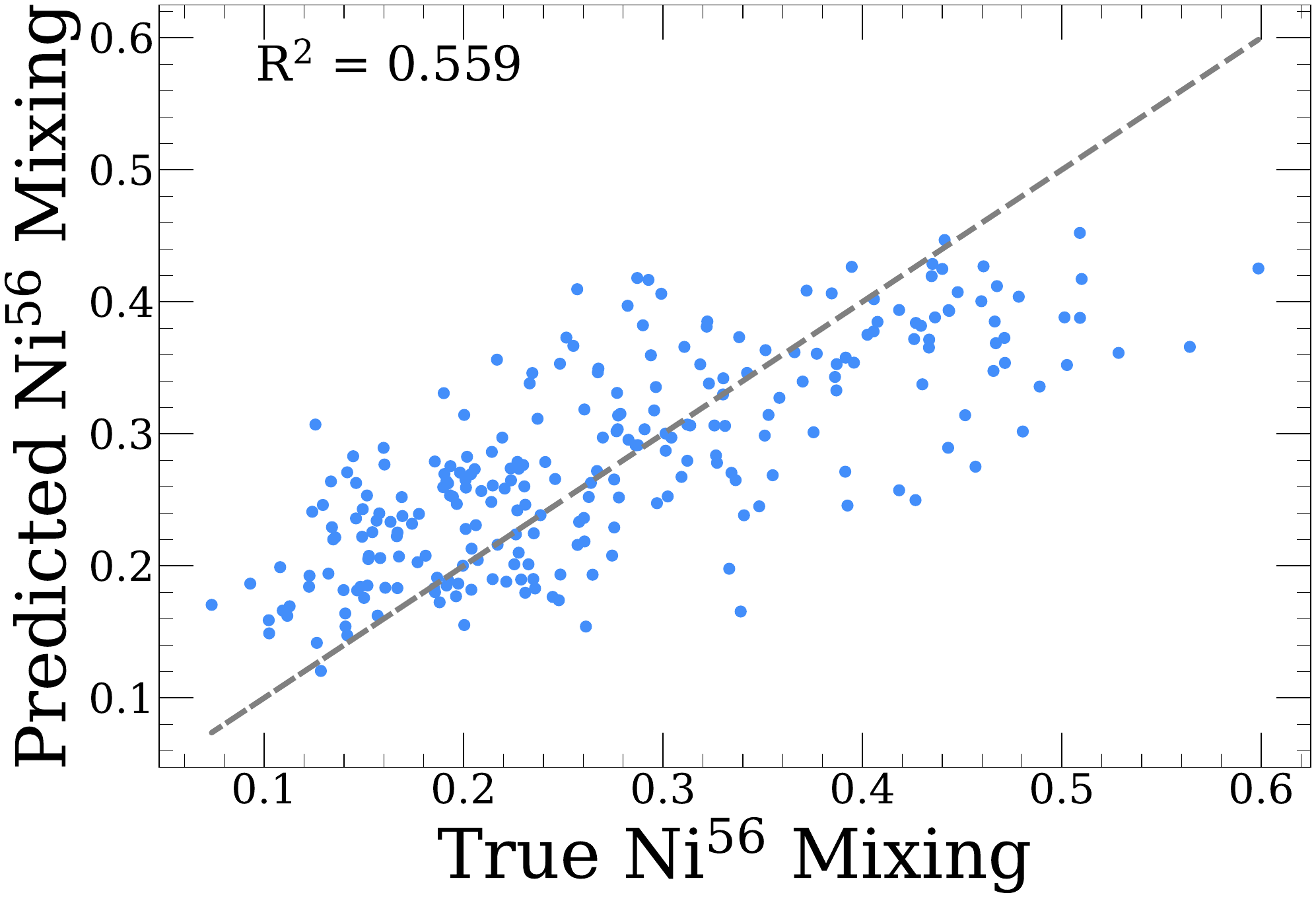}
    \includegraphics[width = 5.9cm, height = 4.15cm]{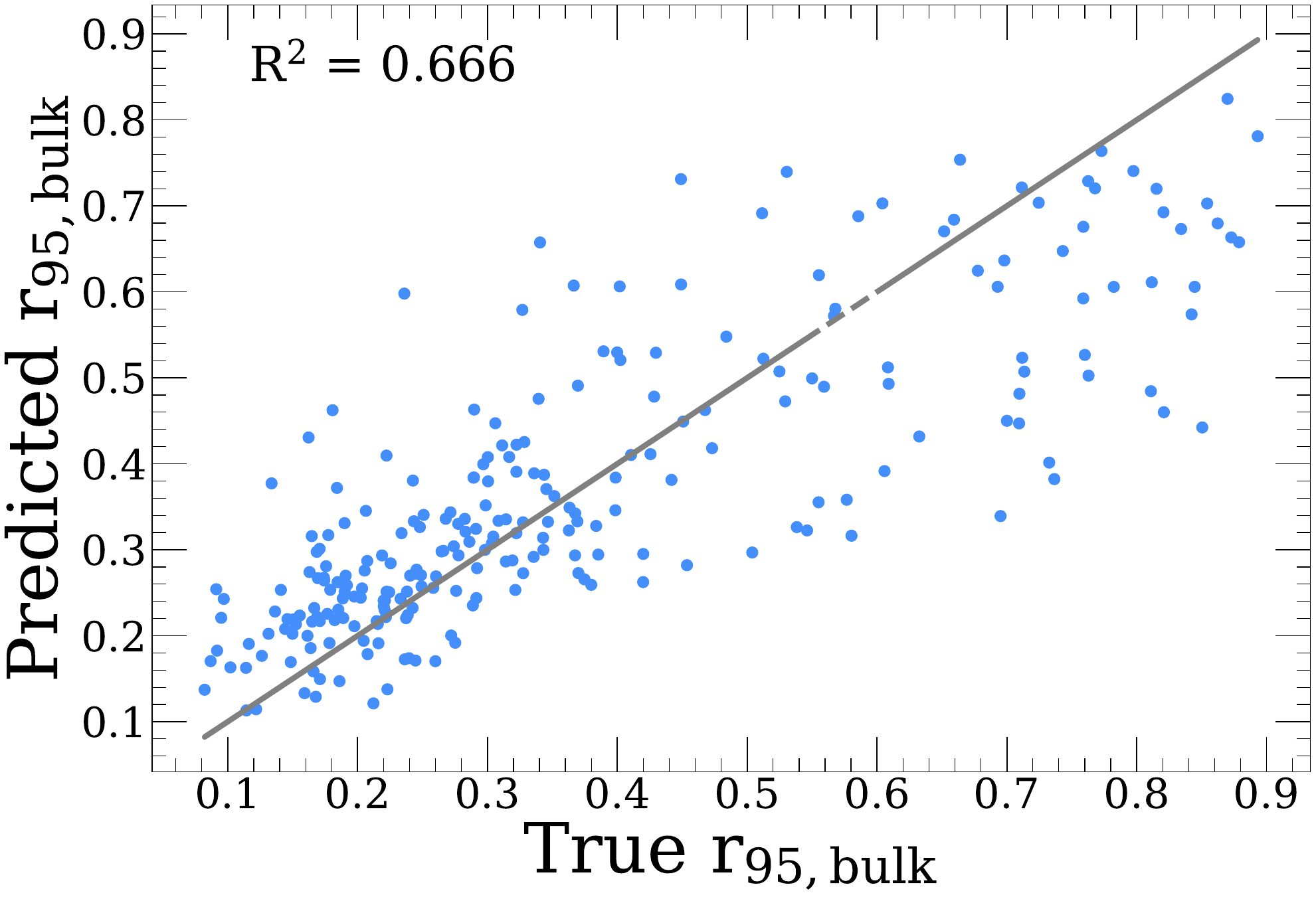}
    \includegraphics[width = 5.9cm, height = 4.15cm]{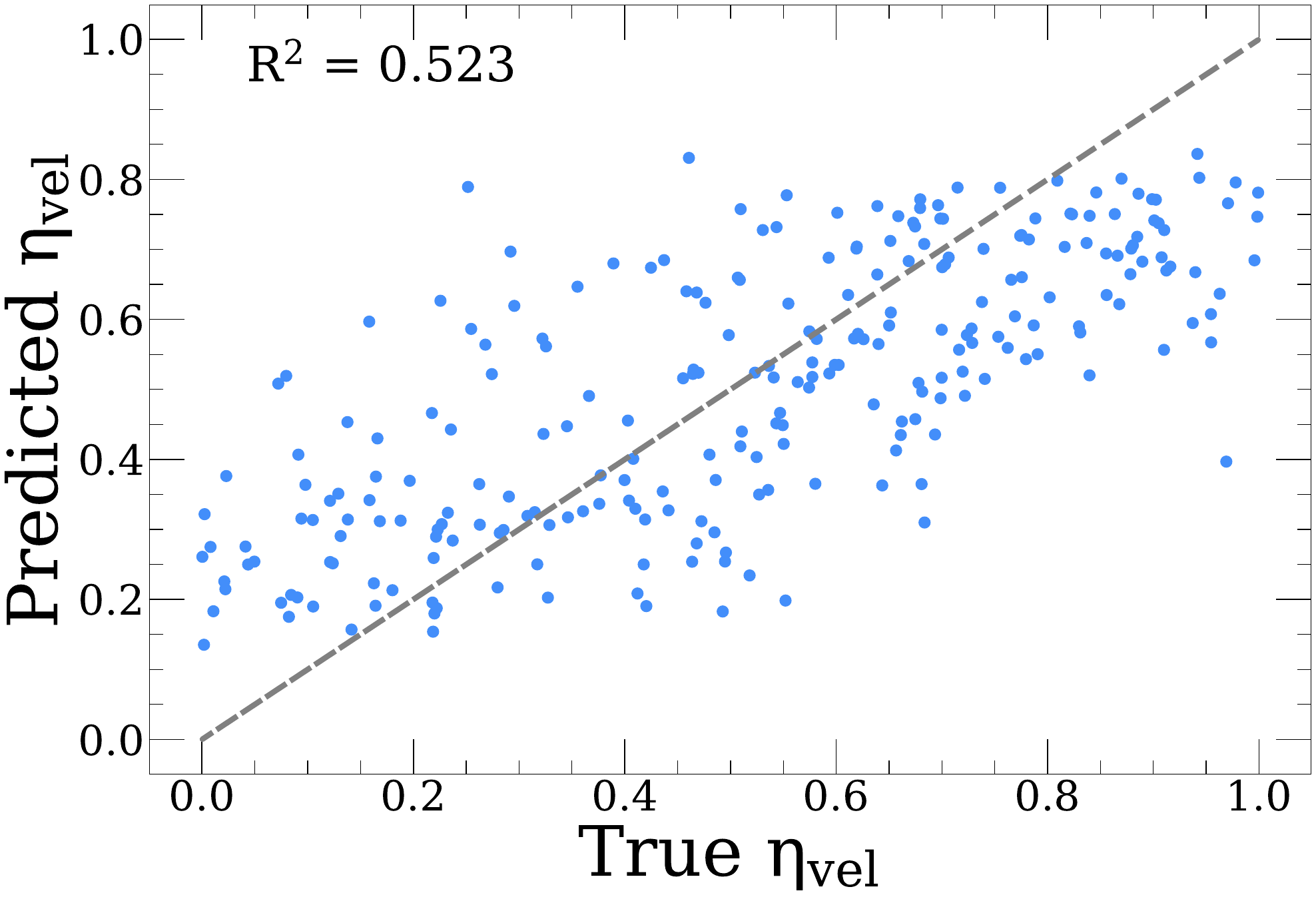}
    \includegraphics[width = 5.9cm, height = 4.15cm]{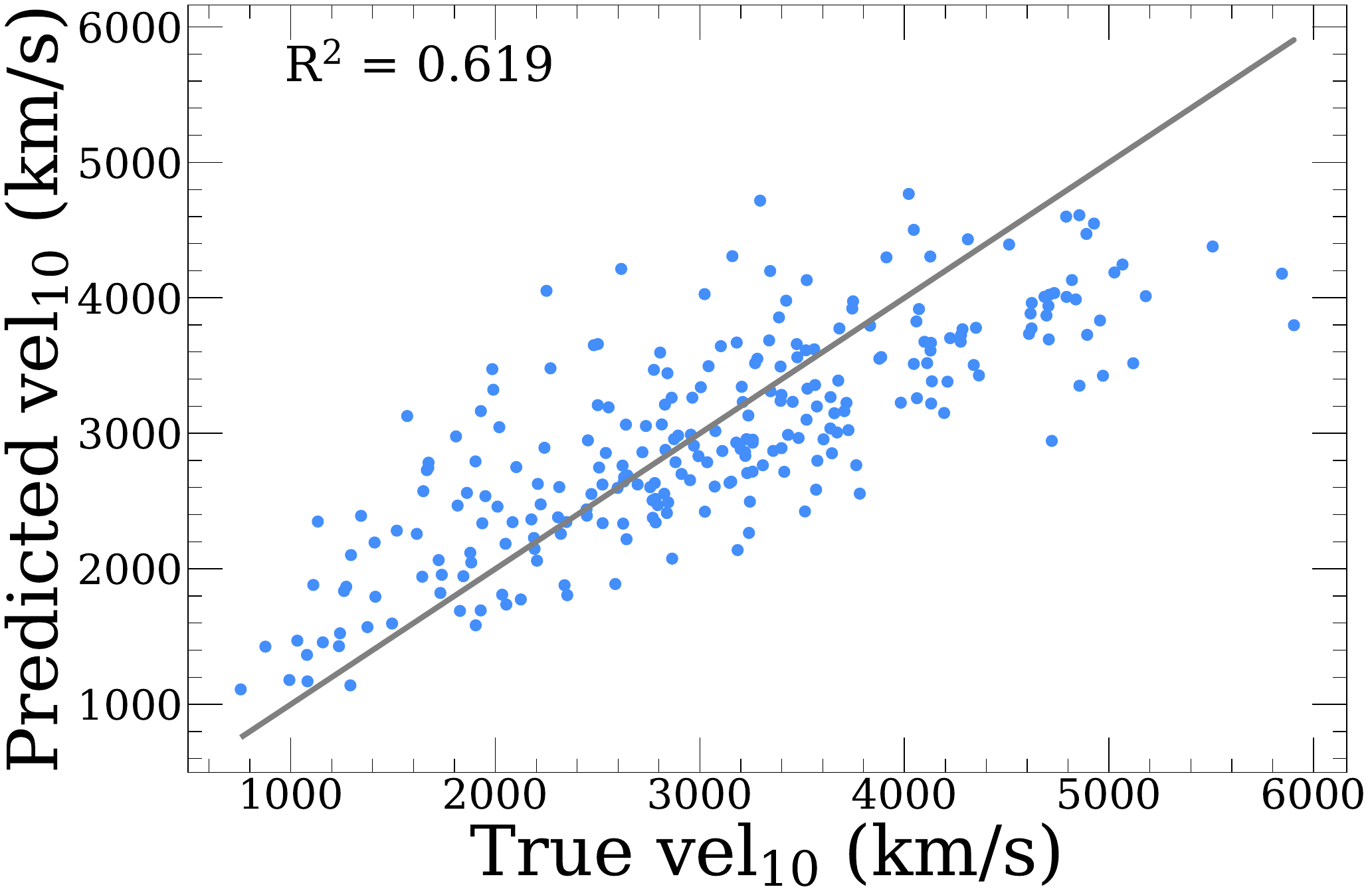}
    \includegraphics[width = 5.9cm, height = 4.15cm]{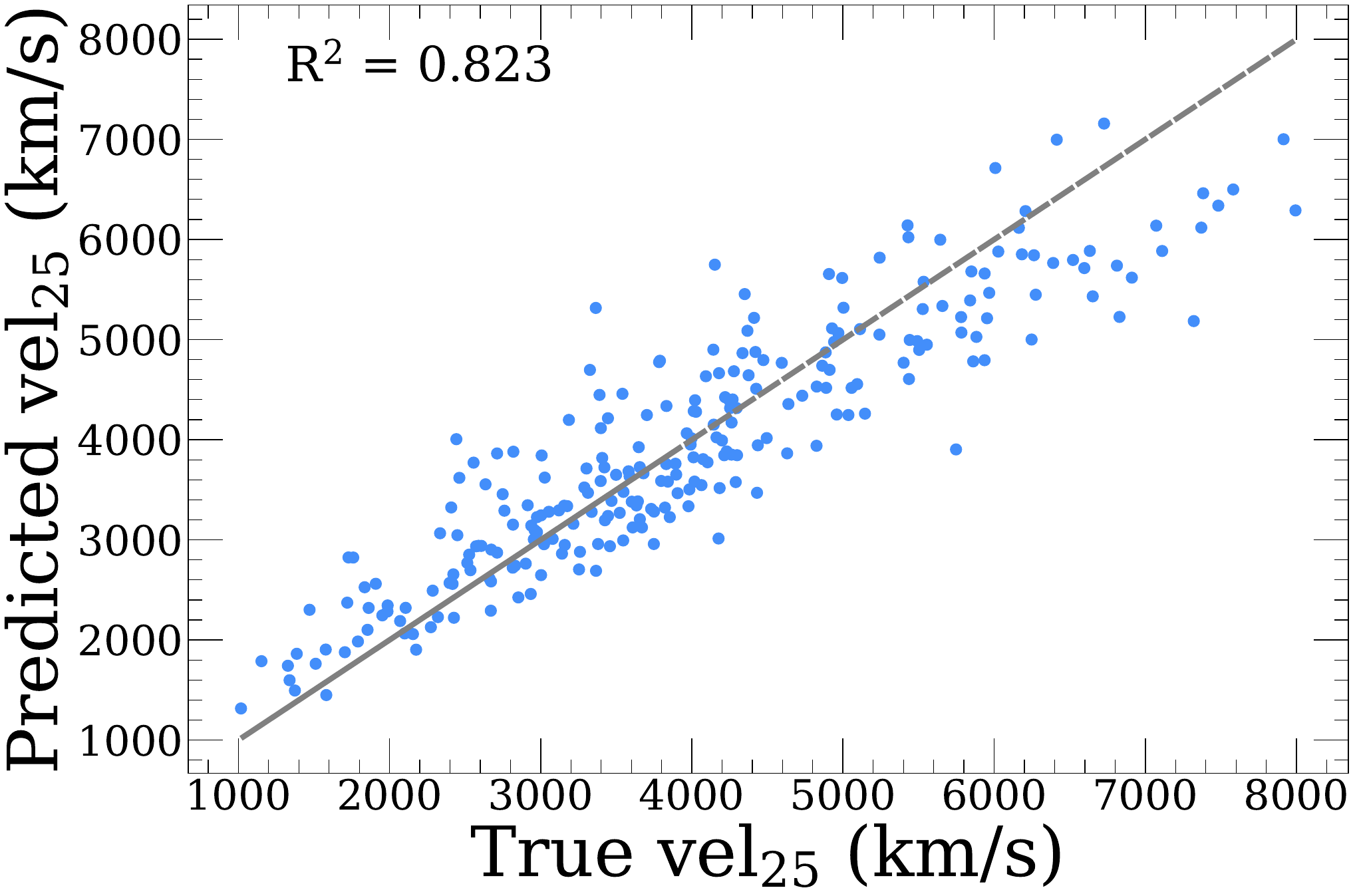}
    \includegraphics[width = 5.9cm, height = 4.15cm]{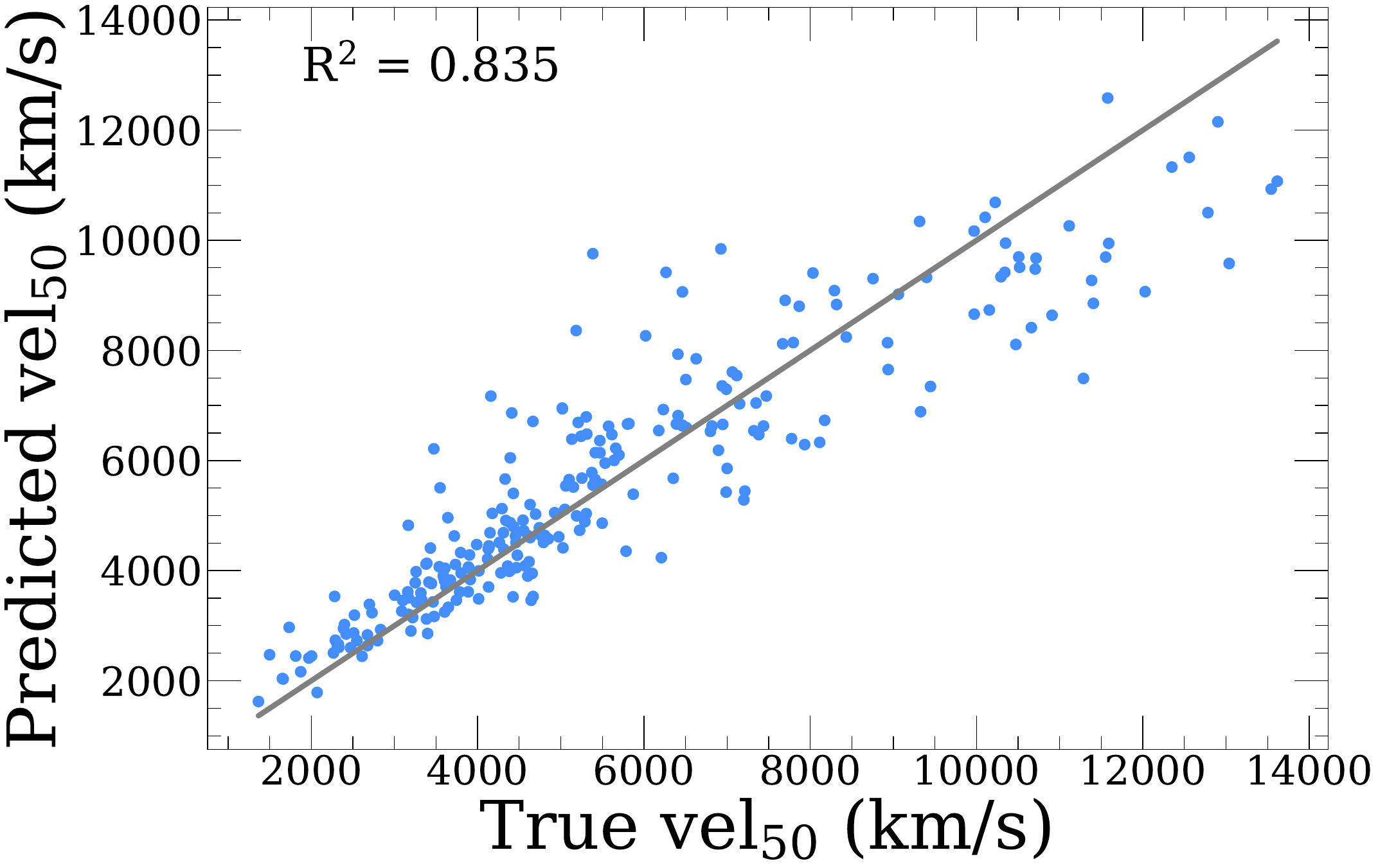}
    \includegraphics[width = 5.9cm, height = 4.15cm]{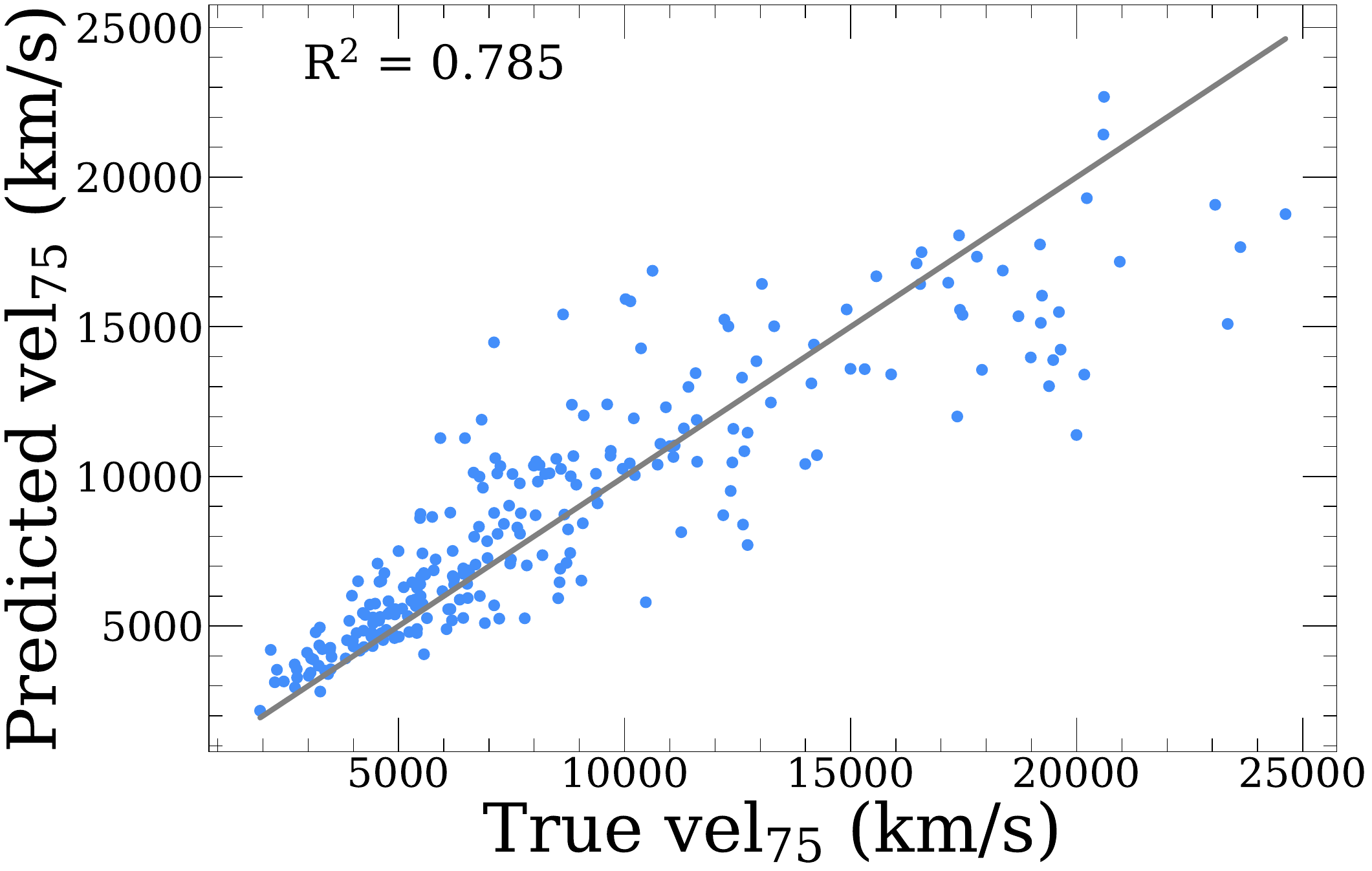}
    \includegraphics[width = 5.9cm, height = 4.15cm]{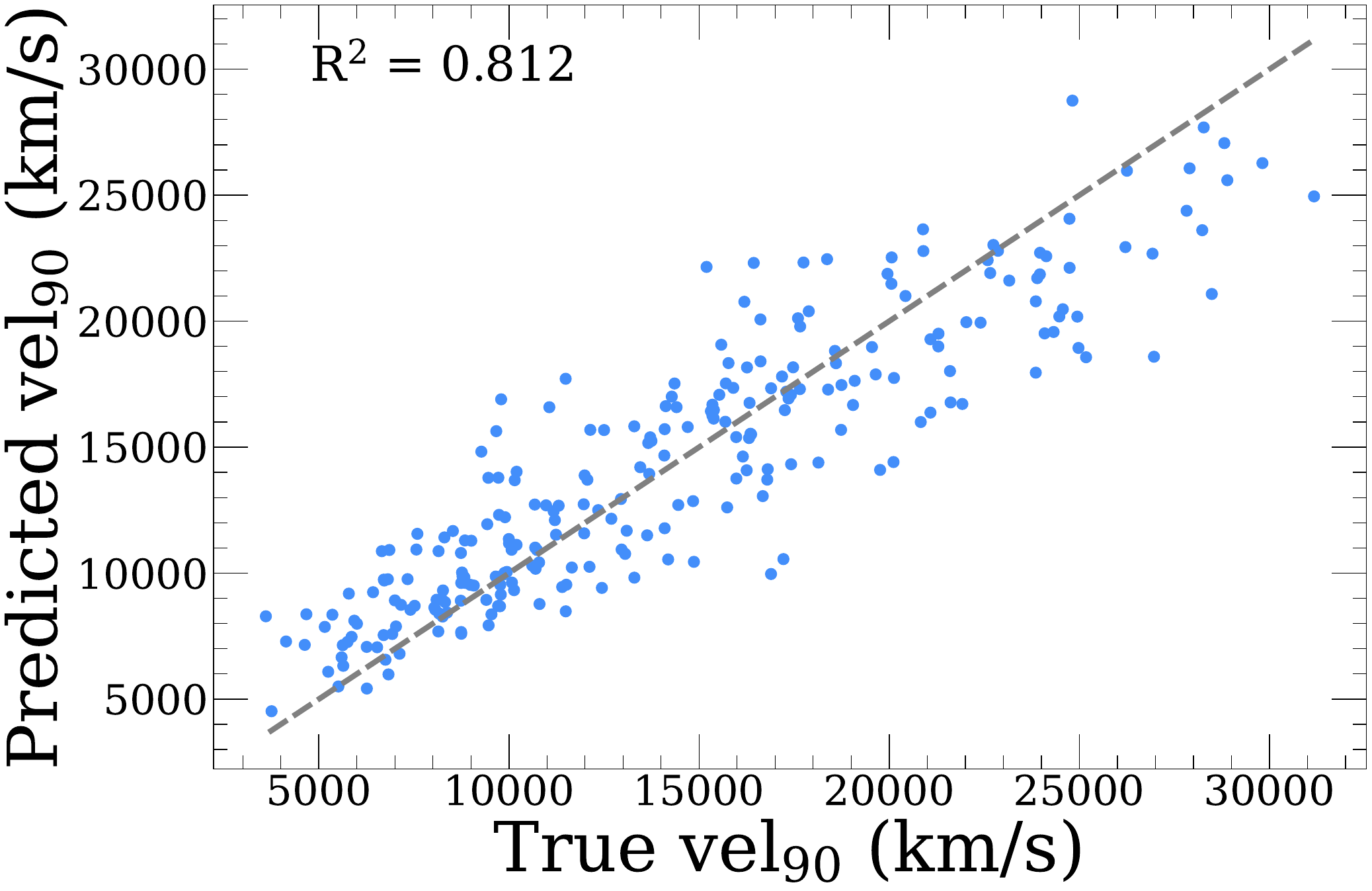}
    \caption{Comparison of physical parameters of the ejecta profile as inferred from our random forest regression methodology vs the true values. The \mni, \mop, \gamnn, \velt, \velf, and \vels~parameters are all well-captured by the inference fit, suggesting that the \NI~mass distribution, \opacity~mass distribution, and velocity distributions can all be potentially constrained by the parameters of this fit.}
    \label{fig:bol_mass_inference}
\end{figure*}

Over the grid of $1$,$000$ \lcs~detailed in Section~\ref{sec:grid}, we run random forest regression on the 15 observed parameters (\tr, \td, and peak luminosity in each of $ugriz$) to derive the nine physical parameters. To establish a baseline for interpretability, we include an additional parameter, ``$\rm random\_val$'', which is a random number with no physical meaning. This serves as a lower bound: any observational feature that is measured to be ``less important" than $\rm random\_val$ must be irrelevant to physical inference.  We show the predicted versus true values for each parameter in Figure~\ref{fig:bol_mass_inference}. In the following paragraphs, we will highlight the physical parameters that can be reliably recovered by this inference and discuss which observed parameters contribute to this inference. We also present the R-squared coefficient of determination, a commonly used measure to characterize goodness of fit when performing regression (see e.g. \citealt{draper1998applied}), between the predicted and true physical parameters to characterize the efficacy of regression for each physical parameter. 

In the top row of Figure~\ref{fig:bol_mass_inference}, we show how well \mhe, \mni, and \mop~are recovered by the inference fit. We find, unsurprisingly, that \mhe~cannot be reliably inferred ($R^2 = 0.07$), with no observed features being more important than $\rm random\_val$. Also as expected, the mass of \NI~(top center panel of Figure~\ref{fig:bol_mass_inference}) can be inferred well ($R^2 = 0.96$). Unsurprisingly (because the \mni~correlates with the total energy emitted in the \lc), the most important features for determining \mni~are the peak fluxes in $g$~band, $r$~band, and $i$~band. However, the $g$~band peak flux is by far the strongest parameter in determining \mni. This is likely due to a combination of 1) the range of peak absolute magnitudes in $g$-band is quite large ($-12.4$ to $-18.6$) and 2) because many models are brightest in $g$-band during peak. The bulk mass (see top row of Figure~\ref{fig:bol_mass_inference}) can also be inferred reasonably well ($R^2 = 0.61$). The most important features of determining the bulk mass are $u$-band peak flux and the $g$-band \td. We also note that we are able to capture the total ejecta mass (i.e. \mhe+\mop+\mni) just as well, with $R^2=0.61$.

The mixing extent of \he, \NI, and bulk are shown in the second row of Figure~\ref{fig:bol_mass_inference}. The mixing of \he~(measured by \gamnh; bottom left panel of Figure~\ref{fig:bol_mass_inference}), is somewhat captured by the regression fit over the observed parameters of the \lc~($R^2 = 0.48$). The $z$-band \td~is the only important feature of this fit, likely because of a slight enhancement of the \ion{He}{1} $\lambda10830$ line as \he~is mixed outwards. However, because non-thermal \he~excitation is not modeled in \sed, this inference should be reevaluated in future work (also see Section~\ref{sec:sedona_exp}).  The extent of \NI~mixing (measured by \gamnn; second row center panel of Figure~\ref{fig:bol_mass_inference}) is also somewhat captured ($R^2=0.56$). The value \gamnn~is most strongly impacted by \tr~and \td~in each of the filters, with the $i$-band \tr~and $z$-band \td~being the most important. This suggests that \NI~mixing more strongly influences the timing of the \lc, rather than the peak luminosity of the \lc. This has been noted before \citep{dessart2012, dessart2015, Moriya2020}, where models with more mixed \NI~tend to have the same peak luminosity and a gentler rise to peak (i.e. longer \tr). Almost all other parameters have lower feature importance than $\rm random\_val$. The bulk mixing (\gamnop) is, again, somewhat well-captured, with $R^2=0.67$ (second row right panel of Figure~\ref{fig:bol_mass_inference}). The most important parameter for this fit is the difference between times of peak in $g$-band and in $r$-band. This suggests that changing \gamnop~can dramatically change how long the ejecta takes to cool, likely because mixing the bulk while holding \NI~mixing constant can significantly vary the diffusion time of photons from \NI~decay. Interestingly, neither\tr~nor \td~in any band are important in predicting \gamnop, suggesting that \gamnop~changes the time that a model takes to cool without changing \tr~or \td. 

The third and fourth rows of Figure~\ref{fig:bol_mass_inference} show how well the velocity profile is captured by our simple regression setup. Although \Dp~is only somewhat recovered ($R^2=0.52$), individual percentiles of the velocity distribution are captured with much higher fidelity. Inner ejecta velocities are least constrained, likely because most of the mass is centrally concentrated, and because \tr~and \td~are more sensitive to outer ejecta velocities. The parameters \velte~ ($R^2 = 0.62$), \velt~($R^2=0.82$), \velf~($R^2=0.84$), \vels~($R^2=0.79$), and \veln~($R^2=0.81$) are all very well-captured. The most important features for predicting velocity percentiles are the \tr~in all filters. The \tr~in redder filters tend to be more important features for predicting higher percentile velocities. In particular, the \tr~in the $g$- and $r$-bands are the most important features for predicting \velte, while the \tr~in $r$-band is the most important feature for predicting \velt, and the \tr in $i$-and $z$-bands are the most important features for predicting \velf~and \vels. This pattern is physically intuitive. Because \NI~tends to be constrained to the inner ejecta, changes to the inner (smaller) velocities more directly influence the \NI-enriched layers which also more directly influence the bluer portion of the SED. Therefore, the \tr~of bluer filters are more strongly influenced by changes to the inner velocity profile. 

\subsection{New Fitting Functions for Peak Bolometric Luminosity and \NI~Mass}
Next, we construct an equation using least-squares regression to predict the peak bolometric luminosity from the physical parameters in Equation~\ref{eq:reconstructed_arnett_rule}. 

\begin{multline} 
\label{eq:reconstructed_arnett_rule}
    L_{\rm peak, bol} = 10^{40.1} \left(\frac{m_{\rm Ni^{56}}}{\rm M_\odot}\right)^{0.81}\times \left(\frac{v_{25}}{\rm 3,500~km~s^{-1}}\right)^{0.29}\times \\ \left(\frac{v_{50}}{\rm 5,000~km~s^{-1}}\right)^{0.04}\times \left(\frac{v_{75}}{\rm 10,000~km~s^{-1}}\right)^{0.12} \rm erg/s
\end{multline}
The bolometric peak luminosity scales as $\propto m_{\rm Ni^{56}}^{0.8}$, which is comparable to the Arnett rule, where this exponent is equal to $1$. We find this equation reproduces the bolometric peak luminosity with $R^2 = 0.92$.

We also construct equations to infer \mni~from the bolometric light curve properties:
\begin{multline} 
\label{eq:mni_bolometric} 
    m_{\rm Ni} =0.04\left(\frac{L_{\rm peak,~bol}}{\rm 10^{42}~erg~s^{-1}}\right)^{1.04}\times \left(\frac{t_{\rm  rise,bol}}{\rm 15~d}\right)^{0.30} \\ \times \left(\frac{t_{\rm  decline,bol}}{20~\rm d}\right)^{0.41} ~\rm{M_\odot}
\end{multline}


Using Equation~\ref{eq:mni_bolometric}, we find a mean fractional difference between predicted and true \mni~of $\sim6\%$ (and $R^2=0.96$). In contrast, the Arnett Rule has a larger mean fractional difference of $\sim25\%$. As a proof of concept, we apply this equation to the bolometric light curve properties of SN1994I, SN2007gr and iPTF13bvn, finding \mni$\sim0.4-0.6$ for the three events, at the lower end of our matched-model inference technique. 

In this Section, we have shown that it is possible to constrain many physical parameters using simple, observable features from the \sed~models. These inferences can potentially be improved if 1) a more detailed inference algorithm were used, 2) more attention were given to interpolating between models in the nine-dimensional physical space, and 3) inference were done on the entire observed \lc~or even the entire observed SED. All three of these improvements will be implemented in the next study presented by the authors of this work.

\section{Expanding Range of \NI~Mass Distributions}
\label{sec:expanded_grid}
In this Section, we probe how varying the \NI~mass distribution beyond the range of \NI~mass distributions from the W21 grid impacts the \lcs. Prior work has suggested that \NI~mixing can drive the observed diversity in SESNe, and furthermore that \NI~mixing can change the shape and color of the resulting light curve \citep{Khatami2019, afsariardchi2021,Rodriguez2024}. To test this, we explore whether changing the distribution of \NI~alone can significantly alter an SESN \lc. Because the range of \NI~mass profiles presented in W21 is constrained to a tight range of \NI~mass distributions (where \gamnn~only varies by $\sim$0.3), we introduce a larger range of generated \NI~mass profiles and present their \lcs. We randomly choose 200 models from the original grid of $1$,$000$ samples, and for each model simulate four additional \NI~distributions (thereby generating a second grid of 800 samples). These new distributions are generated using zero-centered Gaussian distributions of varying widths in velocity space. This setup allows us to isolate and better understand the qualitative impact of \NI~mixing on the resulting \lcs. 

In exploring this grid, we find two dominant \lc~behaviors resulting from mixing the \NI~farther out into shallower layers of the ejecta. First, in most models where \NI~is confined to within the bulk of the ejecta, we observe that increased mixing leads to a faster rise and decline, but no substantial change in the peak luminosity. Second, in a smaller subset of models in which \NI~is mixed far enough out such that a significant fraction is ahead of the \opacity~mass, we observe a more substantial change in luminosity with little change to the time of the peak. Our results suggest that \NI~mixing, has minimal impact on the bolometric luminosity of the light curve unless it is mixed well into the outer layers, consistent with the findings of e.g., \cite{dessart2012, dessart2015, Moriya2020, Woosley2021}.

\begin{figure*}
    \centering
    \includegraphics[width=0.49\linewidth]{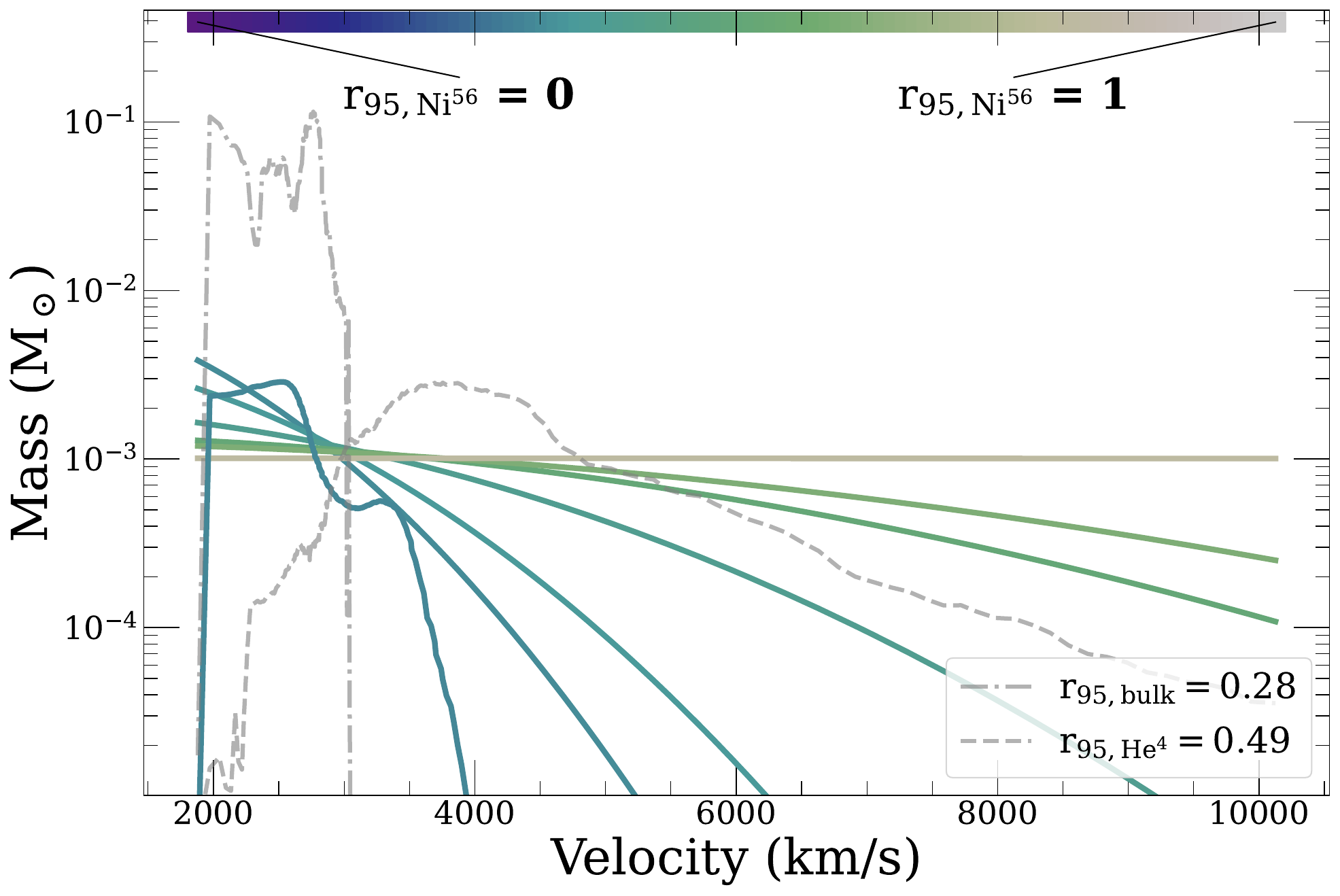}
    \includegraphics[width=0.49\linewidth]{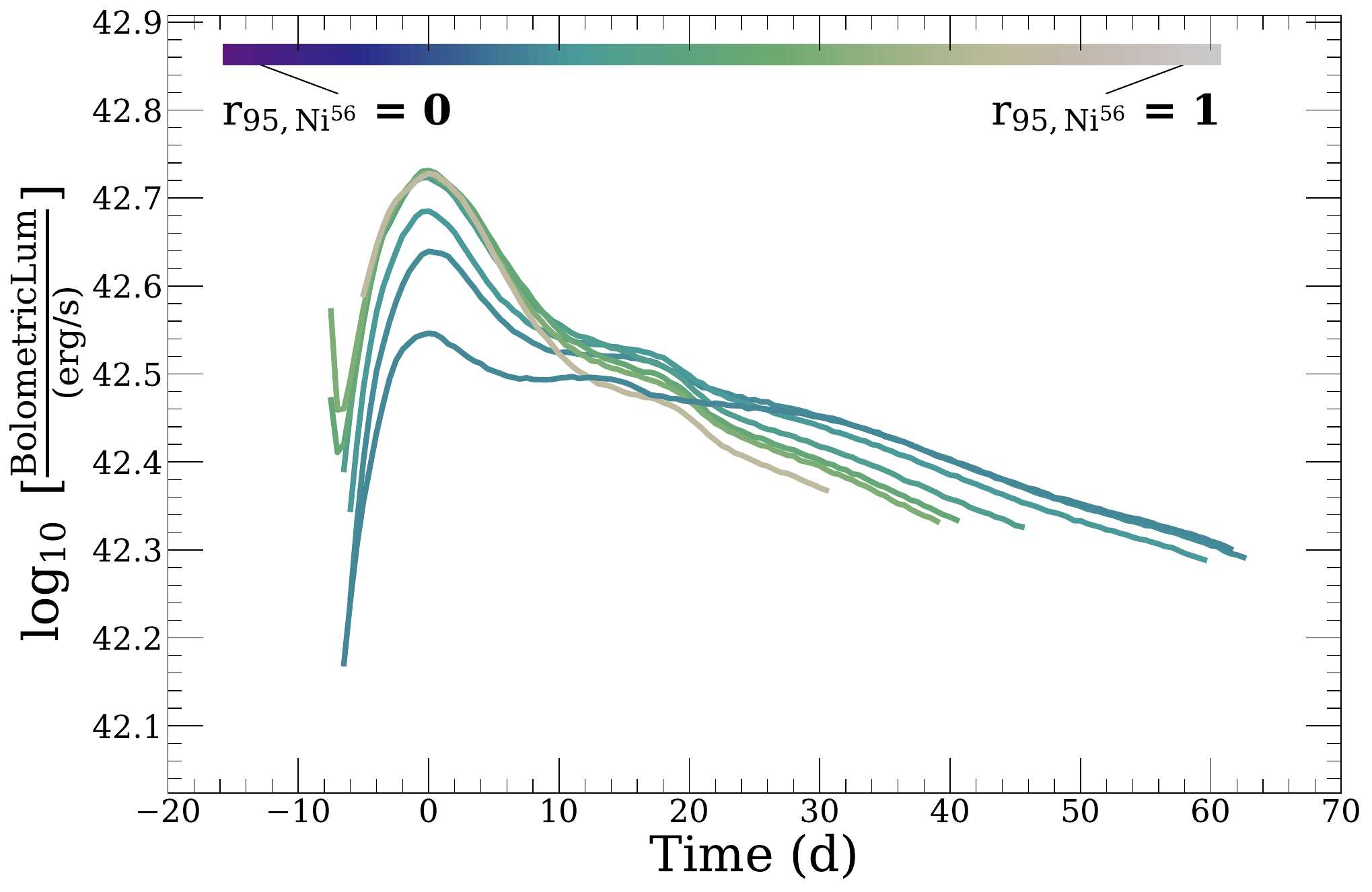}
    \includegraphics[width=0.49\linewidth]{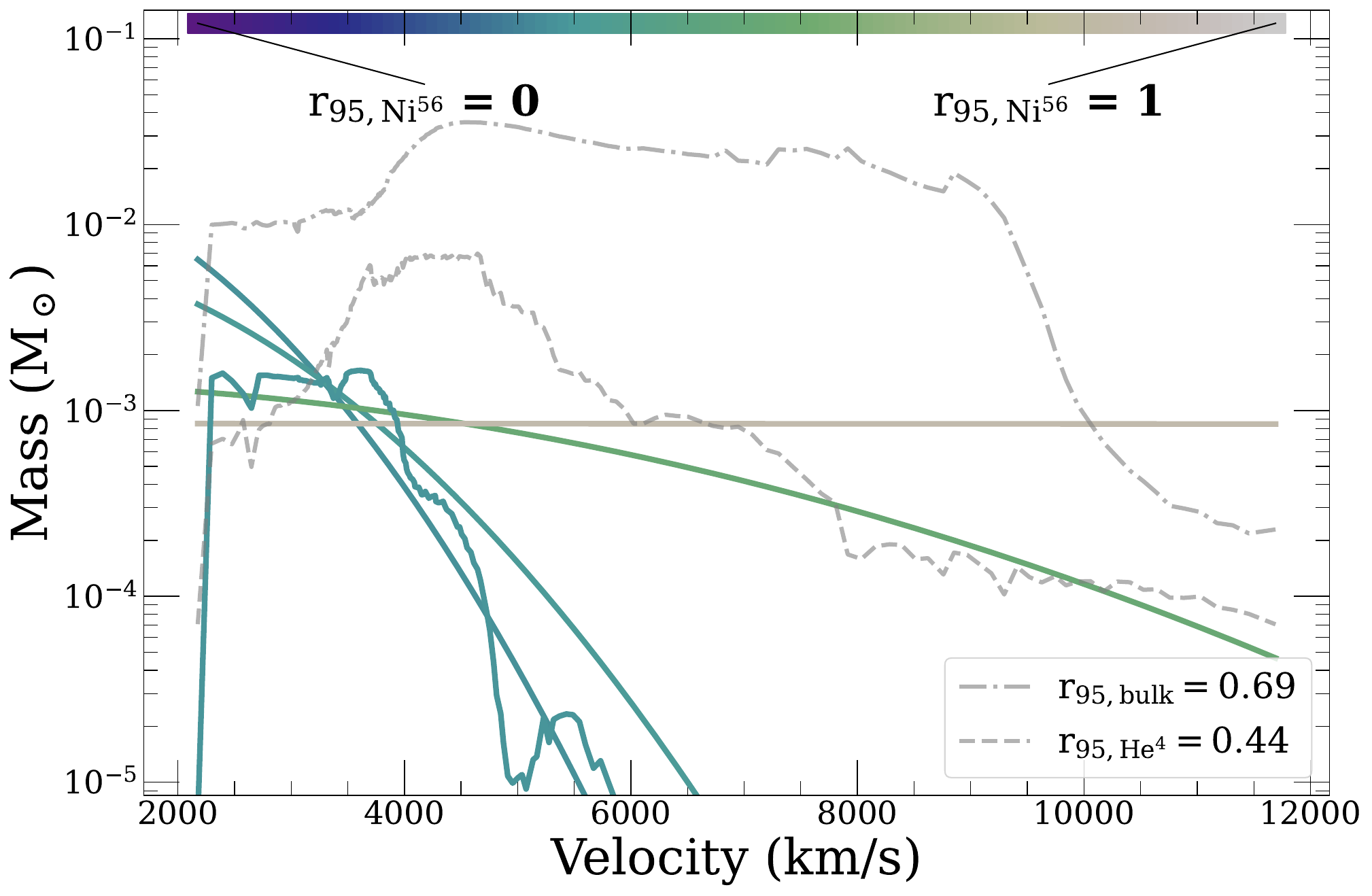}
    \includegraphics[width=0.49\linewidth]{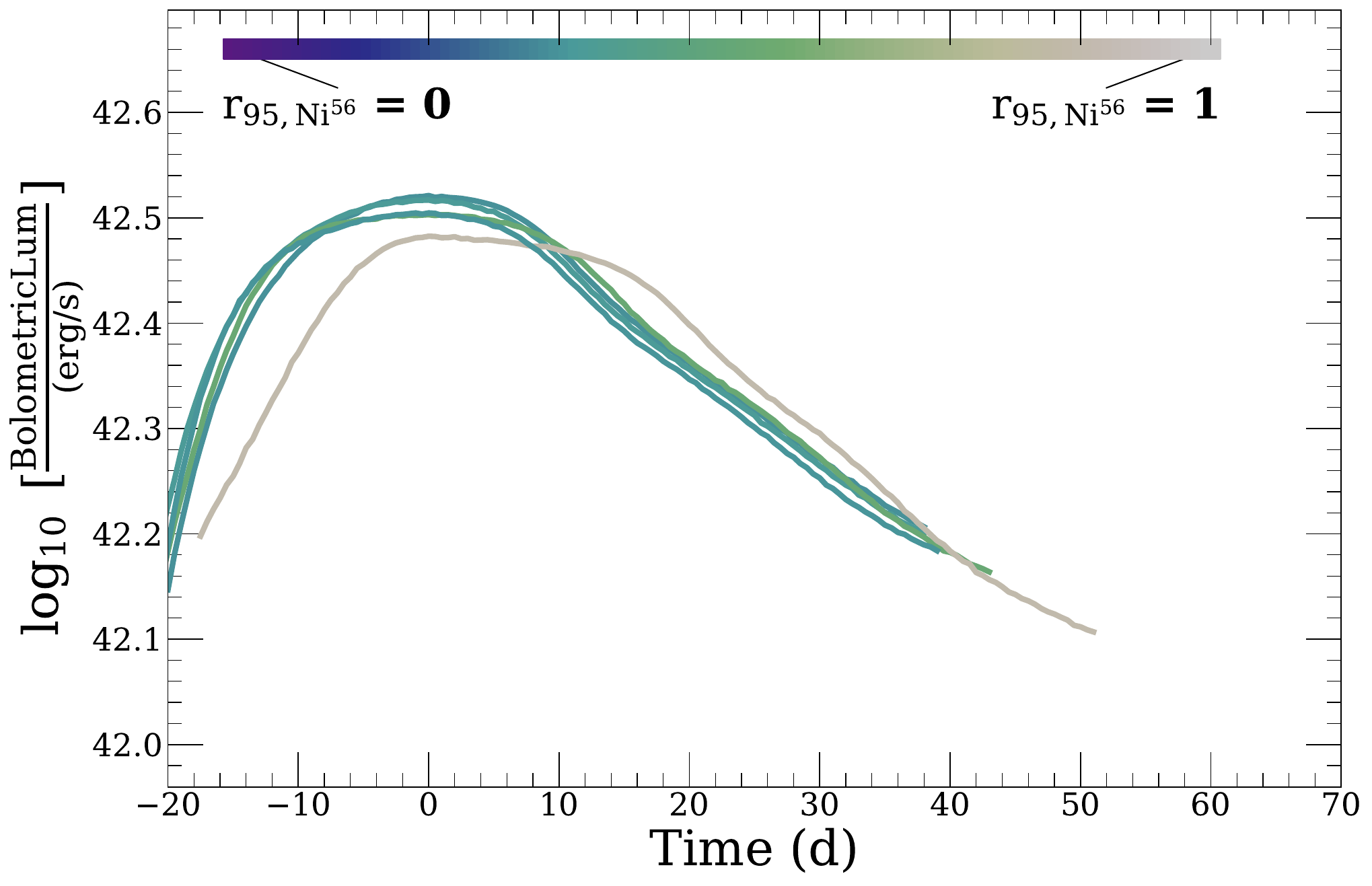}
    \caption{Two SESN ejecta models in which (top left) the \opacity~distribution is sufficiently narrow such that \NI~can be mixed out beyond the \opacity~distribution vs (bottom left) a \opacity~distribution sufficiently wide that even extreme mixing \NI~does not allow it to surpass the \opacity~mass extent. When \NI~can be mixed beyond the \opacity~mass distribution, mixing correlates strongly and positively with peak luminosity (top right). In contrast, if \NI~is contained within the \opacity~mass distribution, the peak luminosity is largely unchanged (bottom right).}
    \label{fig:nickel_mixing_plot}
\end{figure*}

In Figure~\ref{fig:nickel_mixing_plot}, we present two representative models from our extended grid demonstrating each of these scenarios. The top row highlights a model in which the \gamnn~is larger than \gamnop~for all of the \NI~distributions generated. In this model, as \gamnn~is increased, the peak luminosity increases by almost 55\%, while the peak time is largely unchanged. The row of Figure~\ref{fig:nickel_mixing_plot} show a model where \gamnn~is smaller than \gamnop~for all but the most extreme mixing. In this model, mixing \NI~into outer regions of the ejecta produces little change in the peak luminosity, but it does reduce the time from explosion to peak. 

Highly mixed ejecta profiles, where much of the \NI~is beyond the bulk mass, are not usually produced in studies where ejecta profiles are constructed from grids of evolved stars, such as in \cite{dessart2012, dessart2015, dessart2016, woosley2019, ertl2020, Woosley2021}. In those works, mixing within the ejecta profile is emulated by convolving the ejecta profile with a boxcar function, where all of the elements are mixed simultaneously. Because of this, the \NI, which is concentrated in the core of the pre-mixed ejecta, is unlikely to be mixed to be beyond the rest of the ejecta material. However, ejecta profiles with such highly mixed \NI~can be generated if mixed \NI~distributions are analytically generated from an equation (e.g. a Gaussian distribution), as in \cite{Khatami2019} and in our extended grid.

In \cite{Khatami2019}, different \NI~distributions are generated by mixing \NI~outwards to an arbitrary radius. \cite{Khatami2019} explore a broad range of \NI~mixing from strongly centrally located \NI~to \NI~that has been mixed all the way to the outer ejecta profile--well beyond typical SESN models. Just as in the grid we present in this section, the \NI~distribution is allowed to vary independently of the other elements in the ejecta profile. Because of this, \cite{Khatami2019} is able to get ejecta profiles where \NI~can be beyond much of the other elements in the ejecta. As such, \cite{Khatami2019} find a strong dependence of the peak bolometric luminosity on the extent of \NI~mixing, as in the top row of Figure~\ref{fig:nickel_mixing_plot}. Here, we have shown that this strong dependency is not found if \NI~is never mixed to be substantially beyond the other ejecta elements.

\section{Conclusions}
\label{sec:conclusions}

In this paper we have explored how the physical ejecta properties of hydrogen-free SNe, powered only by the radioactive decay of \NI, impact their resulting \lcs. We use \sed, a radiative transfer code, to evolve one-dimensional SN ejecta to generate time-dependent spectral energy distributions and multiband light curves. Our study builds on the set of 81 SESN models from W21 that resulted from evolving helium stars from the zero-age main sequence until core collapse. From the W21 SESN ejecta profiles, we construct a grid of $1$,$000$ ejecta profiles. We explicitly break the correlations that physical parameters of ejecta to isolate how each physical parameter independently influences the \lc. We list our key findings and contributions here:

\begin{itemize}
    \item We parameterize the velocity and mass distributions of \he, \NI, and remaining elements using 256 discrete layers. We find specifying only these four distributions is sufficient to reproduce SESN ejecta profiles and multiband light curves presented originally in W21. To reduce dimensionality of our problem, we use autoencoders (a class of neural networks) to compress these distributions into a nine-dimensional parameter space consisting of the inner velocity, outer velocity, \mhe, \mni, \mop, and four latent variables that describe the shapes of the velocity, \he, \NI, and \opacity~mass distributions. Light curves generated using this reduced representation match the those from the full parameterization (which uses $\sim10^4$ numbers) to within $\sim1$\% accuracy.
    \item We systematically explore how each physical parameter influences the resulting SESN light curves by breaking physical correlations in the W21 grid. This allows us to isolate how ejecta mass, \mni, and the velocity structure shape the bolometric and multiband light curve evolution. Broadly, we recover trends predicted by Arnett-like models. However, we also find that rise and decline times in redder filters correlate most strongly with outer ejecta velocities, while bluer filters are more sensitive to inner layers. We find that many parameters exhibit degenerate effects and interact in complex ways, highlighting the need for robust inference frameworks with radiative transfer models.

    \item We generate a grid of $1$,$000$ \lcs~by uniformly sampling the nine physical parameters. Within this grid, we identify \lcs~that closely match multiband observations of SN1994I, SN2007gr, and iPTF13bvn, suggesting that the grid is able to generate realistic \lcs. The corresponding models also have physical parameters consistent with those inferred by detailed studies of these SESNe. While some SESNe certainly require additional power sources, our results demonstrate that a broader range of SESN diversity than previously appreciated may be explainable through 56Ni-powered explosions.

    \item To further investigate the role of mixing, we present an additional grid of $800$ simulations in which the \NI~distribution is varied while holding all other ejecta properties fixed. We find that \NI~mixing has minimal impact on the peak luminosity if the \NI~remains largely interior to the bulk of the ejecta mass. However, when \NI~is mixed beyond the bulk mass, it can significantly boost peak luminosity. Multiband lightcurves are more sensitive to modest \NI~mixing than the bolometric light curve. Our findings underscore that it is the \textit{relative} distribution of \NI~with respect to the total mass (not mixing alone) that drives changes to the light curve. Our results further highlight the importance of flexible, independent control over individual element distribution when interpreting SN light curves.

    \item We demonstrate that simple properties of SESN light curves, such as peak luminosity and rise/decline timescales, can be used to infer key physical parameters of the ejecta. Using our large grid of models, we perform simple regression of the ejecta properties from the light curve. We find these properties can be used to constrain \mni, \mop, extent to which \NI~is mixed, and the velocity profile of SESN ejecta profiles.

\end{itemize}
We have shown that detailed physical inference from multiband SESN light curves is possible, even when using simple observable parameters. With the rapidly increasing discovery rate of Type Ibc SNe thanks to surveys like the Vera C Rubin Observatory's Legacy Survey of Space and Time, fast methodologies for extracting physical inference from broadband light curves are essential. The grid of radiative transfer-based models presented here offers a foundation for such efforts. In the next paper of this series, we will present a more sophisticated inference framework that utilizes the complete multiband light curves. By applying it to a statistical sample of SESNe, we aim to constrain the distribution of \NI~masses across the SESN population, explore systematic differences between Type Ib and Ic explosions, characterize the diversity of their ejecta profiles, and assess how often additional power sources are necessary to explain SESN \lcs.

\section*{Acknowledgments}
The authors of this work are grateful to David Khatami and Dan Kasen for many initial discussions about MCRT and about setting up \sed. V.A.V. and A.P. acknowledge the Scialog Early Science with the LSST program, which allowed them to begin a new collaboration. V.A.V. acknowledges support through the David and Lucile Packard Foundation, the Research Corporation for Scientific Advancement (through a Cottrell Fellowship), the National Science Foundation under AST-2433718, AST-2407922 and AST-2406110, as well as an Aramont Fellowship for Emerging Science Research. This work is supported by the National Science Foundation under Cooperative Agreement PHY-2019786
(The NSF AI Institute for Artificial Intelligence and
Fundamental Interactions, http://iaifi.org/). M.R.D. acknowledges support from the NSERC through grant RGPIN-2019-06186, the Canada Research Chairs Program, and the Dunlap Institute at the University of Toronto.

\software{\texttt{numpy} \citep{numpy_cite}, \texttt{astropy}, \citep{astropy:2013, astropy:2018, astropy:2022}, \texttt{matplotlib} \citep{matplotlib}, \texttt{scipy} \citep{2020SciPy}, \sed~\citep{Kasen2006}}

\bibliography{sample631}{}
\bibliographystyle{aasjournal}

\end{document}